\documentclass[a4paper,12pt]{article}

\usepackage{amsmath}
\usepackage{amssymb}
\usepackage[dvips]{graphicx}
\usepackage[dvips]{psfrag}

\newcommand{\id}{{1\!\!1}} 
\newcommand {\be}{\begin{equation}}
\newcommand {\ee}{\end{equation}}
\newcommand {\bea}{\begin{eqnarray}}
\newcommand {\eea}{\end{eqnarray}}
\newcommand {\nn}{\nonumber}
\newcommand {\tr}{{\rm tr}}
\newcommand {\Tr}{\mbox{Tr}}

\newcommand {\Det}{\mbox{Det}}

\newcommand {\dd}{\mbox{d}}
\newcommand {\der}{\partial}

\newcommand{\cN}{{\cal N}}
\newcommand{\cO}{{\cal O}}
\newcommand{\cZ}{{\cal Z}}
\newcommand{\cG}{{\cal G}}
\newcommand{\cW}{{\cal W}}

\newcommand{\Slash}[1]{{\ooalign{\hfil/\hfil\crcr$#1$}}}
\newcommand{\thth}[1]{\left. #1 \right|_{\theta\theta}}
\newcommand{\tbtb}[1]{\left. #1 \right|_{\bar{\theta}\bar{\theta}}}
\newcommand{\tttt}[1]{\left. #1 \right|_{\theta\theta\bar{\theta}\bar{\theta}}}
\newcommand{\cD}{{\cal D}}
\newcommand{\twm}{\widetilde{m}}
\newcommand{\whPhi}{\widehat{\Phi}}

\newcommand{\wt}{\widetilde}
\newcommand{\hD}{\widehat{D}}
\newcommand{\hgamma}{\widehat{\gamma}}
\newcommand{\hP}{\widehat{P}}
\newcommand {\bA}{{\tt A}}

\newcommand{\e}{{\rm e}}
\newcommand{\hF}{\widehat{F}}

\begin{document}
\thispagestyle{empty} \addtocounter{page}{-1}
\begin{flushright}
UT-Komaba/08-18\\
OIQP-08-11\\
%
\end{flushright} 
\vspace*{1cm}

\begin{center}
{\large \bf  Ginsparg-Wilson Formulation of 2D $\cN =(2,2)$ SQCD}
\vskip0.4cm
{\large \bf with Exact Lattice Supersymmetry} \\
\vspace*{2cm}
Yoshio Kikukawa$^*$ and Fumihiko Sugino$^\dagger$\\
\vskip0.7cm
{}$^*${\it Institute of Physics, University of Tokyo, }\\
\vspace*{1mm}
{\it Komaba, Meguro-ku, Tokyo 153-8902, Japan}\\
\vspace*{0.2cm}
{\tt kikukawa@hep1.c.u-tokyo.ac.jp}\\
\vskip0.4cm
{}$^\dagger${\it Okayama Institute for Quantum Physics, } \\
\vspace*{1mm}
{\it Kyoyama 1-9-1, Okayama 700-0015, Japan}\\
\vspace*{0.2cm}
{\tt fumihiko\_sugino@pref.okayama.lg.jp}\\
\end{center}
\vskip2cm
\centerline{\bf Abstract}
\vspace*{0.3cm}
{\small 
In this paper, we introduce the overlap Dirac operator, which satisfies the Ginsparg-Wilson relation, to the 
matter sector of two-dimensional $\cN=(2,2)$ lattice supersymmetric QCD (SQCD) with preserving one of the supercharges. 
It realizes the exact chiral flavor symmetry on the lattice, to make possible to define the lattice action 
for general number of the flavors of fundamental and anti-fundamental matter multiplets and for general twisted masses. 
Furthermore, superpotential terms can be introduced with exact holomorphic or anti-holomorphic structure on the lattice. 
We also consider the lattice formulation of matter multiplets charged only under the central ${\rm U}(1)$ (the overall ${\rm U}(1)$) 
of the gauge group $G={\rm U}(N)$, and then construct lattice models for gauged linear sigma models with 
exactly preserving one supercharge and their chiral flavor symmetry.

}
\vspace*{1.1cm}



\newpage

\section{Introduction}
Lattice formulations for supersymmetric field theories, which preserve a part of their supersymmetry, 
have been constructed -- in~\cite{sakai-sakamoto,catterall,catterall4,kikukawa-nakayama} for two-dimensional Wess-Zumino models 
or sigma models without gauge symmetry, 
in~\cite{kaplan,NRS,harada-pinsky,sugino,sugino2,sugino3,sugino4,catterall2,ohta-takimi,tsuchiya,unsal} 
for pure super Yang-Mills (SYM) theories, 
and in~\cite{endre-kaplan,matsuura,sugino_sqcd} for two-dimensional SYM theories coupled with matter 
multiplets\footnote{ Lattice models in~\cite{kaplan,NRS,ohta-takimi,tsuchiya,unsal,endre-kaplan,matsuura} are constructed 
by deconstruction starting from the corresponding matrix models, 
while models in~\cite{sugino,sugino2,sugino3,sugino4,catterall2,sugino_sqcd} are latticized preserving the exact form with respect to 
the twisted supercharges (topological field theory form)~\cite{witten}. 
Relations among these constructions are discussed in~\cite{unsal-takimi,damgaard-m}. 
For a recent review, see~\cite{giedt}. Also, for difficulty on realizing full supersymmetry at the lattice level, 
see for example \cite{kato-ss,bergner}. }. 
For the remaining supercharges, which are broken by the latticization, 
ref.~\cite{kanamori-suzuki} has served a clear numerical evidence of restoration of full supersymmetry in the continuum limit 
in the two-dimensional $\cN=(2,2)$ lattice SYM model~\cite{sugino2}.  
 
On the other hand, there has been a great progress in lattice theories that makes possible exact realization of the chiral symmetry 
on the lattice~\cite{luscher2}, 
since the overlap Dirac operator satisfying the Ginsparg-Wilson relation~\cite{ginsparg-wilson} was 
found~\cite{hasenfratz,neuberger,neuberger-yk}.
Recent results obtained through dynamical simulations of lattice QCD 
with overlap fermions are impressive \cite{Fukaya:2007fb,Fukaya:2007yv,Aoki:2008tq}.
Thus, it will be desirable to apply the Ginsparg-Wilson formulation to the supersymmetric lattice models, 
so that both of the supersymmetry and the chiral structure is preserved on the lattice. 
Although it has been done in~\cite{kikukawa-nakayama} for two-dimensional Wess-Zumino models, 
there seems no literature for supersymmetric theories with gauge symmetry to our knowledge. 
In this paper, we will do it for two-dimensional supersymmetric QCD (SQCD) with $n_+$ fundamental and $n_-$ anti-fundamental 
matter multiplets of the gauge group $G={\rm U}(N)$ or ${\rm SU}(N)$. 

As discussed in~\cite{kikukawa-nakayama}, thanks to the exact chiral symmetry, superpotential terms introduced to the models 
can preserve holomorphic or anti-holomorphic structure on the lattice. 
It helps much to decrease the number of parameters to be fine-tuned in the continuum limit. 
In ref.~\cite{sugino_sqcd}, a lattice model for two-dimensional SQCD was constructed 
with one of the supercharges $Q$ preserved. 
The Wilson terms were introduced there to suppress spices doublers in the matter multiplets. 
Since the Wilson terms break the chiral (flavor) structure, 
the lattice action was defined for the restricted cases of $n_+=n_-$ and for 
the holomorphic twisted masses of fundamental matters equal to those of anti-fundamental matters. 
Here, by using the Ginsparg-Wilson formulation, the chiral flavor symmetry is exactly realized and the lattice models can be 
constructed for general $n_\pm$ and general twisted masses. 
Furthermore, superpotentials can be introduced with the exact holomorphic or anti-holomorphic structure preserved 
on the lattice. 

This paper is organized as follows. 
In the next section, we rewrite the result of~\cite{sugino_sqcd} in a doublet notation, which is convenient to introduce 
the overlap Dirac operator $\hD$ satisfying the Ginsparg-Wilson relation. In section~\ref{sec:lat_sqcd_hD}, 
by using $\hD$, we present two 
lattice formulations (formulations I and II) of the SQCD with $Q$ supersymmetry and the chiral flavor symmetry preserved. 
There are two choices of the chiral projectors dependent on $\hD$: $\hP_\pm$ and $\bar{P}_\pm$. 
The former is used in formulation I, and the latter in formulation II. 
In the followings, we will focus on formulation II, because it gives a simpler expression for the lattice action than 
formulation I. 
Also, we define the path-integral measure, which is gauge and $Q$-supersymmetry invariant and has a desirable property 
under various symmetry transformations.  
In section~\ref{sec:mat_det}, we consider the lattice formulation for matter multiplets in the $\det^q$-representation, 
which are charged only under the central ${\rm U}(1)$ (the overall ${\rm U}(1)$) of $G={\rm U}(N)$. 
Combining the results in sections~\ref{sec:lat_sqcd_hD} and \ref{sec:mat_det}, we construct lattice models of gauged linear 
sigma models, where the fundamental matters and the $\det^q$-matters couple to the SYM sector, in section~\ref{sec:GLS}. 
The summary of the results obtained so far and discussions on future subjects are presented in section~\ref{sec:summary}. 
It seems quite nontrivial to realize the chiral symmetry of the SYM sector\footnote{Based on 
dimensional reduction from four dimensions, a lattice model of two-dimensional ${\cal N}=(2,2)$ SYM with compact Higgs scalars is 
constructed in~\cite{suzuki-taniguchi}. 
There, although the supersymmetry is not exact on the lattice, 
the chiral symmetry is realized by introducing the overlap Dirac operator.} with preserving $Q$. 
Appendix~\ref{app:pauli} serves to explain the cancellation of the ``Pauli terms'' in the lattice SQCD action, 
and appendix~\ref{app:adm_det} to derive the admissibility condition for the $\det^q$-matters. 
In appendix~\ref{app:tensor}, we argue on an attempt to introduce the overlap Dirac operator $\hD$ to the SYM sector.        

Throughout this paper, we focus on the gauge group $G={\rm U}(N)$ or ${\rm SU}(N)$. 
The lattice we work with is the two-dimensional regular lattice with the spacing $a$, and 
the lattice sites are labelled by $x \in {\bf Z}^2$. 
The gauge field $A_{\mu}(x)$ is promoted to the variable $U_\mu(x) = \e^{iaA_{\mu}(x)}$ on the link $(x, x+\hat{\mu})$.
All the other fields are distributed on lattice sites. 

\section{2D ${\cal N}= (2,2)$ SQCD in Doublet Notation} 
\label{sec:2DSQCD_db}
\setcounter{equation}{0}
$\cN =(2,2)$ SQCD in two dimensions is derived from four-dimensional $\cN =1$ SQCD 
by dimensional reduction. 
The field contents are the dimensional reduction of a four-dimensional vector multiplet $V$, 
$n_+$ chiral supermultiplets belonging to the fundamental representation 
$\Phi_{+I} =(\phi_{+I}, \psi_{+I}, F_{+I})$ $(I=1, \cdots, n_+)$, and 
$n_-$ chiral supermultiplets belonging to the anti-fundamental representation 
$\Phi_{-I'} =(\phi_{-I'}, \psi_{-I'}, F_{-I'})$ $(I' =1, \cdots, n_-)$. 
After the dimensional reduction, $V$ contains the gauge fields $A_{\mu}$, the adjoint Higgs scalars $\phi, \bar{\phi}$, 
the gaugino fields $\lambda, \bar{\lambda}$, and the auxiliary field $D$.   
In the presence of twisted masses $\twm_{+I}, \twm_{-I'}$, the action in Euclidean two dimensions is written as 
\bea
S^{(E)}_{\rm 2DSQCD} & = & S^{(E)}_{\rm 2DSYM} + S^{(E)}_{{\rm mat}, +\twm} + S^{(E)}_{{\rm mat}, -\twm}, 
\label{2dSQCD_action}\\
S^{(E)}_{\rm 2DSYM} & = & \frac{1}{g^2} \int \dd^2x \, \tr \left(\frac12 F_{\mu\nu}F_{\mu\nu} 
                                +\cD_\mu\phi\cD_\mu\bar{\phi} +\frac14[\phi, \bar{\phi}]^2 -D^2 \right. \nn \\
 & & \left. \hspace{2.5cm} +4\bar{\lambda}_R\cD_z\lambda_R + 4\bar{\lambda}_L\cD_{\bar{z}}\lambda_L 
                      +2\bar{\lambda}_R[\bar{\phi}, \lambda_L] +2\bar{\lambda}_L [\phi, \lambda_R] \frac{}{} \right), 
\label{SE_2DSYM} 
\eea
\bea
S^{(E)}_{{\rm mat}, +\twm} & = & \int \dd^2x \sum_{I=1}^{n_+}\left[\cD_\mu\phi_{+I}^\dagger\cD_\mu\phi_{+I} 
+\frac12\phi_{+I}^\dagger\{\phi -\twm_{+I}, \bar{\phi} -\twm_{+I}^*\}\phi_{+I} \right.  \nn \\
 & & \hspace{1.5cm}-F_{+I}^\dagger F_{+I} -\phi_{+I}^\dagger D\phi_{+I} 
                      +2\bar{\psi}_{+IR}\cD_z\psi_{+IR} + 2\bar{\psi}_{+IL}\cD_{\bar{z}}\psi_{+IL} \nn \\
 & & \hspace{1.5cm} +\bar{\psi}_{+IL} \left( \phi -\twm_{+I}\right) \psi_{+IR} 
                               +\bar{\psi}_{+IR}  \left(\bar{\phi} -\twm_{+I}^*\right) \psi_{+IL} \nn \\
 & & \hspace{1.5cm} \left. -i\sqrt{2}\left(\phi_{+I}^\dagger(\lambda_L\psi_{+IR}-\lambda_R\psi_{+IL}) 
 +(-\bar{\psi}_{+IR}\bar{\lambda}_L+\bar{\psi}_{+IL}\bar{\lambda}_R)\phi_{+I}\right)\right], \nn \\
 & & 
\label{SE_mat+twm}
\\
S^{(E)}_{{\rm mat}, -\twm} & = &  \int \dd^2x
\sum_{I'=1}^{n_-}\left[\cD_\mu\phi_{-I'}\cD_\mu\phi_{-I'}^\dagger 
+\frac12\phi_{-I'}\{\phi -\twm_{-I'}, \bar{\phi} -\twm_{-I'}^* \}\phi_{-I'}^\dagger \right. \nn \\
 & & \hspace{1.5cm} -F_{-I'}F_{-I'}^\dagger +\phi_{-I'} D\phi_{-I'}^\dagger 
                            +2\psi_{-I'R}\cD_z\bar{\psi}_{-I'R} +2\psi_{-I'L}\cD_{\bar{z}}\bar{\psi}_{-I'L}  \nn \\
 & & \hspace{1.5cm}  +\psi_{-I'R} \left( \phi -\twm_{-I'}\right) \bar{\psi}_{-I'L}
                                 +\psi_{-I'L} \left( \bar{\phi} -\twm_{-I'}^*\right) \bar{\psi}_{-I'R} \nn \\
 & & \hspace{1.5cm} \left. -i\sqrt{2}\left((-\psi_{-I'L}\lambda_R +\psi_{-I'R}\lambda_L)\phi_{-I'}^\dagger 
 +\phi_{-I'} (\bar{\lambda}_R\bar{\psi}_{-I'L} -\bar{\lambda}_L\bar{\psi}_{-I'R})\right) \right].  \nn \\
 & & 
\label{SE_mat-twm} 
\eea
Here, $\cD_z = \frac12(\cD_0-i\cD_1)$, $\cD_{\bar{z}}=\frac12 (\cD_0+i\cD_1)$ are covariant derivatives of the gauge field 
$A_\mu$, $F_{\mu\nu}$ is the field strength tensor $F_{\mu\nu}=\der_\mu A_\nu -\der_\nu A_\mu +i[A_\mu, A_\nu]$, 
and the spinor indices $R$, $L$ are used instead of 1, 2, respectively. 
In the presence of general twisted masses, ${\rm U}(n_+) \times {\rm U}(n_-)$ flavor symmetry breaks down to 
${\rm U}(1)^{n_+} \times {\rm U}(1)^{n_-}$. 

We denote as $Q$ one of the supersymmetry of the action\footnote{$Q$ is given by 
a sum of the supercharges $Q_L$ and $\bar{Q}_R$: $Q \equiv -(Q_L +\bar{Q}_R)/\sqrt{2}$. 
For the full supersymmetry transformation, 
for example see appendix A in ref.~\cite{sugino_sqcd}. }, which will be preserved on the lattice, 
\bea 
 & & QA_\mu =\psi_\mu, \qquad Q\psi_\mu = i\cD_\mu\phi, \nn \\
 & & Q\phi =0, \nn \\
 & & Q\bar{\phi} =\eta, \qquad Q\eta = [\phi, \bar{\phi}], \nn \\
 & & Q\chi = iD +iF_{01}, \qquad QD = -QF_{01} -i[\phi, \chi], 
\label{Q_cont_SYM} 
\eea
\bea
 & & Q\phi_{+I} = -\psi_{+IL}, \qquad Q\psi_{+IL} = -(\phi-\twm_{+I})\phi_{+I}, \nn \\
 & & Q\psi_{+IR} = (\cD_0 +i\cD_1)\phi_{+I} + F_{+I}, \nn \\
 & & QF_{+I} = (\cD_0 +i\cD_1)\psi_{+IL} +(\phi-\twm_{+I})\psi_{+IR} -i(\psi_0+i\psi_1)\phi_{+I}, \nn \\
 & & Q\phi_{+I}^\dagger = -\bar{\psi}_{+IR}, \qquad Q\bar{\psi}_{+IR} = \phi_{+I}^\dagger (\phi-\twm_{+I}), \nn \\
 & & Q\bar{\psi}_{+IL} = (\cD_0-i\cD_1)\phi_{+I}^\dagger + F_{+I}^\dagger, \nn \\
 & & QF_{+I}^\dagger = (\cD_0-i\cD_1)\bar{\psi}_{+IR} -\bar{\psi}_{+IL}(\phi-\twm_{+I}) +i\phi_{+I}^\dagger (\psi_0-i\psi_1), 
\label{Qsusy_cont_mat+_twm}
\eea
\bea 
 & & Q\phi_{-I'} = -\psi_{-I'L}, \qquad Q\psi_{-I'L} = -\phi_{-I'}(\phi-\twm_{-I'}), \nn \\
 & & Q\psi_{-I'R} = (\cD_0+i\cD_1)\phi_{-I'} +F_{-I'}, \nn \\
 & & QF_{-I'} = (\cD_0+i\cD_1)\psi_{-I'L} -\psi_{-I'R}(\phi-\twm_{-I'}) +i\phi_{-I'}(\psi_0+i\psi_1), \nn \\
 & & Q\phi_{-I'}^\dagger = -\bar{\psi}_{-I'R}, \qquad Q\bar{\psi}_{-I'R} =-(\phi-\twm_{-I'})\phi_{-I'}^\dagger, \nn \\
 & & Q\bar{\psi}_{-I'L} = (\cD_0 -i\cD_1)\phi_{-I'}^\dagger + F_{-I'}^\dagger, \nn \\
 & & QF_{-I'}^\dagger = (\cD_0 -i\cD_1)\bar{\psi}_{-I'R} + (\phi-\twm_{-I'})\bar{\psi}_{-I'L} -i(\psi_0-i\psi_1)\phi_{-I'}^\dagger, 
\label{Qsusy_cont_mat-_twm}
\eea
where we renamed the gaugino fields as 
\bea
& & \psi_0 \equiv \frac{1}{\sqrt{2}}(\lambda_L +\bar{\lambda}_R), \qquad \psi_1 \equiv \frac{i}{\sqrt{2}}(\lambda_L-\bar{\lambda}_R), 
\nn \\
& & \chi \equiv \frac{1}{\sqrt{2}}(\lambda_R-\bar{\lambda}_L), \qquad \eta \equiv -i\sqrt{2}(\lambda_R +\bar{\lambda}_L).  
\label{rename}
\eea
$Q$ is nilpotent up to the combination of the infinitesimal gauge transformation with the (complexified) parameter $\phi$ and 
the infinitesimal flavor rotations with the (complexified) parameters $\twm_{+I}, \twm_{-I'}$ acting as 
\bea
\delta\Phi_{+I} = -\twm_{+I}\Phi_{+I}, &  & \delta\Phi_{+I}^\dagger = \twm_{+I}\Phi_{+I}^\dagger, \nn \\
\delta\Phi_{-I'} = \twm_{-I'}\Phi_{-I'}, & & \delta\Phi_{-I'}^\dagger =-\twm_{-I'}\Phi_{-I'}^\dagger. 
\eea

\subsection{Continuum Theory}
To apply the Ginsparg-Wilson formulation to the SQCD system, we will introduce a doublet notation to the matter multiplets. 
First, we add some matter multiplets to the given contents to prepare the same number of the fundamental and anti-fundamental fields 
($n_0 \equiv \max (n_+, n_-)$), and combine them as the doublets: 
\bea
 & & \Phi_I \equiv \left(\begin{array}{c} \phi_{+I} \\ \phi_{-I}^\dagger \end{array} \right), \qquad  
           \Phi_I^\dagger \equiv \left(\phi_{+I}^\dagger, \phi_{-I}\right), \nn \\
 & &  \Psi_{uI} \equiv \left(\begin{array}{c} \psi_{+IL} \\ \bar{\psi}_{-IR} \end{array}\right), \qquad 
           \Psi_{dI} \equiv \left(\begin{array}{c} \bar{\psi}_{-IL} \\ \psi_{+IR} \end{array} \right),  \nn \\
 & &  \Psi_{uI}^\dagger \equiv \left( \bar{\psi}_{+IL}, \psi_{-IR} \right), \qquad 
           \Psi_{dI}^\dagger \equiv \left( \psi_{-IL}, \bar{\psi}_{+IR} \right), \nn \\
 & &  F_I \equiv \left(\begin{array}{c} F_{+I} \\ F_{-I}^\dagger \end{array} \right), \qquad 
            F_I^\dagger \equiv \left( F_{+I}^\dagger, F_{-I}\right) \qquad (I =1, \cdots, n_0). 
\eea
The upper and down components of each doublet have the same gauge transformation property. 
We define the $\gamma$-matrices in terms of the Pauli matrices as 
\be
\gamma_0 \equiv \sigma_1, \qquad \gamma_1 \equiv \sigma_2, \qquad \gamma_3 \equiv -i\gamma_0\gamma_1 = \sigma_3, 
\ee
and use the notation 
\be
\bar{\Psi}_{uI} \equiv \Psi_{uI}^\dagger \gamma_0, \qquad 
\bar{\Psi}_{dI} \equiv \Psi_{dI}^\dagger \gamma_0. 
\ee

The fundamental or anti-fundamental degrees of freedom in (\ref{SE_mat+twm}) and (\ref{SE_mat-twm}) are extracted by acting 
the chiral projectors $P_\pm = \frac12 (1\pm \gamma_3)$ to the doublets.  
The matter-part actions $S^{(E)}_{{\rm mat}, \pm\twm}$ are rewritten as 
\bea
S^{(E)}_{{\rm mat}, +\twm} & = & \int \dd^2x \sum_{I=1}^{n_+}\left[-\Phi_I^\dagger P_+\cD_\mu\cD_\mu P_+\Phi_I  
+\frac12\Phi_I^\dagger P_+ \{\phi -\twm_{+I}, \bar{\phi} -\twm_{+I}^*\} P_+\Phi_I \right.  \nn \\
 & & \hspace{1.5cm} -F_I^\dagger P_+F_I -\Phi_I^\dagger P_+ D P_+\Phi_I 
                      +\bar{\Psi}_{uI}P_- \Slash{\cD} P_+\Psi_{uI} -\bar{\Psi}_{dI} P_+\Slash{\cD}^\dagger P_-\Psi_{dI} \nn \\
 & & \hspace{1.5cm} +\bar{\Psi}_{uI}P_- \left( \phi -\twm_{+I}\right) P_-\Psi_{dI} 
                            +\bar{\Psi}_{dI}P_+ \left(\bar{\phi} -\twm_{+I}^*\right) P_+\psi_{uI} \nn \\
 & & \hspace{1.5cm} -i\bar{\Psi}_{uI} P_- \gamma_\mu\psi_\mu P_+\Phi_I 
                              -i\Phi_I^\dagger P_+ \gamma_\mu\psi_\mu P_-\Psi_{dI} \nn \\
 & & \hspace{1.5cm} \left. -\bar{\Psi}_{dI} P_+ \left(\frac12 \eta + i\chi\right) P_+ \Phi_I 
                             -\Phi_I^\dagger P_+ \left(\frac12 \eta -i\chi\right) P_+\Psi_{uI} \right],  \\
S^{(E)}_{{\rm mat}, -\twm} & = &  \int \dd^2x
\sum_{I'=1}^{n_-}\left[-\Phi_{I'}^\dagger P_-\cD_\mu\cD_\mu P_-\Phi_{I'} 
+\frac12\Phi_{I'}^\dagger P_- \{\phi -\twm_{-I'}, \bar{\phi}-\twm_{-I'}^* \} P_-\Phi_{I'} \right. \nn \\
 & & \hspace{1.5cm} -F_{I'}^\dagger P_- F_{I'} +\Phi_{I'}^\dagger P_- D P_-\Phi_{I'} 
                            +\bar{\Psi}_{uI'} P_+ \Slash{\cD} P_-\Psi_{uI'} +\bar{\Psi}_{dI'} P_- \Slash{\cD}^\dagger P_+{\Psi}_{dI'}  \nn \\
 & & \hspace{1.5cm}  +\bar{\Psi}_{uI'} P_+ \left( \phi -\twm_{-I'}\right) P_+\Psi_{dI'}
                             +\bar{\Psi}_{dI'} P_- \left( \bar{\phi} -\twm_{-I'}^*\right) P_-\Psi_{uI'} \nn \\
 & & \hspace{1.5cm} -i\bar{\Psi}_{uI'} P_+ \gamma_\mu\psi_\mu P_-\Phi_{I'} 
                              -i\Phi_{I'}^\dagger P_- \gamma_\mu\psi_\mu P_+\Psi_{dI'} \nn \\
 & & \hspace{1.5cm} \left. -\bar{\Psi}_{dI'} P_- \left( \frac12\eta -i\chi\right) P_- \Phi_{I'} 
                                   -\Phi_{I'}^\dagger P_- \left(\frac12 \eta +i\chi\right) P_-\Psi_{uI'} \right].   
\eea

The actions $S^{(E)}_{\rm 2DSQCD}$, $S^{(E)}_{{\rm mat}, \pm\twm}$ separately have 
a R-symmetry under the ${\rm U(1)}_V$ transformation 
\bea
 & & \left(\begin{array}{c} \psi_0 \\ \psi_1\end{array}\right) \to 
          \left(\begin{array}{cc} \cos\alpha & \sin\alpha \\ -\sin\alpha & \cos\alpha \end{array}\right)  
                                                      \left(\begin{array}{c} \psi_0 \\ \psi_1\end{array}\right) , \quad 
\left(\begin{array}{c} \chi \\ \frac12 \eta \end{array}\right) \to 
          \left(\begin{array}{cc} \cos\alpha & -\sin\alpha \\ \sin\alpha & \cos\alpha \end{array}\right)  
                                                      \left(\begin{array}{c} \chi \\ \frac12 \eta \end{array}\right), \nn \\
 & & \Psi_{uI} \to \e^{-i\alpha\gamma_3} \, \Psi_{uI}, \qquad 
                        \bar{\Psi}_{uI} \to \bar{\Psi}_{uI} \, \e^{-i\alpha\gamma_3},  \qquad 
                 \Psi_{dI} \to \e^{i\alpha\gamma_3}\,\Psi_{dI}, \qquad \bar{\Psi}_{dI} \to \bar{\Psi}_{dI} \, \e^{i\alpha\gamma_3}, \nn \\
 & & F_I \to \e^{-i2\alpha\gamma_3} \, F_I, \qquad F_I^\dagger \to F_I^\dagger \, \e^{i2\alpha\gamma_3} 
\label{U(1)_V_cont} 
\eea    
with the others unchanged.                                          
In the case of the twisted masses set to zero, the ${\rm U}(1)_A$ transformation 
\bea
 & &  \phi \to \e^{i2\alpha}\, \phi, \qquad \bar{\phi} \to \e^{-i2\alpha} \, \bar{\phi},  \qquad 
         \psi_\mu \to \e^{i\alpha} \, \psi_\mu, \qquad \chi \to \e^{-i\alpha}\, \chi, \qquad \eta \to \e^{-i\alpha}\, \eta,  \nn \\
 & & \Psi_{uI} \to \e^{i\alpha} \, \Psi_{uI}, \qquad \bar{\Psi}_{uI} \to \bar{\Psi}_{uI} \, \e^{-i\alpha}, \qquad 
       \Psi_{dI} \to \e^{-i\alpha} \, \Psi_{dI}, \qquad \bar{\Psi}_{dI} \to \bar{\Psi}_{dI} \, \e^{i\alpha} 
\label{U(1)_A_cont}       
\eea
with the others unchanged also becomes another R-symmetry of the classical actions.        

The $Q$ transformation (\ref{Qsusy_cont_mat+_twm}), (\ref{Qsusy_cont_mat-_twm}) is written as\footnote{ 
$\Phi_I^\dagger \Slash{\cD}^\dagger$ is understood as $\left(\Slash{\cD} \Phi_I\right)^\dagger$.} 
\bea
 & & Q \Phi_I =  -\Psi_{uI}, \qquad 
        Q \Psi_{uI} = -\left(\phi -\twm_{+I} P_+ -\twm_{-I} P_-\right) \Phi_I, \nn \\
 & & Q \Psi_{dI} = \Slash{\cD} \Phi_I + \gamma_0 F_I, \nn \\
 & & Q (\gamma_0 F_I) = \left(\phi -\twm_{+I} P_- -\twm_{-I} P_+\right) \Psi_{dI} + \Slash{\cD} \Psi_{uI} -i\gamma_\mu \psi_\mu \Phi_I, 
         \nn \\
 & & Q \Phi_I^\dagger = -\bar{\Psi}_{dI}, \qquad 
        Q \bar{\Psi}_{dI} = \Phi_I^\dagger \left(\phi -\twm_{+I} P_+ -\twm_{-I} P_-\right), \nn \\
 & & Q \bar{\Psi}_{uI} = \Phi_I^\dagger \Slash{\cD}^\dagger + F_I^\dagger \gamma_0, \nn \\
 & & Q (F_I^\dagger \gamma_0) = -\bar{\Psi}_{uI} \left(\phi -\twm_{+I} P_- -\twm_{-I} P_+\right) + \bar{\Psi}_{dI} \Slash{\cD}^\dagger 
                      + i\Phi_I^\dagger \gamma_\mu \psi_\mu,   
\label{Q_cont_mat}                      
\eea       
and the nilpotency 
\bea
Q^2 & = & (\mbox{infinitesimal gauge transformation with the parameter $\phi$}) \nn \\
      & & + (\mbox{infinitesimal flavor rotations (\ref{flavor_rot})})
\label{cont_nilpotency}      
\eea
holds with 
\bea
\delta \Phi_I = -\left(\twm_{+I} P_+ + \twm_{-I} P_-\right) \Phi_I, & &    
     \delta \Phi_I^\dagger  = \Phi_I^\dagger \left(\twm_{+I} P_+ + \twm_{-I} P_-\right), \nn \\
\delta \Psi_{uI} = -\left(\twm_{+I} P_+ + \twm_{-I} P_-\right) \Psi_{uI}, & & 
     \delta \bar{\Psi}_{uI} = \bar{\Psi}_{uI} \left(\twm_{+I} P_- + \twm_{-I} P_+\right), \nn \\
\delta \Psi_{dI} = -\left(\twm_{+I} P_-  + \twm_{-I} P_+\right) \Psi_{dI}, & & 
     \delta \bar{\Psi}_{dI} = \bar{\Psi}_{dI}\left( \twm_{+I} P_+ + \twm_{-I} P_-\right), \nn \\
\delta F_I = -\left(\twm_{+I} P_+ + \twm_{-I} P_-\right) F_I, & & 
     \delta F_I^\dagger  = F_I^\dagger \left(\twm_{+I} P_+ + \twm_{-I} P_-\right).  
\label{flavor_rot}
\eea
Notice that (\ref{Q_cont_mat}) for each $I$ splits into four irreducible parts consisting of 
\bea
\{ P_+\Phi_I, P_+\Psi_{uI}, P_-\Psi_{dI}, P_+F_I \},  & & \{ \Phi_I^\dagger P_+, \bar{\Psi}_{dI}P_+, \bar{\Psi}_{uI}P_-, F_I^\dagger P_+ \}, 
\nn \\ 
\{ P_-\Phi_I, P_-\Psi_{uI}, P_+\Psi_{dI}, P_-F_I \},  & & \{ \Phi_I^\dagger P_-, \bar{\Psi}_{dI}P_-, \bar{\Psi}_{uI}P_+, F_I^\dagger P_- \}, 
\nn
\eea
respectively. 
We reproduce (\ref{Qsusy_cont_mat+_twm}) and (\ref{Qsusy_cont_mat-_twm}) in terms of the doublets 
with the chiral projections from (\ref{Q_cont_mat}). 

The action can be expressed as the $Q$-exact form:
\bea
S^{(E)}_{\rm 2DSYM} & = & Q\frac{1}{g^2} \int \dd^2x \, \tr \left[-i\chi (F_{01}-D) +\frac14 \eta [\phi, \bar{\phi}] 
-i\psi_\mu\cD_\mu\bar{\phi}\right], 
\label{S_cont_SYM}\\
S^{(E)}_{{\rm mat}, +\twm} & = & Q \int \dd^2x \sum_{I=1}^{n_+} \frac12 \left[ 
        \bar{\Psi}_{uI} P_-\left(\Slash{\cD} P_+ \Phi_I -P_I \gamma_0 F_I\right) 
+\left(\Phi_I^\dagger P_+ \Slash{\cD}^\dagger -F_I^\dagger \gamma_0 P_-\right) P_- \Psi_{dI} \right. \nn \\
  & & \hspace{2.5cm} -\Phi_I^\dagger P_+ \left(\bar{\phi} -\twm_{+I}^*\right) P_+ \Psi_{uI} 
                                 +\bar{\Psi}_{dI} P_+ \left(\bar{\phi} -\twm_{+I}^*\right) P_+ \Phi_I \nn \\
  & & \hspace{2.5cm} \left. +2i \Phi_I^\dagger P_+ \chi P_+ \Phi_I\right], \\
S^{(E)}_{{\rm mat}, -\twm} & = & Q \int \dd^2x \sum_{I'=1}^{n_-} \frac12 \left[   
       \bar{\Psi}_{uI'} P_+\left( \Slash{\cD} P_- \Phi_{I'} -P_+ \gamma_0 F_{I'}\right) 
+\left( \Phi_{I'}^\dagger P_- \Slash{\cD}^\dagger -F_{I'}^\dagger \gamma_0 P_+\right) P_+ \Psi_{dI'} \right. \nn \\
  & & \hspace{2.5cm} -\Phi_{I'}^\dagger P_- \left(\bar{\phi} -\twm_{-I'}^*\right) P_- \Psi_{uI'} 
                              +\bar{\Psi}_{dI'} P_- \left(\bar{\phi} -\twm_{-I'}^*\right) P_- \Phi_{I'} \nn \\
  & & \hspace{2.5cm} \left. -2i \Phi_{I'} P_- \chi P_- \Phi_{I'} \right]. 
\eea 
Since in these formulas $Q$ acts to the gauge invariant expressions with the symmetry under (\ref{flavor_rot}), 
$Q$ invariance of the action is manifestly seen. 

Interaction terms from superpotentials are given by dimensional reduction to the two dimensions of 
\be
\thth{W(\Phi_+, \Phi_-) \frac{}{} } 
+ \tbtb{\bar{W}(\Phi_+^\dagger, \Phi_-^\dagger) \frac{}{} }.   
\ee
We often use the four-dimensional $\cN=1$ superfield notations in~\cite{wess-bagger}.   
It is written as the $Q$ exact form in the doublet notation: 
\bea
S^{(E)}_{\rm pot} & = & Q \int \dd^2x \sum_{i=1}^N\sum_{I=1}^{n_+} 
\left[-\frac{\der W}{\der (P_+\Phi_I)_i} \left(\gamma_0 P_-\Psi_{dI}\right)_i 
            -\left(\bar{\Psi}_{uI} P_- \gamma_0\right)_i\frac{\der\bar{W}}{\der (\Phi_I^\dagger P_+)_i} \right] \nn \\
 & &\hspace{-3mm} +Q \int \dd^2x \sum_{i=1}^N\sum_{I'=1}^{n_-}\left[ 
                      -\frac{\der \bar{W}}{\der (P_-\Phi_{I'})_i} \left(\gamma_0P_+\Psi_{dI'}\right)_i 
           -\left(\bar{\Psi}_{uI'}P_+ \gamma_0\right)_i\frac{\der W}{\der (\Phi_{I'}^\dagger P_-)_i}\right] \nn \\
 & &           
\label{S_cont_pot} 
\eea
with 
\be
W=W(P_+\Phi_I, \Phi_{I'}^\dagger P_-), \qquad \bar{W} =\bar{W} (\Phi_I^\dagger P_+, P_-\Phi_{I'} ). 
\ee 
$(\cdots)_i$ represent independent color degrees of freedom of the projected doublet.   

For the case $G={\rm U}(N)$, the Fayet-Iliopoulos (FI) and $\vartheta$-terms can be introduced to the action:
\bea
S^{(E)}_{{\rm FI}, \, \vartheta} & = & \int \dd^2x \, \tr \left(\kappa D -i\frac{\vartheta}{2\pi} F_{01}\right) \nn \\
 & = &  Q \kappa \int \dd^2x \,\tr \left(-i\chi\right) 
-i \frac{\vartheta-2\pi i\kappa}{2\pi} \int \dd^2x \,\tr \, F_{01}
\label{FI_theta_cont}
\eea
with $\kappa$ being the FI parameter. 
The second term in the r.h.s. is a topological term, and thus $Q$-invariant. 
The first term yields the $\vartheta$-term with the imaginary value $\vartheta =2\pi i \kappa$, that is 
compensated by the second term.

\subsection{Lattice Formulation of SYM Part}
Let us formulate the lattice theory realizing the supersymmetry $Q$. 
The SYM part of the lattice theory is presented in~\cite{sugino2,sugino_sqcd}. 
To summarize, $Q$-supersymmetry can be realized on the lattice as 
 \bea
 & & QU_{\mu}(x) = i\psi_{\mu}(x) U_{\mu}(x), \qquad QU_{\mu}(x)^{-1} = -iU_{\mu}(x)^{-1} \psi_{\mu}(x), \nn \\
 & & Q\psi_{\mu}(x) = i\psi_{\mu}(x)\psi_{\mu}(x) -i\left(\phi(x) -U_{\mu}(x)\phi(x+\hat{\mu}) U_{\mu}(x)^{-1}\right), \nn \\
 & & Q\phi(x) = 0, \nn \\
 & & Q\bar{\phi}(x) = \eta(x), \qquad Q\eta(x) = [\phi(x), \bar{\phi}(x)], \nn \\
 & & Q\chi(x) = iD(x) + \frac{i}{2}\whPhi(x), \qquad 
 QD(x) = -\frac12Q\whPhi(x) -i[\phi(x), \chi(x)]\, ,
\label{Qsusy_lat_SYM} 
\eea
where $\whPhi(x)$ is a lattice counterpart of $2F_{01}(x)$ defined by 
\bea
 & & \Phi(x) = -i(U_{01}(x)-U_{10}(x)), \qquad 
U_{\mu\nu}(x) \equiv U_{\mu}(x)U_{\nu}(x+\hat{\nu})U_{\mu}(x+\hat{\nu})^{-1}U_{\nu}(x)^{-1}, \nn \\
 & & \whPhi(x) \equiv \frac{\Phi(x)}{1-\frac{1}{\epsilon^2}||1-U_{01}(x)||^2}. 
\eea
The norm of an arbitrary $M \times M$ complex matrix $A$ is defined as\footnote{
Notice that the definition of the norm is different by the factor $\frac{1}{\sqrt{M}}$ from that in refs.~\cite{sugino2,sugino_sqcd}.}  $||A||\equiv \sqrt{\frac{1}{M} \, \tr (AA^\dagger)}$, and 
$\epsilon$ is a constant chosen as 
\be
0<\epsilon < \frac{2}{\sqrt{N}}  \qquad \text{for} \quad G={\rm U}(N). 
\label{adm_U(N)}
\ee
In the case $G={\rm SU}(N)$, here and in what follows, $\whPhi(x)$ is understood to be replaced with its traceless part: 
\be
\whPhi_{\rm TL}(x) \equiv \whPhi(x) - \frac{1}{N}\left(\tr \,\whPhi(x)\right)\id_N, 
\ee
and $\epsilon$ is chosen as 
\bea
0 < \epsilon < 2 & & G = {\rm SU}(2), \nn \\
0 < \epsilon < \frac{2\sqrt{2}}{\sqrt{3}} & & G={\rm SU}(3), \nn \\
0 < \epsilon < \sqrt{2} & & G={\rm SU}(4), \nn \\
0 < \epsilon < 2\sin \left(\frac{\pi}{N}\right) & & G={\rm SU}(N) \quad (N\ge 5). 
\label{epsilon_vac}
\eea
The transformation (\ref{Qsusy_lat_SYM}) is defined for the lattice gauge fields satisfying 
the admissibility condition~\cite{HJL,luscher}: 
\be
||1-U_{01}(x)||<\epsilon, 
\label{admissibility}
\ee
and $Q$ is nilpotent up to the infinitesimal gauge transformation with the parameter $\phi(x)$ on the lattice. 

The lattice action can be expressed as the $Q$-exact form: 
\bea
S^{\rm LAT}_{\rm 2DSYM} & = & Q\frac{1}{g_0^2} \sum_x\tr\left[\chi(x)\left(-\frac{i}{2}\whPhi(x) +iD(x)\right) 
+\frac14\eta(x)[\phi(x), \bar{\phi}(x)] \right. \nn \\
 & & \hspace{2cm} \left. +i\sum_{\mu=0}^1\psi_\mu(x)\left(\bar{\phi}(x)-U_\mu(x)\bar{\phi}(x+\hat{\mu})U_\mu(x)^{-1}\right)\right] 
\label{S_lat_SYM1} 
\eea
for the admissible gauge fields satisfying (\ref{admissibility}) for $\forall x$, and 
\be
 S^{\rm LAT}_{\rm 2DSYM} =+\infty \qquad \mbox{otherwises}.
\label{S_lat_SYM2} 
\ee
$Q$ invariance of the action is manifest.

For the case $G={\rm U}(N)$, the FI and topological $\vartheta$-terms can be introduced to the action 
as\footnote{By introducing the Ginsparg-Wilson Dirac operator $\hD$,  
we may also use (\ref{FI_theta_lat_hD}) instead of (\ref{FI_theta_lat}) as discussed in section~\ref{sec:summary}.} : 
\be
S^{\rm LAT}_{{\rm FI}, \, \vartheta} = Q \kappa \sum_x \tr \left(-i\chi(x)\right) 
-\frac{\vartheta-2\pi i\kappa}{2\pi} \sum_x \tr \, \ln U_{01}(x), 
\label{FI_theta_lat}
\ee
where the second term is $Q$-invariant by its topological nature~\cite{sugino_sqcd}. 
In order for the logarithm of the plaquette fields to be well-defined, it is sufficient to choose $\epsilon$ as
\be
0< \epsilon <\frac{1}{\sqrt{N}} \qquad \mbox{for $G={\rm U}(N)$ with $\vartheta$-term}. 
\label{adm_theta}
\ee 

The path-integral measure with respect to the SYM variables is given by  
\bea
\left(\dd\mu_{\rm 2DSYM}\right) & \equiv & \prod_x \left[\prod_{\mu=0}^1 \dd U_{\mu}(x)\right] \nn \\
 & & \times \prod_{A} \dd\psi^{A}_0(x) \, \dd\psi^{A}_1(x) \, \dd \chi^{A}(x) \, \dd\eta^{A}(x) \, \dd\phi^{A}(x) \, 
 \dd\bar{\phi}^{A}(x) \, \dd D^{A}(x), 
\label{measure_SYM}
\eea
where $\dd U_{\mu}(x)$ is the Haar measure of the gauge group $G$, the index $A$ labels the generators of $G$, 
and the lattice fields with the index $A$ represent the expansion coefficients by the generators of $G$: 
\be
(\text{field})(x) = \sum_{A} (\text{field})^{A}(x) \, T^{A}, \qquad 
\tr \left(T^{A} T^{B}\right) = \frac12 \, \delta^{AB}.
\ee

\subsection{Lattice Formulation of Matter Part with Wilson-Dirac Operator}
The construction of the matter action in~\cite{sugino_sqcd} can be rewritten in the doublet notation. 
First, the covariant forward (backward) difference operators $D_\mu$ ($D_\mu^*$) act as  
\bea
aD_\mu\Phi_{I}(x) & = & U_{\mu}(x)\Phi_{I}(x+\hat{\mu}) -\Phi_{I}(x), \nn \\
aD_\mu^*\Phi_{I}(x) & = & \Phi_{I}(x) - U_{\mu}(x-\hat{\mu})^{-1}\Phi_{I}(x-\hat{\mu}) 
\eea
with the same for $\Psi_{uI}, \Psi_{dI}, F_I$. Also, 
\bea
aD_\mu \Phi_{I}(x)^\dagger & = & \Phi_{I}(x+\hat{\mu})^\dagger U_{\mu}(x)^{-1} -\Phi_{I}(x)^\dagger, \nn \\
aD_\mu^*\Phi_{I}(x)^\dagger & = & \Phi_{I}(x)^\dagger -\Phi_{I}(x-\hat{\mu})^\dagger U_{\mu}(x-\hat{\mu}) 
\eea
with the same for $\bar{\Psi}_{uI}, \bar{\Psi}_{dI}, F_I^\dagger$. 
These operations are trivial to the Dirac indices and commute with the $\gamma$-matrices. 

Next, using the Wilson-Dirac operator 
\bea
D_W & \equiv & \sum_{\mu=0}^1 \left(\gamma_\mu D_\mu^S - rD_\mu^A\right) 
        =  \left(\begin{array}{cc} -\sum_{\mu=0}^1 rD_\mu^A &  D_0^S -i D_1^S \\ 
                                                              D_0^S +i D_1^S & -\sum_{\mu=0}^1 rD_\mu^A \end{array} \right), 
\label{D_W}\\
D_W^\dagger & = & \sum_{\mu=0}^1 \left(-\gamma_\mu D_\mu^S - rD_\mu^A\right) =\gamma_3 D_W \gamma_3  
\eea
with   
\be
D_\mu^S \equiv \frac12 (D_\mu + D_\mu^*), \qquad D_\mu^A \equiv \frac12 (D_\mu -D_\mu^*), 
\label{DS_DA}
\ee
the $Q$ supersymmetry (\ref{Q_cont_mat}) can be expressed on the lattice 
as\footnote{We here flip the sign of the Wilson parameter $r \to -r$ 
in ref.~\cite{sugino_sqcd}. Since poles of the lattice propagators depend on $r$ only through the form $r^2$ there 
(eqs.~(4.27)--(4.30) in \cite{sugino_sqcd}), 
the flip of the sign does not change physical results.} 
\bea
Q\Phi_I(x) & = & -\Psi_{uI}(x), \nn \\
Q\Psi_{uI}(x) & = & -\phi(x) \Phi_I(x) +\twm_{+I}P_+\Phi_I(x) + \twm_{-I}P_-\Phi_I(x), \nn \\
Q\Psi_{dI}(x) & = & aD_W \Phi_I(x) + \gamma_0 F_I(x), \nn \\
Q\gamma_0F_I(x) & = & \phi(x)\Psi_{dI}(x) +aD_W\Psi_{uI}(x) 
-\twm_{+I}\gamma_0P_-\Psi_{dI}(x) -\twm_{-I}\gamma_0 P_+\Psi_{dI}(x)\nn \\
 & & -i\sum_{\mu=0}^1\left(\frac{\gamma_\mu -r}{2}\psi_\mu(x)U_\mu(x)\Phi_I(x) \right. \nn \\
 & & \left. \hspace{1.5cm} + \frac{\gamma_\mu +r}{2} U_\mu(x-\hat{\mu})^{-1} \psi_\mu(x-\hat{\mu}) \Phi_I(x-\hat{\mu})\right), \\
Q\Phi_I(x)^\dagger & = & -\bar{\Psi}_{dI}(x), \nn \\
Q\bar{\Psi}_{dI}(x) & = & \Phi_I(x)^\dagger \phi(x) -\twm_{+I}\Phi_I(x)^\dagger P_+ -\twm_{-I}\Phi_I(x)^\dagger P_-, \nn \\
Q\bar{\Psi}_{uI}(x) & = & \Phi_I(x)^\dagger aD_W^\dagger + F_I(x)^\dagger \gamma_0, \nn \\
QF_I(x)^\dagger \gamma_0 & = & -\bar{\Psi}_{uI}(x) \phi(x) +\bar{\Psi}_{dI}(x)aD_W^\dagger 
+\twm_{+I}\bar{\Psi}_{uI}(x) P_- + \twm_{-I}\bar{\Psi}_{uI}(x) P_+\nn \\
 & & +i\sum_{\mu=0}^1\left(\Phi_I(x+\hat{\mu})^\dagger U_\mu(x)^{-1}\psi_\mu(x)\frac{\gamma_\mu -r}{2} \right. \nn \\
 & & \left. \hspace{1.5cm} +\Phi_I(x-\hat{\mu})^\dagger \psi_\mu(x-\hat{\mu})U_\mu(x-\hat{\mu}) \frac{\gamma_\mu +r}{2}\right). 
\label{QSUSY_DW_twm} 
\eea
Differently from the continuum case, (\ref{QSUSY_DW_twm}) is not closed among chirally projected variables 
due to the Wilson terms, which connect variables with different chiralities. 
Thus, we can not realize on the lattice the $Q$ supersymmetry (\ref{Qsusy_cont_mat+_twm}) and 
(\ref{Qsusy_cont_mat-_twm}) for general $n_\pm$ by means of the projection $P_\pm$ to (\ref{QSUSY_DW_twm}). 
This is nothing but the situation in \cite{sugino_sqcd}, where in order for the $Q$ transformation to be closed and nilpotent, 
it is required to take $n_+ = n_- (\equiv n)$ and $\twm_{+I} = \twm_{-I} (\equiv \twm_I)$. 
(Note that the anti-holomorphic twisted masses $\twm_{\pm I}^*$ still can be freely chosen. 
We could discuss the case $n_+ \neq n_-$ in~\cite{sugino_sqcd} by sending some $\twm_{+I}^*$ or $\twm_{-I'}^*$ to the infinity 
to decouple the corresponding matter multiplets.) 
Then, we can write the $Q$ invariant lattice action in~\cite{sugino_sqcd} as  
\bea
S^{\rm LAT}_{{\rm mat}, \twm} & = & Q\sum_x\sum_{I=1}^n \frac12 \left[ \frac{}{} 
\bar{\Psi}_{uI}(x)\left(aD_W\Phi_I(x) -\gamma_0 F_I(x)\right) \right. \nn \\
 & & \hspace{2cm} +\left(\Phi_I(x)^\dagger aD_W^\dagger -F_I(x)^\dagger \gamma_0\right)\Psi_{dI}(x) \nn \\
 & &  \hspace{2cm} -\Phi_I(x)^\dagger \left( \bar{\phi}(x)-\twm_{+I}^* P_+ -\twm_{-I}^* P_-\right) 
\Psi_{uI}(x) \nn \\
 & & \hspace{2cm} + \bar{\Psi}_{dI}(x) \left( \bar{\phi}(x) -\twm_{+I}^* P_+ -\twm_{-I}^* P_- \right) \Phi_I(x) \nn \\ 
 & &  \hspace{2cm} \left. \frac{}{} + 2i\Phi_I(x)^\dagger\gamma_3\chi(x)\Phi_I(x) \right].  
 \label{eq:action-wilson-matter}
\eea     
For general twisted masses, the lattice action preserves only the diagonal ${\rm U}(1)^n$ of 
the flavor symmetry of the continuum action ${\rm U}(1)^{n_+}\times {\rm U}(1)^{n_-}$ due to the Wilson terms. 

For the superpotential terms, 
\bea
S^{\rm LAT}_{\rm pot} & = & Q\sum_x\sum_{i=1}^N\sum_{I=1}^n\left[-\frac{\der W}{\der (P_+\Phi_I(x))_i} (\gamma_0P_-\Psi_{dI}(x))_i 
-(\bar{\Psi}_{uI}(x)P_- \gamma_0)_i\frac{\der\bar{W}}{\der (\Phi_I(x)^\dagger P_+)_i} 
\right. \nn \\
& &  \hspace{2.5cm} \left.  -\frac{\der \bar{W}}{\der (P_-\Phi_I(x))_i} (\gamma_0P_+\Psi_{dI}(x))_i 
    -(\bar{\Psi}_{uI}(x)P_+ \gamma_0)_i\frac{\der W}{\der (\Phi_I(x)^\dagger P_-)_i} \right]. \nn \\
     & & 
\eea
This expression does not have exact holomorphic or anti-holomorphic structure, 
because the Wilson terms in $D_W$ originating from $Q\Psi_{dI}, Q\bar{\Psi}_{uI}$ mix holomorphic and anti-holomorphic 
variables.

\setcounter{equation}{0}
\section{Lattice formulations with Overlap Dirac Operator}
\label{sec:lat_sqcd_hD}
In this section, we show that it is possible to construct $Q$-exact lattice action of the matter part 
for general $n_\pm, \twm_{\pm I}$ by employing the overlap Dirac operator $\hD$ satisfying 
the Ginsparg-Wilson relation~\cite{ginsparg-wilson}: 
\be
\gamma_3 \hD + \hD \gamma_3 =a \hD\gamma_3\hD.  
\label{GW_relation}
\ee
The explicit form of $\hD$ has been explicitly given by Neuberger \cite{neuberger, neuberger-yk} as\footnote{We use 
the symbol $\hD$ for the overlap operator to distinguish it from the auxiliary field $D(x)$.} 
\be
\hD \equiv \frac{1}{a}\left( 1-X\frac{1}{\sqrt{X^\dagger X}}\right), \qquad X = 1-aD_W.  
\ee
In order for $\hD$ to express the propagation of physical modes with spices doublers decoupled, 
we have to take $r> 1/2$. 
It can be seen, for example, from the computation of chiral anomaly~\cite{kikukawa-yamada,hsuzuki}.  
In what follows, $r$ is fixed to the standard value $r=1$. 

We present  two lattice formulations (formulations I and II) for the matter part of the action with
 different structures in chiral projection. 
Let us suppose to 
replace the Wilson-Dirac operator $D_W$ in the action of the matter part, 
eq.~(\ref{eq:action-wilson-matter}),  by the overlap Dirac operator $\hD$ and 
to decompose the quadratic part  into the chiral components. 
The first possibility is to put the chiral projector $P_\pm$ as in 
\begin{eqnarray}
&& \bar \Psi_{uI}(x) P_\pm \left(a \hD \Phi_I(x) - \gamma_0 F_I(x) \right)
+\left(\Phi_I(x)^\dagger a \hD^\dagger -F_I(x)^\dagger \gamma_0\right)  P_\pm \Psi_{dI}(x).
\end{eqnarray}
This leads us to the standard chiral decomposition in the Ginsparg-Wilson formulation,  
\bea
P_\pm \hD = \hD \hP_\mp, \qquad \hD^\dagger P_\pm = \hP_\mp \hD^\dagger, \qquad \hP_\pm^\dagger = \hP_\pm, 
\eea
where the chiral projectors are  defined by
\bea
\hP_\pm & \equiv & \frac{1 \pm \hgamma_3}{2}, \qquad \hgamma_3 \equiv \gamma_3 (1-a\hD) . 
\label{Phat} 
\eea
In this case, the bosonic variables $\Phi_I(x)$ are projected by $\hP_\pm$.

One another possibility is to replace $P_\pm$ above by  new chiral projectors $\bar P_\pm$
with the properties
\bea
\bar{P}_\pm \hD = \hD P_\mp, \qquad \hD^\dagger \bar{P}_\pm = P_\mp \hD^\dagger, \qquad \bar{P}_\pm^\dagger = \bar{P}_\pm, 
\eea
where the chiral projectors are  defined by
\bea
\bar{P}_\pm & \equiv & \frac{1\pm \bar{\gamma}_3}{2} , \qquad \bar{\gamma}_3 \equiv (1-a\hD)\gamma_3. 
\label{Pbar}
\eea
In this case, the bosonic variables $\Phi_I(x)$ are projected by $P_\pm$, while the auxiliary fields
$\gamma_0 F_I(x)$ are projected by $\bar P_\pm$. 

With these chiral decompositions, the question is then how to define the $Q$ supersymmetry 
which is closed within the chiral components.  We will discuss this issue in the following sections.

\subsection{Formulation I}
Here, we use the projectors $\hP_\pm$ to construct the lattice formulation. 
Let us pick as chiral and anti-chiral 
variables\footnote{For example, $\Phi_I^\dagger \hP_+(x)$ is understood as $(\hP_+\Phi_I)(x)^\dagger$.} 
in $S^{\rm LAT}_{{\rm mat}, +\twm}$
\bea
 & & \hP_+\Phi_I, \qquad  \hP_+\Psi_{uI}, \qquad P_-\Psi_{dI}, \qquad P_+F_I, 
\label{chiral_fields_+} \\
 & & \Phi_I^\dagger\hP_+,  \qquad \bar{\Psi}_{dI}\hP_+, \qquad \bar{\Psi}_{uI}P_-,  \qquad F_I^\dagger P_+ 
\qquad (I=1,\cdots, n_+),   
\label{anti-chiral_fields_+}
\eea  
and as chiral and anti-chiral variables in $S^{\rm LAT}_{{\rm mat}, -\twm}$ 
\bea
 & & \Phi_{I'}^\dagger\hP_-, \qquad \bar{\Psi}_{dI'}\hP_-, \qquad \bar{\Psi}_{uI'}P_+, \qquad F_{I'}^\dagger P_-,  
\label{chiral_fields_-} \\
& & \hP_-\Phi_{I'}, \qquad \hP_-\Psi_{uI'}, \qquad  P_+\Psi_{dI'}, \qquad P_-F_{I'} 
\qquad (I'=1, \cdots, n_-). 
\label{anti-chiral_fields_-}
\eea

As a first trial, we assume the transformation (\ref{QSUSY_DW_twm}) with $D_W$ simply replaced by $\hD$. 
Then, we have, for example,  
\bea
Q(\hP_+\Phi_I(x)) & = &  \hP_+ (Q\Phi_I(x)) + (Q\hP_+) \Phi_I(x) \nn \\
& = & -\hP_+\Psi_{uI}(x) +(Q\hP_+)\hP_+\Phi_I(x) + (Q\hP_+)\hP_-\Phi_I(x).
\eea
Note that $Q\hP_\pm$ generally do not vanish since $\hP_\pm$ involve the link variables. 
Due to the last term in the r.h.s., the transformation does not close 
among the chiral variables (\ref{chiral_fields_+}). 

Instead, we regard (\ref{chiral_fields_+}), (\ref{anti-chiral_fields_+}), (\ref{chiral_fields_-}), (\ref{anti-chiral_fields_-}) 
as fundamental contents of the theory, and let us define 
their transformation by starting with 
\bea
Q(\hP_+\Phi_I(x))  & = &  -\hP_+\Psi_{uI}(x) +(Q\hP_+)\hP_+\Phi_I(x), \nn \\
Q(\Phi_I^\dagger \hP_+(x)) & = & -\bar{\Psi}_{dI}\hP_+(x) +\Phi_I^\dagger\hP_+(Q\hP_+)(x), \nn \\
Q(\hP_-\Phi_{I'}(x)) & = & -\hP_-\Psi_{uI'}(x) +(Q\hP_-)\hP_-\Phi_{I'}(x), \nn \\
Q(\Phi_{I'}^\dagger \hP_-(x)) & = & -\bar{\Psi}_{dI'}\hP_-(x) +\Phi_{I'}^\dagger\hP_-(Q\hP_-)(x). 
\eea
It turns out that the $Q$ supersymmetry transformation can be consistently determined as a closed form 
among the (anti-)chiral variables, satisfying the nilpotency. 
Concretely, we have 
\bea
Q(\hP_+\Phi_I(x)) & = & -\hP_+\Psi_{uI}(x) +(Q\hP_+)\hP_+\Phi_I(x), \nn \\
Q(\hP_+\Psi_{uI}(x)) & = & -(\hP_+\phi -\twm_{+I}) \hP_+\Phi_I(x) +(Q\hP_+)\hP_+\Psi_{uI}(x) -(Q\hP_+)^2\hP_+\Phi_I(x), \nn \\
Q(P_-\Psi_{dI}(x)) & = & a\hD\hP_+\Phi_I(x) +\gamma_0 P_+ F_I(x), \nn \\
Q(\gamma_0 P_+ F_I(x)) & = & (\phi(x) -\twm_{+I}) P_-\Psi_{dI}(x) +a\hD\hP_+\Psi_{uI}(x) -P_-Q(a\hD)\hP_+\Phi_I(x) \nn \\
Q(\Phi_I^\dagger \hP_+(x)) & = & -\bar{\Psi}_{dI}\hP_+(x) +\Phi_I^\dagger\hP_+(Q\hP_+)(x), \nn \\
Q(\bar{\Psi}_{dI}\hP_+(x)) & = & \Phi_I^\dagger\hP_+(\phi\hP_+ -\twm_{+I})(x)  -\bar{\Psi}_{dI} \hP_+(Q\hP_+)(x) 
                                            +\Phi_I^\dagger\hP_+(Q\hP_+)^2(x), \nn \\
Q(\bar{\Psi}_{uI}(x)P_-) & = & \Phi_I^\dagger\hP_+(x)a\hD^\dagger + F_I(x)^\dagger P_+\gamma_0, \nn \\
Q(F_I(x)^\dagger P_+\gamma_0) & = & -\bar{\Psi}_{uI}(x) P_-(\phi(x) -\twm_{+I}) +\bar{\Psi}_{dI}\hP_+(x)a\hD^\dagger 
                                                   -\Phi_I^\dagger \hP_+(x) Q(a\hD^\dagger)P_-  ,    \nn \\
\label{Qmodified_+twm}     \\
Q(\hP_-\Phi_{I'}(x)) & = & -\hP_-\Psi_{uI'}(x) +(Q\hP_-)\hP_-\Phi_{I'}(x), \nn \\
Q(\hP_-\Psi_{uI'}(x)) & = & -(\hP_-\phi -\twm_{-I'}) \hP_-\Phi_{I'}(x) +(Q\hP_-)\hP_-\Psi_{uI'}(x) -(Q\hP_-)^2\hP_-\Phi_{I'}(x), \nn \\
Q(P_+\Psi_{dI'}(x)) & = & a\hD\hP_-\Phi_{I'}(x) +\gamma_0 P_- F_{I'}(x), \nn \\
Q(\gamma_0 P_- F_{I'}(x)) & = & (\phi(x) -\twm_{-I'}) P_+\Psi_{dI'}(x) +a\hD\hP_-\Psi_{uI'}(x) -P_+Q(a\hD)\hP_-\Phi_{I'}(x) , \nn \\
Q(\Phi_{I'}^\dagger \hP_-(x)) & = & -\bar{\Psi}_{dI'}\hP_-(x) +\Phi_{I'}^\dagger\hP_-(Q\hP_-)(x), \nn \\
Q(\bar{\Psi}_{dI'}\hP_-(x)) & = & \Phi_{I'}^\dagger\hP_- (\phi\hP_- -\twm_{-I'})(x) -\bar{\Psi}_{dI'} \hP_-(Q\hP_-)(x) 
                                            +\Phi_{I'}^\dagger\hP_-(Q\hP_-)^2(x), \nn \\
Q(\bar{\Psi}_{uI'}(x)P_+) & = & \Phi_{I'}^\dagger\hP_-(x)a\hD^\dagger + F_{I'}(x)^\dagger P_-\gamma_0, \nn \\
Q(F_{I'}(x)^\dagger P_-\gamma_0) & = & -\bar{\Psi}_{uI'}(x) P_+ (\phi(x) -\twm_{-I'}) +\bar{\Psi}_{dI'}\hP_-(x)a\hD^\dagger 
                                                   -\Phi_{I'}^\dagger \hP_-(x) Q(a\hD^\dagger)P_+ .     \nn \\ 
\label{Qmodified_-twm}                                           
\eea
The nilpotency holds as 
\bea
Q^2 & = & (\mbox{infinitesimal gauge transformation with the parameter $\phi(x)$}) \nn \\
      & & + (\mbox{infinitesimal flavor rotations (\ref{flavor_rot_+}) and (\ref{flavor_rot_-})})
\eea
with 
\bea
\delta (\hP_+\Phi_I) = -\twm_{+I}\hP_+\Phi_I, & &    \delta (\Phi_I^\dagger \hP_+) = \twm_{+I}\Phi_I^\dagger \hP_+, \nn \\
\delta (\hP_+\Psi_{uI}) = -\twm_{+I} \hP_+\Psi_{uI}, & & \delta (\bar{\Psi}_{uI} P_-) = \twm_{+I} \bar{\Psi}_{uI}P_-, \nn \\
\delta (P_- \Psi_{dI}) = -\twm_{+I} P_-\Psi_{dI}, & & \delta (\bar{\Psi}_{dI}\hP_+) = \twm_{+I}\bar{\Psi}_{dI}\hP_+, \nn \\
\delta (P_+ F_I) = -\twm_{+I} P_+ F_I, & & \delta (F_I^\dagger P_+) = \twm_{+I} F_I^\dagger P_+, 
\label{flavor_rot_+}
\eea
\bea
\delta (\Phi_{I'}^\dagger \hP_-) = \twm_{-I'}\Phi_{I'}^\dagger\hP_-, & & \delta (\hP_-\Phi_{I'}) = -\twm_{-I'} \hP_-\Phi_{I'}, \nn \\
\delta (\bar{\Psi}_{uI'}P_+) = \twm_{-I'}\bar{\Psi}_{uI'} P_+, & & \delta (\hP_-\Psi_{uI'}) = -\twm_{-I'}\hP_-\Psi_{uI'}, \nn \\
\delta (\bar{\Psi}_{dI'}\hP_-) = \twm_{-I'}\bar{\Psi}_{dI'}\hP_-, & & \delta (P_+\Psi_{dI'}) = -\twm_{-I'} P_+\Psi_{dI'}, \nn \\
\delta (F_{I'}^\dagger P_-) = \twm_{-I'} F_{I'}^\dagger P_-, & & \delta (P_-F_{I'}) = -\twm_{-I'} P_-F_{I'}. 
\label{flavor_rot_-}
\eea     
We used the identity 
\be
\hP_\pm (Q \hP_\pm) \hP_\pm =0,
\label{identity_hat}
\ee
which is derived from the $Q$ transformation of $\hP_\pm^2=\hP_\pm$. 
Differently from the situation in the previous section, we here have no requirement to $n_\pm$ nor to the twisted masses 
for the $Q$ supersymmetry being closed and nilpotent.  

For the case $n_+ =n_-$, from the sum of the first formulas both of (\ref{Qmodified_+twm}) and (\ref{Qmodified_-twm}), 
we have 
\bea
Q\Phi_I(x)  & = &  -\Psi_{uI}(x) +\left[(Q\hP_+)\hP_+ + (Q\hP_-) \hP_-\right] \Phi_I(x)  \nn \\
               & = & -\Psi_{uI}(x) -\frac12 Q(a\hD^\dagger) (1-a\hD)\Phi_I(x).  
\eea
The second term in the r.h.s. is $\cO(a)$ but indicates nontrivial difference from the naive transformation. 

Using the identity 
\be
\frac{1}{\sqrt{X^\dagger X}} = \int_{-\infty}^\infty\frac{\dd t}{\pi} \, \frac{1}{t^2+X^\dagger X}, 
\ee
$Q(a\hD)$ can be expressed as 
\bea
Q(a\hD) & = & Q(aD_W) \frac{1}{\sqrt{X^\dagger X}}  \nn \\
                     & & -X\int_{-\infty}^\infty\frac{dt}{\pi} \, 
\frac{1}{t^2+X^\dagger X} \left( Q(aD_W^\dagger ) X + X^\dagger Q(aD_W)\right) \frac{1}{t^2+X^\dagger X}, 
\label{QhD}
\eea
where the action of $Q(aD_W)$ or $Q(aD_W^\dagger)$ are explicitly given by 
\bea
Q(aD_W){\cal X}(x) &= & \sum_{\mu=0}^1 \left[\frac{\gamma_\mu -r}{2}\, i\psi_\mu(x)U_\mu(x){\cal X}(x+\hat{\mu}) \right. \nn \\
             & & \hspace{1cm} \left. +\frac{\gamma_\mu +r}{2}\, iU_\mu(x-\hat{\mu})^{-1}\psi_\mu(x-\hat{\mu}){\cal X}(x-\hat{\mu})\right], 
                                 \nn \\
Q(aD_W^\dagger){\cal X}(x) &= & \sum_{\mu=0}^1 \left[-\frac{\gamma_\mu +r}{2}\, i\psi_\mu(x)U_\mu(x){\cal X}(x+\hat{\mu}) \right. \nn \\
             & & \hspace{1cm} \left. -\frac{\gamma_\mu -r}{2}\, iU_\mu(x-\hat{\mu})^{-1}\psi_\mu(x-\hat{\mu}){\cal X}(x-\hat{\mu})\right]      
\eea
with ${\cal X}(x)$ being any doublet belonging to the fundamental representation of $G$, 
for example $\Phi_I(x)$, $\Psi_{uI}(x)$, $\Psi_{dI}(x)$, $F_I(x)$, and their chiral projections. 

It can be seen that each of (\ref{chiral_fields_+}) and (\ref{chiral_fields_-}) forms ``chiral multiplet'' for the $Q$ transformation, 
and each of (\ref{anti-chiral_fields_+}) and (\ref{anti-chiral_fields_-}) forms ``anti-chiral multiplet''. 
The matter-part action is given as the $Q$-exact form: 
\bea
S^{\rm LAT}_{{\rm mat}, +\twm} & = & Q\sum_x \sum_{I=1}^{n_+} \frac12 \left[ 
       \bar{\Psi}_{uI}(x) P_- \left(a\hD \hP_+\Phi_I(x) -\gamma_0 P_+F_I(x)\right) \right. \nn \\
       & & \hspace{2cm} +\left(\Phi_I^\dagger \hP_+(x) \, a\hD^\dagger -F_I(x)^\dagger P_+\gamma_0\right) P_-\Psi_{dI}(x) \nn \\
       & & \hspace{2cm} -\Phi_I^\dagger \hP_+(x) \, \left(\bar{\phi}(x)-\twm_{+I}^*\right) \hP_+\Psi_{uI}(x) \nn \\
       & & \hspace{2cm}  +\bar{\Psi}_{dI}\hP_+(x) \, \left(\bar{\phi}(x)-\twm_{+I}^*\right) \hP_+\Phi_I(x) \nn \\
       & & \hspace{2cm} \left. +2i\Phi_I^\dagger \hP_+(x) \, \chi(x) \hP_+\Phi_I(x) \right], 
\label{S_mat+twm_hD}\\
S^{\rm LAT}_{{\rm mat}, -\twm} & = &  Q\sum_x \sum_{I'=1}^{n_-} \frac12 \left[  
       \bar{\Psi}_{uI'}(x) P_+ \left(a\hD \hP_-\Phi_{I'}(x) -\gamma_0 P_-F_{I'}(x)\right) \right. \nn \\
       & & \hspace{2cm} +\left(\Phi_{I'}^\dagger \hP_-(x) \, a\hD^\dagger -F_{I'}(x)^\dagger P_-\gamma_0\right) P_+\Psi_{dI'}(x) \nn \\
       & & \hspace{2cm} -\Phi_{I'}^\dagger \hP_-(x) \, \left(\bar{\phi}(x)-\twm_{-I'}^*\right) \hP_-\Psi_{uI'}(x) \nn \\
       & & \hspace{2cm} +\bar{\Psi}_{dI'}\hP_-(x) \, \left(\bar{\phi}(x)-\twm_{-I'}^*\right) \hP_-\Phi_{I'}(x) \nn \\
       & & \hspace{2cm} \left. -2i\Phi_{I'}^\dagger \hP_-(x) \, \chi(x) \hP_-\Phi_{I'}(x) \right], 
\label{S_mat-twm_hD}       
\eea
which becomes, after the $Q$ operation in the r.h.s.,
\bea
S^{\rm LAT}_{{\rm mat}, +\twm} & = &\sum_x \sum_{I=1}^{n_+} \left[ 
         a^2 \Phi_I^\dagger \hP_+(x) \, \hD^\dagger \hD\hP_+\Phi_I(x) -\left(F_I(x)^\dagger P_+\right)\left(P_+F_I(x)\right) 
                           \right. \nn \\
  & & +\bar{\Psi}_{uI}(x)P_- \, a\hD \hP_+\Psi_{uI}(x) -\bar{\Psi}_{dI} \hP_+(x) \, a\hD^\dagger P_-\Psi_{dI}(x) \nn \\
 & & +\frac12 \Phi_I^\dagger \hP_+(x) \left\{ \phi\hP_+ -\twm_{+I}, \bar{\phi}\hP_+ -\twm_{+I}^*\right\} \hP_+ \Phi_I(x) \nn \\
 & & -\Phi_I^\dagger \hP_+(x) \left(D(x) +\frac12 \whPhi(x)\right) \hP_+\Phi_I(x) \nn \\
 & & +\bar{\Psi}_{uI}(x)P_- \left(\phi(x) -\twm_{+I}\right) P_-\Psi_{dI}(x) 
                         + \bar{\Psi}_{dI} \hP_+(x) \left(\bar{\phi}(x) -\twm_{+I}^*\right)\hP_+\Psi_{uI}(x) \nn \\
 & & -\bar{\Psi}_{uI}(x) P_- \, Q(a\hD) \hP_+\Phi_I(x) + \Phi_I^\dagger \hP_+(x) \, Q(a\hD^\dagger )P_- \Psi_{dI}(x) \nn \\
 & & -\bar{\Psi}_{dI} \hP_+(x) \left( \frac12 \eta(x) +i\chi(x)\right) \hP_+ \Phi_I(x) \nn \\
  & &  -\Phi_I^\dagger \hP_+(x) \left(\frac12 \eta(x) -i\chi(x)\right) \hP_+\Psi_{uI}(x) \nn \\
 & & -\frac12 \Phi_I^\dagger \hP_+(x) \left\{ (Q\hP_+), \bar{\phi}\right\} \hP_+\Psi_{uI}(x) 
             -\frac12 \bar{\Psi}_{dI} \hP_+(x) \left\{ (Q\hP_+), \bar{\phi}\right\} \hP_+\Phi_I(x) \nn \\
 & & +\frac12 \Phi_I^\dagger \hP_+(x) \left\{ (Q\hP_+)^2, \bar{\phi}\right\} \hP_+ \Phi_I(x) 
        \left. +i \Phi_I^\dagger \hP_+(x) \left[ (Q\hP_+), \chi\right] \hP_+ \Phi_I(x)  \right], \nn \\
 & &  
\label{S_mat+twm_hD2}\\             
S^{\rm LAT}_{{\rm mat}, -\twm} & = &\sum_x \sum_{I'=1}^{n_-} \left[
       a^2 \Phi_{I'}^\dagger \hP_-(x) \, \hD^\dagger \hD\hP_-\Phi_{I'}(x) -\left(F_{I'}(x)^\dagger P_-\right)\left(P_-F_{I'}(x)\right)
                           \right. \nn \\
 & & +\bar{\Psi}_{uI'}(x) P_+ \, a\hD \hP_-\Psi_{uI'}(x) -\bar{\Psi}_{dI'}\hP_-(x) \, a\hD^\dagger P_+ \Psi_{dI'}(x) \nn \\
 & & +\frac12 \Phi_{I'}^\dagger \hP_-(x) \left\{ \phi\hP_- -\twm_{-I'}, \bar{\phi}\hP_- -\twm_{-I'}^*\right\} \hP_- \Phi_{I'}(x) \nn \\
 & & +\Phi_{I'}^\dagger \hP_-(x) \left(D(x) +\frac12 \whPhi(x)\right) \hP_-\Phi_{I'}(x) \nn \\
 & & +\bar{\Psi}_{uI'}(x) P_+ \left(\phi(x)-\twm_{-I'}\right) P_+\Psi_{dI'}(x) 
        +\bar{\Psi}_{dI'}\hP_-(x) \left(\bar{\phi}(x)-\twm_{-I'}^*\right) \hP_-\Psi_{uI'}(x) \nn \\
 & & -\bar{\Psi}_{uI'}(x) P_+ \, Q(a\hD)\hP_-\Phi_{I'}(x) + \Phi_{I'}^\dagger \hP_-(x) \, Q(a\hD^\dagger) P_+\Psi_{dI'}(x) \nn \\
 & &  -\bar{\Psi}_{dI'}\hP_-(x) \left(\frac12\eta(x) -i\chi(x)\right) \hP_-\Phi_{I'}(x) \nn \\
 & & -\Phi_{I'}^\dagger \hP_-(x) \left(\frac12\eta(x) +i\chi(x)\right) \hP_-\Psi_{uI'}(x) \nn \\
 & & -\frac12 \Phi_{I'}^\dagger \hP_-(x) \left\{ (Q\hP_-), \bar{\phi}\right\} \hP_-\Psi_{uI'}(x) 
        -\frac12 \bar{\Psi}_{dI'} \hP_-(x) \left\{ (Q\hP_-), \bar{\phi}\right\} \hP_- \Phi_{I'}(x) \nn \\
 & & +\frac12 \Phi_{I'}^\dagger \hP_-(x) \left\{ (Q\hP_-)^2, \bar{\phi}\right\} \hP_- \Phi_{I'}(x) 
         \left. -i \Phi_{I'}^\dagger \hP_-(x) \left[ (Q\hP_-), \chi \right] \hP_- \Phi_{I'}(x)  \right].  \nn \\
 & & 
\label{S_mat-twm_hD2} 
\eea
The last four terms both in (\ref{S_mat+twm_hD2}) and (\ref{S_mat-twm_hD2}) are lattice artifacts having 
no counterparts in the continuum theory. 
As explained in appendix~\ref{app:pauli}, the boson kinetic kernel $\hD^\dagger \hD$ 
cancels with the ``Pauli terms'' containing $\whPhi(x)$ in the fourth line 
and leaves the covariant Laplacian $-\cD_\mu\cD_\mu$ in the continuum limit. 

The Ginsparg-Wilson formulation realizes the exact chiral flavor symmetry on the lattice. 
The action has the symmetry ${\rm U}(1)^{n_+} \times {\rm U}(1)^{n_-}$ for general twisted masses, 
which maximally enhances to ${\rm U}(n_+) \times {\rm U}(n_-)$ 
for the case $\twm_{\pm 1}=\cdots =\twm_{\pm n_\pm}$ and $\twm^*_{\pm 1}=\cdots =\twm^*_{\pm n_\pm}$.   

This formulation (formulation I) has somewhat involved interaction terms containing $\hP_\pm$ or $Q\hP_\pm$. 
We will develop another formulation in the next subsection (formulation II), which leads to simpler interactions.

\subsection{Formulation II}
In the formulation here, we use the projectors $\bar{P}_\pm$ instead of $\hP_\pm$. 
Chiral and anti-chiral variables appearing in $S^{\rm LAT}_{{\rm mat}, +\twm}$ are defined as 
\bea
 & & P_+\Phi_I, \qquad P_+\Psi_{uI}, \qquad \bar P_-\Psi_{dI}, \qquad \bar P_-\gamma_0 F_I, 
\label{chiral_fields_+_a} \\
 & & \Phi_I^\dagger P_+, \qquad \bar{\Psi}_{dI} P_+, \qquad \bar{\Psi}_{uI} \bar P_-, \qquad  F_I^\dagger \gamma_0 \bar P_-, 
\label{anti-chiral_fields_+_a}
\eea
and chiral and anti-chiral variables in $S^{\rm LAT}_{{\rm mat}, -\twm} $ are defined as 
\bea
 & & \Phi_{I'}^\dagger P_-, \qquad \bar{\Psi}_{dI'} P_-, \qquad \bar{\Psi}_{uI'} \bar P_+, \qquad F_{I'}^\dagger \gamma_0 \bar P_+, 
\label{chiral_fields_-_a} \\
  & & P_-\Phi_{I'}, \qquad P_-\Psi_{uI'}, \qquad \bar P_+\Psi_{dI'}, \qquad \bar P_+ \gamma_0 F_{I'} .
\label{anti-chiral_fields_-_a}
\eea
Then, the $Q$ supersymmetry transformation 
\bea
Q(P_+\Phi_I(x)) & = & -P_+\Psi_{uI}(x) , \nn \\
Q(P_+\Psi_{uI}(x)) & = & -(\phi(x) -\twm_{+I}) P_+\Phi_I(x) , \nn \\
Q(\bar{P}_-\Psi_{dI}(x)) & = & a\hD P_+\Phi_I(x) +\bar{P}_-\gamma_0 F_I(x) +(Q\bar{P}_-)\bar{P}_-\Psi_{dI}(x) , \nn \\
Q(\bar{P}_-\gamma_0 F_I(x)) & = & (\bar{P}_-\phi -\twm_{+I}) \bar{P}_-\Psi_{dI}(x) +a\hD P_+\Psi_{uI}(x) 
                                             -\bar{P}_-Q(a\hD)P_+\Phi_I(x) \nn \\
                                       &   & +(Q\bar{P}_-)\bar{P}_-\gamma_0 F_I(x) +(Q\bar{P}_-)^2\bar{P}_-\Psi_{dI}(x) \nn \\      
Q(\Phi_I(x)^\dagger P_+) & = & -\bar{\Psi}_{dI}(x)P_+ , \nn \\
Q(\bar{\Psi}_{dI}(x)P_+) & = & \Phi_I(x)^\dagger P_+(\phi(x) -\twm_{+I}) , \nn \\
Q(\bar{\Psi}_{uI}\bar{P}_-(x)) & = & \Phi_I(x)^\dagger P_+ a\hD^\dagger + F_I^\dagger \gamma_0 \bar{P}_-(x) 
                                                 -\bar{\Psi}_{uI}\bar{P}_- (Q\bar{P}_-)(x) , \nn \\
Q(F_I^\dagger \gamma_0\bar{P}_-(x)) & = & -\bar{\Psi}_{uI} \bar{P}_-(\phi\bar{P}_- -\twm_{+I})(x) +\bar{\Psi}_{dI}(x)P_+ a\hD^\dagger 
                                                   -\Phi_I(x)^\dagger P_+ Q(a\hD^\dagger)\bar{P}_-     \nn \\
                                 & & +F_I^\dagger \gamma_0\bar{P}_-(Q\bar{P}_-)(x) -\bar{\Psi}_{uI}\bar{P}_-(Q\bar{P}_-)^2(x) , 
\label{Qmodified_+twm_a}     
\eea
\bea
Q(P_-\Phi_{I'}(x)) & = & -P_-\Psi_{uI'}(x) , \nn \\
Q(P_-\Psi_{uI'}(x)) & = & -(\phi(x) -\twm_{-I'}) P_-\Phi_{I'}(x) , \nn \\
Q(\bar{P}_+\Psi_{dI'}(x)) & = & a\hD P_-\Phi_{I'}(x) +\bar{P}_+\gamma_0 F_{I'}(x) +(Q\bar{P}_+)\bar{P}_+ \Psi_{dI'}(x), \nn \\
Q(\bar{P}_+\gamma_0 F_{I'}(x)) & = & (\bar{P}_+\phi -\twm_{-I'}) \bar{P}_+\Psi_{dI'}(x) +a\hD P_-\Psi_{uI'}(x) 
                                                   -\bar{P}_+Q(a\hD)P_-\Phi_{I'}(x) \nn \\
                          & & +(Q\bar{P}_+)\bar{P}_+\gamma_0 F_{I'}(x) + (Q\bar{P}_+)^2 \bar{P}_+\Psi_{dI'}(x)  , \nn \\
Q(\Phi_{I'}(x)^\dagger P_-) & = & -\bar{\Psi}_{dI'}(x)P_- , \nn \\
Q(\bar{\Psi}_{dI'}(x)P_-) & = & \Phi_{I'}(x)^\dagger P_- (\phi(x) -\twm_{-I'}) , \nn \\
Q(\bar{\Psi}_{uI'}\bar{P}_+(x)) & = & \Phi_{I'}(x)^\dagger P_- a\hD^\dagger + F_{I'}^\dagger \gamma_0 \bar{P}_+(x) 
                                                  -\bar{\Psi}_{uI'}\bar{P}_+(Q\bar{P}_+)(x), \nn \\
Q(F_{I'}^\dagger \gamma_0\bar{P}_+(x)) & = & -\bar{\Psi}_{uI'}\bar{P}_+ (\phi\bar{P}_+ -\twm_{-I'})(x) 
                                     +\bar{\Psi}_{dI'}(x)P_- a\hD^\dagger -\Phi_{I'}(x)^\dagger P_- Q(a\hD^\dagger)\bar{P}_+  \nn \\
                                      & & +F_{I'}^\dagger \gamma_0\bar{P}_+(Q\bar{P}_+)(x) -\bar{\Psi}_{uI'}\bar{P}_+(Q\bar{P}_+)^2(x) ,  
\label{Qmodified_-twm_a}                                           
\eea
is nilpotent in the sense of 
\bea
Q^2 & = & (\mbox{infinitesimal gauge transformation with the parameter $\phi(x)$}) \nn \\
      & & + (\mbox{infinitesimal flavor rotations (\ref{flavor_rot_+_a}) and (\ref{flavor_rot_-_a})})
\eea
with 
\bea
\delta (P_+\Phi_I) = -\twm_{+I}P_+\Phi_I, & &    \delta (\Phi_I^\dagger P_+) = \twm_{+I}\Phi_I^\dagger P_+, \nn \\
\delta (P_+\Psi_{uI}) = -\twm_{+I} P_+\Psi_{uI}, & & \delta (\bar{\Psi}_{uI} \bar{P}_-) = \twm_{+I} \bar{\Psi}_{uI}\bar{P}_-, \nn \\
\delta (\bar{P}_- \Psi_{dI}) = -\twm_{+I} \bar{P}_-\Psi_{dI}, & & \delta (\bar{\Psi}_{dI} P_+) = \twm_{+I}\bar{\Psi}_{dI} P_+, \nn \\
\delta (\bar{P}_-\gamma_0 F_I) = -\twm_{+I} \bar{P}_-\gamma_0 F_I, 
                                      & & \delta (F_I^\dagger \gamma_0 \bar{P}_-) = \twm_{+I} F_I^\dagger \gamma_0 \bar{P}_-, 
\label{flavor_rot_+_a}
\eea
\bea
\delta (\Phi_{I'}^\dagger P_-) = \twm_{-I'}\Phi_{I'}^\dagger P_-, & & \delta (P_-\Phi_{I'}) = -\twm_{-I'} P_-\Phi_{I'}, \nn \\
\delta (\bar{\Psi}_{uI'} \bar{P}_+) = \twm_{-I'}\bar{\Psi}_{uI'} \bar{P}_+, & & \delta (P_-\Psi_{uI'}) = -\twm_{-I'}P_-\Psi_{uI'}, \nn \\
\delta (\bar{\Psi}_{dI'} P_-) = \twm_{-I'}\bar{\Psi}_{dI'} P_-, & & \delta (\bar{P}_+\Psi_{dI'}) = -\twm_{-I'} \bar{P}_+\Psi_{dI'}, \nn \\
\delta (F_{I'}^\dagger \gamma_0 \bar{P}_+) = \twm_{-I'} F_{I'}^\dagger \gamma_0 \bar{P}_+, 
                                                & & \delta (\bar{P}_+ \gamma_0 F_{I'}) = -\twm_{-I'} \bar{P}_+ \gamma_0 F_{I'}. 
\label{flavor_rot_-_a}
\eea      
Similarly to (\ref{identity_hat}), we have 
\be
\bar{P}_\pm (Q\bar{P}_\pm) \bar{P}_\pm =0.
\label{identity_bar}
\ee 

Under the $Q$ transformation, each of (\ref{chiral_fields_+_a}) and (\ref{chiral_fields_-_a}) forms ``chiral multiplet'', 
and each of (\ref{anti-chiral_fields_+_a}) and (\ref{anti-chiral_fields_-_a}) forms ``anti-chiral multiplets''. 
The matter-part action can be written as the $Q$-exact form:
\bea
S^{\rm LAT}_{{\rm mat}, \twm} & = & S^{\rm LAT}_{{\rm mat}, +\twm} + S^{\rm LAT}_{{\rm mat}, -\twm},   \nn \\
S^{\rm LAT}_{{\rm mat}, +\twm} & = & Q\sum_x \sum_{I=1}^{n_+} \frac12 \left[ 
       \bar{\Psi}_{uI}\bar P_-(x) \left(a\hD P_+\Phi_I(x) -\bar{P}_-\gamma_0 F_I(x)\right) \right. \nn \\
       & & \hspace{2cm} +\left(\Phi_I(x)^\dagger P_+ \, a\hD^\dagger -F_I^\dagger\gamma_0 \bar{P}_-(x) \right) \bar P_-\Psi_{dI}(x) 
              \nn \\
       & & \hspace{2cm} -\Phi_I(x)^\dagger P_+ \, \left(\bar{\phi}(x) -\twm_{+I}^*\right) P_+\Psi_{uI}(x) \nn \\
       & & \hspace{2cm} +\bar{\Psi}_{dI}(x) P_+ \, \left(\bar{\phi}(x) -\twm_{+I}^*\right) P_+\Phi_I(x) \nn \\
       & & \hspace{2cm} \left. +2i\Phi_I(x)^\dagger P_+ \, \chi(x) P_+\Phi_I(x) \right], 
\label{S_mat+twm_hD_a}\\
S^{\rm LAT}_{{\rm mat}, -\twm} & = &  Q\sum_x \sum_{I'=1}^{n_-} \frac12 \left[  
       \bar{\Psi}_{uI'}\bar P_+(x) \left(a\hD P_-\Phi_{I'}(x) -\bar{P}_+\gamma_0 F_{I'}(x)\right) \right. \nn \\
       & & \hspace{2cm} +\left(\Phi_{I'}(x)^\dagger P_- \, a\hD^\dagger -F_{I'}^\dagger\gamma_0 \bar{P}_+(x)\right) \bar P_+\Psi_{dI'}(x) 
              \nn \\
       & & \hspace{2cm} -\Phi_{I'}(x)^\dagger P_- \, \left(\bar{\phi}(x) -\twm_{-I'}^*\right) P_-\Psi_{uI'}(x)  \nn \\
       & & \hspace{2cm} +\bar{\Psi}_{dI'}(x) P_- \, \left(\bar{\phi}(x) -\twm_{-I'}^*\right)  P_-\Phi_{I'}(x) \nn \\
       & & \hspace{2cm} \left. -2i\Phi_{I'}(x)^\dagger P_- \, \chi(x) P_-\Phi_{I'}(x) \right]. 
\label{S_mat-twm_hD_a}       
\eea
The last three terms both in (\ref{S_mat+twm_hD_a}) and (\ref{S_mat-twm_hD_a}) yield interactions 
without the projectors depending on $\hD$. This is a main difference from (\ref{S_mat+twm_hD}) and (\ref{S_mat-twm_hD}) 
in formulation I, and it is expected to give a simpler action.  
In fact, the $Q$ operation in the r.h.s. leads to 
\bea
S^{\rm LAT}_{{\rm mat}, +\twm} & = &\sum_x \sum_{I=1}^{n_+} \left[ 
         a^2 \Phi_I(x)^\dagger P_+\hD^\dagger \hD P_+\Phi_I(x) 
                   -\left(F_I^\dagger \gamma_0 \bar{P}_-(x)\right)\left(\bar{P}_-\gamma_0 F_I(x)\right) \right. \nn \\
  & & +\bar{\Psi}_{uI}\bar{P}_-(x) \, a\hD P_+\Psi_{uI}(x) -\bar{\Psi}_{dI}(x) P_+ \, a\hD^\dagger \bar{P}_-\Psi_{dI}(x) \nn \\
 & & +\frac12 \Phi_I(x)^\dagger P_+ \left\{ \phi(x) -\twm_{+I}, \bar{\phi}(x) -\twm_{+I}^*\right\} P_+ \Phi_I(x) \nn \\
 & & -\Phi_I(x)^\dagger P_+ \left(D(x) +\frac12 \whPhi(x)\right) P_+\Phi_I(x) \nn \\
 & & +\bar{\Psi}_{uI}\bar{P}_-(x) \left(\phi(x) -\twm_{+I}\right) \bar{P}_-\Psi_{dI}(x) 
                         + \bar{\Psi}_{dI}(x) P_+ \left(\bar{\phi}(x) -\twm_{+I}^*\right)P_+\Psi_{uI}(x) \nn \\
 & & -\bar{\Psi}_{uI} \bar{P}_-(x) \, Q(a\hD) P_+\Phi_I(x) + \Phi_I(x)^\dagger P_+ \, Q(a\hD^\dagger )\bar{P}_- \Psi_{dI}(x) \nn \\
 & & -\bar{\Psi}_{dI}(x) P_+ \left(\frac12\eta(x) +i\chi(x)\right) P_+ \Phi_I(x) \nn \\
 & & -\Phi_I(x)^\dagger P_+ \left(\frac12\eta(x) -i\chi(x)\right) P_+\Psi_{uI}(x) \nn \\
 & &   \left. +\bar{\Psi}_{uI}\bar{P}_-(x) \, (Q\bar{P}_-)^2 \bar{P}_-\Psi_{dI}(x) \right], 
\label{S_mat+twm_hD_a2}
\eea
\bea             
S^{\rm LAT}_{{\rm mat}, -\twm} & = &\sum_x \sum_{I'=1}^{n_-} \left[
       a^2 \Phi_{I'}(x)^\dagger P_- \hD^\dagger \hD P_-\Phi_{I'}(x) 
                -\left(F_{I'}^\dagger \gamma_0 \bar{P}_+(x)\right)\left(\bar{P}_+\gamma_0 F_{I'}(x)\right) \right. \nn \\
 & & +\bar{\Psi}_{uI'} \bar{P}_+(x) \, a\hD P_-\Psi_{uI'}(x) -\bar{\Psi}_{dI'}(x) P_- \, a\hD^\dagger \bar{P}_+ \Psi_{dI'}(x) \nn \\
 & & +\frac12 \Phi_{I'}(x)^\dagger P_- \left\{ \phi(x) -\twm_{-I'}, \bar{\phi}(x) -\twm_{-I'}^*\right\} P_- \Phi_{I'}(x) \nn \\
 & & +\Phi_{I'}(x)^\dagger P_- \left(D(x) +\frac12 \whPhi(x)\right) P_-\Phi_{I'}(x) \nn \\
 & & +\bar{\Psi}_{uI'} \bar{P}_+(x) \left(\phi(x)-\twm_{-I'}\right) \bar{P}_+\Psi_{dI'}(x) 
        +\bar{\Psi}_{dI'}(x) P_- \left(\bar{\phi}(x)-\twm_{-I'}^*\right) P_-\Psi_{uI'}(x) \nn \\
 & & -\bar{\Psi}_{uI'} \bar{P}_+(x) \, Q(a\hD) P_-\Phi_{I'}(x) + \Phi_{I'}(x)^\dagger P_- \, Q(a\hD^\dagger) \bar{P}_+\Psi_{dI'}(x) \nn \\
 & & -\bar{\Psi}_{dI'}(x) P_- \left(\frac12\eta(x) -i\chi(x)\right) P_-\Phi_{I'}(x) \nn \\
 & &  -\Phi_{I'}(x)^\dagger P_- \left(\frac12\eta(x) +i\chi(x)\right) P_-\Psi_{uI'}(x) \nn \\
 & &  \left. + \bar{\Psi}_{uI'}\bar{P}_+(x) \, (Q\bar{P}_+)^2 \bar{P}_+\Psi_{dI'}(x)  \right],  
\label{S_mat-twm_hD_a2} 
\eea
where the last terms both in (\ref{S_mat+twm_hD_a2}) and (\ref{S_mat-twm_hD_a2}) are lattice artifacts. 
As in formulation I, the kinetic kernel $\hD^\dagger\hD$ cancels with the ``Pauli terms'' 
leaving the covariant Laplacian in the continuum limit. 
The flavor symmetry ${\rm U}(1)^{n_+} \times {\rm U}(1)^{n_-}$ for general twisted masses is preserved.

Since the construction here gives a simpler expression, 
we will mainly develop formulation II in what follows, 
although the subsequent discussion would be possible also in formulation I. 

\paragraph{Superpotentials}
We can latticize the superpotential terms as 
\bea
S^{\rm LAT}_{\rm pot} 
& = & Q\sum_x\sum_{i=1}^N\sum_{I=1}^{n_+}\left[-\frac{\der W}{\der (P_+\Phi_I(x))_i} \left(\gamma_0 \bar P_-\Psi_{dI}(x)\right)_i 
-\left(\bar{\Psi}_{uI} \bar P_-(x) \gamma_0\right)_i\frac{\der\bar{W}}{\der (\Phi_I(x)^\dagger P_+)_i} 
\right] \nn \\
& & \hspace{-3mm} + Q\sum_x\sum_{i=1}^N\sum_{I'=1}^{n_-}\left[
      -\frac{\der \bar{W}}{\der (P_-\Phi_{I'}(x))_i} \left(\gamma_0 \bar P_+\Psi_{dI'}(x)\right)_i 
    -\left(\bar{\Psi}_{uI'} \bar P_+(x) \gamma_0\right)_i\frac{\der W}{\der (\Phi_{I'}(x)^\dagger P_-)_i} \right] \nn \\
     & & 
\label{S_pot_a}     
\eea
with 
\be
W=W(P_+\Phi_I, \Phi_{I'}^\dagger P_-), \qquad \bar{W} =\bar{W} (\Phi_I^\dagger P_+, P_-\Phi_{I'}). 
\ee
$(\cdots)_i$ represent independent color degrees of freedom of the projected doublet by $P_\pm$ or $\bar{P}_\pm$. 
Note that (\ref{S_pot_a}) has holomorphic or anti-holomorphic structure on the lattice, 
i.e. terms containing $W$ depend only on 
the chiral variables (\ref{chiral_fields_+_a}) and (\ref{chiral_fields_-_a}), 
and terms containing $\bar{W}$ depend only on the anti-chiral variables (\ref{anti-chiral_fields_+_a}) 
and (\ref{anti-chiral_fields_-_a}), besides the SYM variables which come in via $\bar{P}_\pm$ or $Q\bar{P}_\pm$. 
Similarly to the continuum case, the holomorphy tempts us to expect that (\ref{S_pot_a}) receives no radiative correction 
on lattice perturbative computations concerning the matter sector.

\subsection{Transformation Properties of Path-integral Measures}
The path-integral measure with respect to the matter variables is defined by 
\bea 
\left(\dd \mu_{\rm mat} \right) & = & \left(\prod_{I=1}^{n_+} \dd \mu_{{\rm mat}, +I} \right) 
                                                            \left(\prod_{I'=1}^{n_-} \dd \mu_{{\rm mat}, -I'} \right) \nn \\
\dd \mu_{{\rm mat}, +I}  & \equiv & \prod_x \prod_{i=1}^N \dd (P_+\Phi_I(x))_i \, 
                                         \dd (\Phi_I(x)^\dagger P_+)_i \, \dd (\bar{P}_-\gamma_0 F_I(x))_i \,
                                         \dd (F_I^\dagger\gamma_0\bar{P}_-(x) )_i \nn \\
   & & \hspace{7mm} \times \dd (P_+\Psi_{uI}(x))_i \, \dd (\bar{\Psi}_{uI}\bar{P}_-(x))_i \, 
                           \dd (\bar{P}_-\Psi_{dI}(x))_i \, \dd (\bar{\Psi}_{dI}(x)P_+)_i , 
    \label{measure_mat+I}                     \\
\dd \mu_{{\rm mat}, -I'}  & \equiv & \prod_x \prod_{i=1}^N \dd\left(P_-\Phi_{I'}(x)\right)_i \, 
                                           \dd (\Phi_{I'}(x)^\dagger P_-)_i \, \dd (\bar{P}_+\gamma_0 F_{I'}(x))_i \, 
                                           \dd (F_{I'}^\dagger\gamma_0\bar{P}_+(x))_i \nn \\
   & & \hspace{7mm} \times \dd (P_-\Psi_{uI'}(x))_i \, \dd (\bar{\Psi}_{uI'}\bar{P}_+(x))_i \, 
                                        \dd (\bar{P}_+\Psi_{dI'}(x))_i \, \dd (\bar{\Psi}_{dI'}(x)P_-)_i.    
\eea
Together with the measure for the SYM sector (\ref{measure_SYM}), 
\be
\dd\mu =  \left(\dd\mu_{\rm 2DSYM} \right) \left( \dd\mu_{\rm mat} \right)
\ee
gives the measure of the total system. 
Here, we examine the transformation properties of the measures under the following various transformations.  

\paragraph{Gauge Transformation}
An element $g(x)=\e^{i\omega(x)} \in G$ with $\omega(x)$ being infinitesimal transforms the variables (\ref{chiral_fields_+_a}) 
and (\ref{anti-chiral_fields_+_a}) as 
\bea
P_+ \Phi_I(x) & \to & g(x) P_+\Phi_I(x) = (1+i\omega(x)P_+) P_+\Phi_I(x), \nn \\
\Phi_I(x)^\dagger P_+ & \to & \Phi_I(x)^\dagger P_+ g(x)^{-1} = \Phi_I(x)^\dagger P_+ (1-iP_+ \omega(x)), \nn \\
\bar{P}_-\gamma_0 F_I(x) & \to & g(x) \bar{P}_-\gamma_0 F_I(x) = (1+i\omega(x) \bar{P}_-) \bar{P}_-\gamma_0 F_I(x) , \nn \\
F_I^\dagger \gamma_0 \bar{P}_-(x) & \to & F_I^\dagger \gamma_0 \bar{P}_-(x) g(x)^{-1} 
                            = F_I^\dagger \gamma_0 \bar{P}_-(1-i\bar{P}_-\omega)(x),  \\
    & &  \nn \\                        
P_+\Psi_{uI}(x) & \to & g(x) P_+\Psi_{uI}(x) = (1+i\omega(x) P_+) P_+\Psi_{uI}(x), \nn \\
\bar{\Psi}_{uI}\bar{P}_-(x) & \to & \bar{\Psi}_{uI} \bar{P}_-(x) g(x)^{-1} = \bar{\Psi}_{uI} \bar{P}_-(1-i\bar{P}_-\omega)(x), \nn \\
\bar{P}_-\Psi_{dI}(x) & \to & g(x) \bar{P}_-\Psi_{dI}(x) = (1+i\omega(x)\bar{P}_-)\bar{P}_-\Psi_{dI}(x), \nn \\
\bar{\Psi}_{dI}(x) P_+ & \to & \bar{\Psi}_{dI}(x) P_+ g(x)^{-1} = \bar{\Psi}_{dI}(x) P_+ (1-iP_+\omega(x)).                             
\eea
For the bosonic variables $P_+ \Phi_I(x)$,  
the $\cO(\omega)$ part of Jacobians from the measures cancels with that from their conjugates. 
For the fermionic variables $P_+\Psi_{uI}(x)$ and $\bar{\Psi}_{dI}(x) P_+$, the cancellation occurs among themselves. 
As to the field variables projected by $\bar P_-$, 
$\bar{\Psi}_{uI}\bar{P}_-(x)$,  $\bar{P}_-\Psi_{dI}(x)$ and $\bar{P}_-\gamma_0 F_I(x)$, 
we specify the path-integral measure by introducing 
the chiral basis~\cite{Narayanan:1993sk,Narayanan:1994gw,Luscher:1998du, Luscher:1999un}\footnote{We 
assume that the lattice size is $La \times La$ with any suitable boundary conditions which preserve the $Q$-supersymmetry.}
\begin{equation}
\bar P_-  v_{ k}(x) = v_{ k}(x) \qquad (k=1, \cdots, N \times L^2).
\end{equation}
The path-integral measure is then expressed by the integrals over the coefficients 
of the expansion of the field variables:
\begin{eqnarray}
&& 
\bar{P}_-\gamma_0 F_I(x) = \sum_k  b_{I k}  \, v_{ k}(x), \qquad 
F_I^\dagger\gamma_0\bar{P}_-(x) = \sum_k  v_{ k}^\dagger(x) \, \bar b_{I k} , 
\nonumber\\
&&
\bar{P}_-\Psi_{dI}(x) = \sum_k  c_{dI k}  \, v_{ k}(x),
\qquad
\bar{\Psi}_{uI}\bar{P}_-(x) = \sum_k  v_{ k}^\dagger(x) \, \bar c_{uI k}, 
\end{eqnarray}
respectively. With this choice of the basis, one can see that 
the gauge field dependence of the path-integral measure exactly cancels out 
for each  flavor $I=1,\cdots, n_+$.  Therefore the measure is defined smoothly and gauge-invariantly for any admissible gauge fields. 
%
Thus, the gauge invariance of $\dd \mu_{{\rm mat}, +I}$ (and of $\dd \mu_{{\rm mat}, -I'}$ from the similar argument) 
is shown. $\left( \dd\mu_{\rm 2dSYM}\right)$ is clearly gauge invariant.

\paragraph{$Q$-supersymmetry Transformation}
Under the $Q$-supersymmetry transformation (\ref{Qmodified_+twm_a}) and (\ref{Qmodified_-twm_a}) 
with a Grassmann parameter $\varepsilon$, 
the variables (\ref{chiral_fields_+_a}) and (\ref{anti-chiral_fields_+_a}) change as 
\bea
P_+\Phi_I(x) & \to & (1+i\varepsilon Q) P_+\Phi_I(x) = P_+\Phi_I(x) + \cdots,   \nn \\
\Phi_I(x)^\dagger P_+ & \to & (1+i\varepsilon Q) \Phi_I(x)^\dagger P_+ = \Phi_I(x)^\dagger P_+ + \cdots, \nn \\
\bar{P}_-\gamma_0 F_I(x) & \to & (1+i\varepsilon Q) \bar{P}_-\gamma_0 F_I(x) =
                                                 \left[1+i\varepsilon (Q\bar{P}_-)\bar{P}_-\right] \bar{P}_-\gamma_0 F_I(x) + \cdots, \nn \\
F_I^\dagger\gamma_0 \bar{P}_-(x) & \to & (1+i\varepsilon Q) F_I^\dagger\gamma_0 \bar{P}_-(x) =
                                                 F_I^\dagger\gamma_0 \bar{P}_- \left[1+i\varepsilon \bar{P}_-(Q\bar{P}_-)\right](x) +\cdots, \\
   & & \nn \\                                   
P_+\Psi_{uI}(x) & \to & (1+i\varepsilon Q) P_+\Psi_{uI}(x) = P_+\Psi_{uI}(x) + \cdots,  \nn \\
\bar{\Psi}_{uI}\bar{P}_-(x) & \to &  (1+i\varepsilon Q) \bar{\Psi}_{uI}\bar{P}_-(x) = 
                                                 \bar{\Psi}_{uI}\bar{P}_- \left[1+i\varepsilon \bar{P}_-(Q\bar{P}_-)\right](x) + \cdots, \nn \\
\bar{P}_-\Psi_{dI}(x) & \to & (1+i\varepsilon Q) \bar{P}_-\Psi_{dI}(x) = 
                                                      \left[1+i\varepsilon (Q\bar{P}_-)\bar{P}_-\right]\bar{P}_-\Psi_{dI}(x) + \cdots, \nn \\
\bar{\Psi}_{dI}(x)P_+ & \to & (1+i\varepsilon Q) \bar{\Psi}_{dI}(x)P_+ = \bar{\Psi}_{dI}(x)P_+ + \cdots,  
\eea 
where ``$\cdots$'' correspond to off-diagonal elements of Jacobi matrices and are irrelevant for the calculation.
For example, the measure $\prod_x \prod_{i=1}^N\dd (\bar{P}_-\gamma_0 F_I(x))_i$ contributes to the Jacobian factor by 
\be
\Det\left[1+i\varepsilon (Q\bar{P}_-)\bar{P}_-\right] = 1+ i\varepsilon \, \Tr\left[(Q\bar{P}_-)\bar{P}_-\right] 
= 1+ i\varepsilon \, \Tr\left[\bar{P}_-(Q\bar{P}_-)\bar{P}_-\right] =1. 
\ee
$\bar{P}_- = \bar{P}_-^2$ and (\ref{identity_bar}) was used. 
(``$\Det$'' and ``$\Tr$'' respectively represent the determinant and the trace with respect to all of the sites, Dirac and color indices.) 
Repeating the same computation, we can show that $\dd\mu_{{\rm mat}, +I}$ is $Q$-invariant and $\dd\mu_{{\rm mat}, -I'}$ 
is also. The $Q$-invariance of $\left(\dd\mu_{\rm 2dSYM}\right)$ is shown in ref.~\cite{sugino4}. 

\paragraph{${\rm U}(1)_A$ Transformation}
The ${\rm U}(1)_A$ transformation given by the first line of (\ref{U(1)_A_cont}) for the SYM variables and  
\bea
P_+\Psi_{uI}(x) & \to & \e^{i\alpha} P_+\Psi_{uI}(x) = (1+i\alpha P_+)P_+\Psi_{uI}(x), \nn \\
\bar{\Psi}_{uI}\bar{P}_-(x) & \to & \bar{\Psi}_{uI}\bar{P}_-(x) \e^{-i\alpha} = \bar{\Psi}_{uI}\bar{P}_-(1-i\alpha\bar{P}_-)(x), \nn \\
\bar{P}_-\Psi_{dI}(x) & \to & \e^{-i\alpha}\bar{P}_-\Psi_{dI}(x) = (1-i\alpha \bar{P}_-) \bar{P}_-\Psi_{dI}(x), \nn \\
\bar{\Psi}_{dI}(x)P_+ & \to & \bar{\Psi}_{dI}(x) P_+ \e^{i\alpha} = \bar{\Psi}_{dI}(x) P_+ (1+i\alpha P_+), 
\label{U(1)_A_+} \\
& & \nn \\
P_- \Psi_{uI'}(x) & \to & \e^{i\alpha} P_- \Psi_{uI'}(x) = (1+i\alpha P_-) P_-\Psi_{uI'}(x), \nn \\
\bar{\Psi}_{uI'}\bar{P}_+(x) & \to & \bar{\Psi}_{uI'}\bar{P}_+(x) \e^{-i\alpha} = \bar{\Psi}_{uI'}\bar{P}_+(1-i\alpha \bar{P}_+)(x), \nn \\
\bar{P}_+\Psi_{dI'}(x) & \to & \e^{-i\alpha}\bar{P}_+\Psi_{dI'}(x) = (1-i\alpha \bar{P}_+) \bar{P}_+ \Psi_{dI'}(x), \nn \\
\bar{\Psi}_{dI'}(x) P_- & \to & \bar{\Psi}_{dI'}(x) P_- \e^{i\alpha} = \bar{\Psi}_{dI'}(x) P_- (1+i\alpha P_-) 
\label{U(1)_A_-}
\eea
for the matter variables (with the others unchanged) 
is a symmetry of the lattice actions (\ref{S_mat+twm_hD_a2}) and (\ref{S_mat-twm_hD_a2}). 
The parameter $\alpha\in {\bf R}$ is assumed to be infinitesimal for the r.h.s. in (\ref{U(1)_A_+}) 
and (\ref{U(1)_A_-}). 
Then, the path-integral measures change as 
\bea
\dd \mu_{{\rm mat}, +I} & \to & \left[1-2i\alpha \Tr (P_+ - \bar{P}_-)\right] \dd\mu_{{\rm mat}, +I}
                                      = \left[1+i\alpha \Tr(\gamma_3 a\hD)\right] \dd\mu_{{\rm mat}, +I}, 
\label{U(1)_A_measure_+I}          \\
\dd \mu_{{\rm mat}, -I'} & \to & \left[1+2i\alpha \Tr (\bar{P}_+ - P_-)\right] \dd\mu_{{\rm mat}, -I'}
                                      = \left[1-i\alpha \Tr(\gamma_3 a\hD)\right] \dd\mu_{{\rm mat}, -I'}.                       
\eea
Thus, 
\be
\left( \dd\mu_{\rm mat}\right) \to \left[1+i\alpha \, (n_+ -n_-) \, \Tr(\gamma_3 a\hD)\right] \left(\dd\mu_{\rm mat}\right) . 
\ee
$\Tr(\gamma_3 a\hD)$ has been computed in the two-dimensional case \cite{kikukawa-yamada}, 
assuming that $A_\mu(x)$ appearing in the expansion of $U_\mu(x) = \e^{iaA_\mu(x)}$ 
are smooth external variables. 
We obtain 
\bea
\tr_{\rm spin}\left(\gamma_3 a\hD\right) (x,x)  & \simeq  & \frac{1}{\pi} a^2 F_{01}(x), 
\label{tr_spin_anomaly} \\
\Tr\, \left(\gamma_3 a\hD \right)  & \simeq  & \frac{1}{\pi} \int \dd^2x \, \tr \, F_{01} \qquad (a\to 0) , 
\label{gamma_hD_cont}
\eea      
where the trace with the suffix ``spin'' means taking the trace with respect to the Dirac indices.                             
This correctly reproduces the ${\rm U}(1)_A$ anomaly obtained in the continuum theory 
or in the lattice perturbation~\cite{sugino_sqcd}. 

On the other hand, $\left(\dd\mu_{\rm 2dSYM}\right)$ is ${\rm U}(1)_A$-invariant.

\paragraph{${\rm U}(n_+) \times {\rm U}(n_-)$ Transformation}
Although the flavor rotational symmetry ${\rm U}(n_+)\times {\rm U}(n_-)$ breaks down to ${\rm U}(1)^{n_+}\times {\rm U}(1)^{n_-}$ 
in the presence of general twisted masses in the matter action, 
we can show that the measure itself maintains the full flavor symmetry. 
Denoting $(U, V) \in {\rm U}(n_+) \times {\rm U}(n_-)$, the transformation acts as 
\bea
{\cal X}_I \to \sum_{J=1}^{n_+} U_{IJ} {\cal X}_J, &  &  {\cal X}_I^\dagger  \to \sum_{J=1}^{n_+} {\cal X}_J^\dagger (U^{-1})_{JI}, \nn \\
{\cal Y}_{I'} \to \sum_{J'=1}^{n_-} V_{I'J'} {\cal Y}_{J'}, &  &  {\cal Y}_{I'}^\dagger  \to \sum_{J'=1}^{n_-} {\cal Y}_{J'}^\dagger (V^{-1})_{J'I'}
\eea
with 
\bea
{\cal X}_I = P_+\Phi_I, \, \bar{P}_-\gamma_0 F_I, \, P_+\Psi_{uI}, \, \bar{P}_-\Psi_{dI}, & & 
{\cal X}_I^\dagger = \Phi_I^\dagger P_+, \, F_I^\dagger \gamma_0 \bar{P}_-, \, \bar{\Psi}_{uI}\bar{P}_-, \, \bar{\Psi}_{dI} P_+, \nn \\
{\cal Y}_{I'} = P_-\Phi_{I'}, \, \bar{P}_+\gamma_0 F_{I'}, \, P_-\Psi_{uI'}, \, \bar{P}_+\Psi_{dI'}, & & 
{\cal Y}_{I'}^\dagger = \Phi_{I'}^\dagger P_-, \, F_{I'}^\dagger \gamma_0\bar{P}_+, \, \bar{\Psi}_{uI'} \bar{P}_+, \, \bar{\Psi}_{dI'} P_-. 
\nn 
\eea
Contribution to the Jacobian from bosons cancels with their conjugates, 
and that from fermions cancels between $u$- and $d$-variables. 
Thus, the invariance of 
$\left(\prod_{I=1}^{n_+} \dd \mu_{{\rm mat}, +I} \right)$ and  
$\left(\prod_{I'=1}^{n_-} \dd \mu_{{\rm mat}, -I'} \right)$ can be seen. 

\paragraph{${\rm U}(1)_V$ Transformation}
The ${\rm U}(1)_V$ transformation is given by the first line of (\ref{U(1)_V_cont}) for the SYM sector and 
\bea
\bar{P}_-\gamma_0 F_I(x) \to  \e^{-i2\alpha} \bar{P}_-\gamma_0 F_I(x), & & 
F_I^\dagger\gamma_0 \bar{P}_-(x)  \to F_I^\dagger\gamma_0 \bar{P}_-(x) \e^{i2\alpha},\nn \\
P_+\Psi_{uI}(x) \to \e^{-i\alpha} P_+\Psi_{uI}(x), & & 
\bar{\Psi}_{uI}\bar{P}_-(x) \to \bar{\Psi}_{uI}\bar{P}_-(x) \e^{i\alpha}, \nn \\                                          
\bar{P}_-\Psi_{dI}(x) \to \e^{-i\alpha}\bar{P}_-\Psi_{dI}(x), & & 
\bar{\Psi}_{dI}(x) P_+ \to \bar{\Psi}_{dI}(x) P_+ \e^{i\alpha}, 
\label{U(1)_V_lat_1}
\\
& & \nn \\
\bar{P}_+\gamma_0 F_{I'}(x) \to \e^{i2\alpha} \bar{P}_+\gamma_0 F_{I'}(x), & & 
F_{I'}^\dagger\gamma_0 \bar{P}_+(x) \to F_{I'}^\dagger\gamma_0 \bar{P}_+(x) \e^{-i2\alpha},\nn \\                                             
P_-\Psi_{uI'}(x) \to  \e^{i\alpha} P_-\Psi_{uI'}(x), & & 
\bar{\Psi}_{uI'}\bar{P}_+(x) \to \bar{\Psi}_{uI'}\bar{P}_+(x) \e^{-i\alpha}, \nn \\            
\bar{P}_+\Psi_{dI'}(x) \to \e^{i\alpha}\bar{P}_+\Psi_{dI'}(x), & & 
\bar{\Psi}_{dI'}(x) P_- \to  \bar{\Psi}_{dI'}(x) P_- \e^{-i\alpha}         
\label{U(1)_V_lat_2}                   
\eea 
for the matter sector. The other variables do not change, and $\alpha\in {\bf R}$. 
Although the lattice action loses the ${\rm U}(1)_V$ symmetry by the latticization, the measures 
$\dd\mu_{{\rm mat}, +I}$, $\dd\mu_{{\rm mat}, -I'}$ can be shown invariant in a similar manner to the case 
of the gauge transformation. 
The invariance of $\left(\dd\mu_{\rm 2DSYM}\right)$ is clear.

\subsection{Admissibility Condition}
We already have the admissibility condition (\ref{admissibility}) with (\ref{epsilon_vac}) or (\ref{adm_theta}) 
in constructing the SYM sector. 
In addition, in order for the overlap Dirac operator $\hD$ to be well-defined, there will arise a further constraint on 
the choice of $\epsilon$. 

Tracing the computation of appendix~C in ref.~\cite{HJL} in the two-dimensional case, we have the bound 
\be
||X^\dagger X|| \ge 1-5\epsilon.   
\ee
It imposes a constraint 
\be
0<\epsilon <\frac15  
\label{epsilon_hD}
\ee
for the well-definedness of $\hD$. 

Comparing it with the condition from the SYM sector, we find the choice of $\epsilon$ as follows. \\
$G={\rm U}(N)$ without $\vartheta$-term :
\bea
0<\epsilon <\frac15 & & \mbox{for } N=1,2,\cdots, 100 \nn \\
0<\epsilon <\frac{2}{\sqrt{N}} & & \mbox{for } N\ge 101, 
\label{adm_U(N)_theta}
\eea
$G={\rm U}(N)$ with $\vartheta$-term :
\bea
0<\epsilon <\frac15 & & \mbox{for } N=1,2,\cdots, 25 \nn \\
0<\epsilon <\frac{1}{\sqrt{N}} & & \mbox{for } N\ge 26, 
\eea
$G={\rm SU}(N)$ :
\bea
0<\epsilon <\frac15 & & \mbox{for } N=2,3,\cdots, 31 \nn \\
0<\epsilon < 2\sin\left(\frac{\pi}{N}\right)  & & \mbox{for } N\ge 32.  
\eea

\setcounter{equation}{0}
\section{Matter Multiplets of $\det^q$-representation}
\label{sec:mat_det}
We will latticize gauged linear sigma models in the next section, that are examples 
of models which couple matter multiplets belonging to different representations of $G$~\cite{witten2,hori-tong}. 
As a preparation, we here consider the matter chiral multiplets charged under the central ${\rm U}(1)$ of 
$G={\rm U}(N)$: 
\bea
\Xi_{+\bA} = \left(\xi_{+\bA}, \zeta_{+\bA}, G_{+\bA} \right) & & (\bA =1, \cdots, \ell_+), \nn \\
\Xi_{-\bA'} = \left(\xi_{-\bA'}, \zeta_{-\bA'}, G_{-\bA'} \right) & & (\bA' =1, \cdots, \ell_-). 
\eea 
They transform under $g(x)\in G$ as 
\bea
\Xi_{+ \bA}(x) \to (\det g(x))^{ q_{\bA}} \, \Xi_{+ \bA}(x),  & & 
\Xi_{+ \bA}(x)^* \to (\det g(x))^{- q_{\bA}} \, \Xi_{+ \bA}(x)^*, \nn \\
\Xi_{- \bA'}(x) \to (\det g(x))^{ -q_{\bA'}} \, \Xi_{- \bA'}(x),  & & 
\Xi_{- \bA'}(x)^* \to (\det g(x))^{ q_{\bA'}} \, \Xi_{- \bA'}(x)^*,
\eea
in other words, under $g(x)=1+i\omega(x)$ with $\omega(x)$ infinitesimal as
\bea
\delta \Xi_{+ \bA}(x) = iq_{\bA} \left(\tr \, \omega(x)\right) \Xi_{+ \bA}(x), & & 
\delta \Xi_{+ \bA}(x)^* = - iq_{\bA} \left(\tr \, \omega(x)\right) \Xi_{+ \bA}(x)^*,  \nn \\
\delta \Xi_{- \bA'}(x) = -iq_{\bA'} \left(\tr \, \omega(x)\right) \Xi_{- \bA'}(x), & & 
\delta \Xi_{- \bA'}(x)^* =  iq_{\bA'} \left(\tr \, \omega(x)\right) \Xi_{- \bA'}(x)^*,
\eea
from which the covariant derivatives are given by 
\bea
\cD_\mu \Xi_{+ \bA} = \left(\der_\mu + iq_{\bA} (\tr \, A_\mu) \right) \Xi_{+ \bA}, & & 
\cD_\mu \Xi_{+ \bA}^* = \left(\der_\mu - iq_{\bA} (\tr \, A_\mu) \right) \Xi_{+ \bA}^*, \nn \\
\cD_\mu \Xi_{- \bA'} = \left(\der_\mu - iq_{\bA'} (\tr \, A_\mu) \right) \Xi_{- \bA'}, & & 
\cD_\mu \Xi_{- \bA'}^* = \left(\der_\mu + iq_{\bA'} (\tr \, A_\mu) \right) \Xi_{- \bA'}^*. 
\label{cD_det}
\eea 
The lagrangian density of the continuum theory is expressed in dimensional reduction of the four-dimensional 
superfield formalism as 
\be
\tttt{\sum_{\bA =1}^{\ell_+} \Xi_{+\bA}^*  \, \e^{q_{\bA} \, \tr \, V -\widetilde{V}'_{+\bA}} \, \Xi_{+\bA}} 
+ \tttt{\sum_{\bA'=1}^{\ell_-} \Xi_{-\bA'} \, \e^{-q_{\bA'} \, \tr \,V +\widetilde{V}'_{-\bA'}} \, \Xi_{-\bA'}^*}
\ee
with twisted masses 
\be
\widetilde{V}'_{+ \bA} \equiv  2\theta_R\bar{\theta}_L \twm'_{+ \bA} + 2\theta_L \bar{\theta}_R \twm'^*_{+ \bA}, \qquad 
\widetilde{V}'_{- \bA'} \equiv  2\theta_R\bar{\theta}_L \twm'_{- \bA'} + 2\theta_L \bar{\theta}_R \twm'^*_{- \bA'}. 
\ee 
The Euclidean action in terms of the component fields becomes 
\bea
S'^{(E)}_{{\rm mat}, \twm'} & = & S'^{(E)}_{{\rm mat}, +\twm'} + S'^{(E)}_{{\rm mat}, -\twm'}, \nn \\
S'^{(E)}_{{\rm mat}, +\twm'} & = & \int \dd^2x \sum_{\bA =1}^{\ell_+} \left[\cD_\mu \xi_{+\bA}^* \, \cD_\mu \xi_{+\bA} 
                                           -G_{+\bA}^*G_{+\bA}\right. \nn \\
  & & \hspace{2cm}  +\xi_{+\bA}^* \left(q_{\bA} \, \tr \, \phi -\twm'_{+\bA}\right) 
                               \left( q_{\bA} \, \tr \, \bar{\phi} -\twm'^*_{+\bA}\right) \xi_{+\bA} 
        -q_{\bA}\xi_{+\bA}^*\left(\tr D\right) \xi_{+\bA} \nn \\
  & & \hspace{2cm} +2\bar{\zeta}_{+\bA R} \, \cD_z \zeta_{+\bA R} + 2\bar{\zeta}_{+\bA L} \, \cD_{\bar{z}} \zeta_{+\bA L} \nn \\
  & & \hspace{2cm} +\bar{\zeta}_{+\bA R}\left(q_{\bA} \, \tr \, \bar{\phi} -\twm'^*_{+\bA}\right) \zeta_{+\bA L} 
        + \bar{\zeta}_{+\bA L} \left(q_{\bA} \, \tr \, \phi -\twm'_{+\bA}\right) \zeta_{+\bA R} \nn \\
  & & \hspace{2cm}  -i\sqrt{2} \, q_{\bA} \left\{\xi_{+\bA}^*\left((\tr \, \lambda_L) \, \zeta_{+\bA R} 
                                    -(\tr \, \lambda_R) \, \zeta_{+\bA L} \right) \right. \nn \\
 & & \hspace{3cm}       \left.\left. +\left(-\bar{\zeta}_{+\bA R} (\tr \, \bar{\lambda}_L) 
                        + \bar{\zeta}_{+\bA L} (\tr \, \bar{\lambda}_R)\right) \xi_{+\bA}\right\} \right], \\
S'^{(E)}_{{\rm mat}, -\twm'} & = &  \int \dd^2x \sum_{\bA' =1}^{\ell_-} \left[\cD_\mu \xi_{-\bA'} \, \cD_\mu \xi_{-\bA'}^* 
                                           -G_{-\bA'}G_{-\bA'}^* \right. \nn \\
  & & \hspace{2cm}   +\xi_{-\bA'} \left(q_{\bA'} \, \tr \, \phi -\twm'_{-\bA'}\right) 
                               \left( q_{\bA'} \, \tr \, \bar{\phi} -\twm'^*_{-\bA'}\right) \xi_{-\bA'}^* 
        +q_{\bA'}\xi_{-\bA'} \left(\tr D\right) \xi_{-\bA'}^* \nn \\          
 & & \hspace{2cm} +2\zeta_{-\bA' R} \, \cD_z \bar{\zeta}_{-\bA' R} + 2\zeta_{-\bA' L} \, \cD_{\bar{z}} \bar{\zeta}_{-\bA' L} \nn \\
 & & \hspace{2cm} +\zeta_{-\bA' L}\left(q_{\bA'} \, \tr \, \bar{\phi} -\twm'^*_{-\bA'}\right) \bar{\zeta}_{-\bA' R} 
        + \zeta_{-\bA' R} \left(q_{\bA'} \, \tr \, \phi -\twm'_{-\bA'}\right) \bar{\zeta}_{-\bA' L} \nn \\
  & & \hspace{2cm}  -i\sqrt{2} \, q_{\bA'} \left\{ \left(-\zeta_{-\bA' L} \, (\tr\, \lambda_R)  
                                     +\zeta_{-\bA' R} \, (\tr \, \lambda_L) \right) \xi_{-\bA'}^* \right. \nn \\
 & & \hspace{3cm}         \left.\left. +\xi_{-\bA'} \left((\tr \, \bar{\lambda}_R) \, \bar{\zeta}_{-\bA' L}  
                              -  (\tr \, \bar{\lambda}_L) \, \bar{\zeta}_{-\bA' R}\right) \right\} \right]. 
\eea 
One of the supersymmetries of the theory, which will be respected in the lattice formulation, is 
\bea
& & Q\xi_{+\bA} = -\zeta_{+\bA L}, \qquad Q\zeta_{+\bA L}=-\left( q_{\bA} \, \tr \,\phi -\twm'_{+\bA}\right) \xi_{+\bA}, \nn \\
& & Q\zeta_{+\bA R} = \left(\cD_0 +i\cD_1\right) \xi_{+\bA} + G_{+\bA}, \nn \\
& & QG_{+\bA} = \left(\cD_0+i\cD_1\right) \zeta_{+\bA L} + \left( q_{\bA} \, \tr \, \phi -\twm'_{+\bA}\right) \zeta_{+\bA R}  
                              -iq_{\bA} \, \tr \, (\psi_0 +i\psi_1) \, \xi_{+\bA}, \nn \\
& & Q\xi_{+\bA}^* = -\bar{\zeta}_{+\bA R}, \qquad 
             Q\bar{\zeta}_{+\bA R} = \xi_{+\bA}^* \left(q_{\bA} \, \tr \, \phi -\twm'_{+\bA}\right), \nn \\
& & Q\bar{\zeta}_{+ \bA L} = \left(\cD_0 -i\cD_1\right) \xi_{+\bA}^* +G_{+\bA}^*, \nn \\
& & QG_{+\bA}^* =  \left(\cD_0 -i\cD_1\right) \bar{\zeta}_{+\bA R} 
                           -\bar{\zeta}_{+\bA L} \left(q_{\bA} \, \tr \, \phi -\twm'_{+\bA}\right) 
                            +iq_{\bA} \, \xi_{+\bA}^* \, \tr \, (\psi_0 -i\psi_1), 
\eea
\bea
& & Q\xi_{-\bA'} = -\zeta_{-\bA' L}, \qquad Q\zeta_{-\bA' L} = -\xi_{-\bA'} \left(q_{\bA'} \, \tr \, \phi -\twm'_{-\bA'}\right), \nn \\
& & Q\zeta_{-\bA' R} = \left(\cD_0+i\cD_1\right) \xi_{-\bA'} +G_{-\bA'}, \nn \\
& & QG_{-\bA'} = \left(\cD_0+i\cD_1\right) \zeta_{-\bA' L} -\zeta_{-\bA' R} \left(q_{\bA'} \, \tr \, \phi -\twm'_{-\bA'}\right) 
                          +iq_{\bA'} \, \xi_{-\bA'} \, \tr \, (\psi_0+i\psi_1), \nn \\
& & Q\xi_{-\bA'}^* = -\bar{\zeta}_{-\bA' R} , \qquad 
                 Q\bar{\zeta}_{-\bA' R} = -\left(q_{\bA'} \, \tr \, \phi -\twm'_{-\bA'}\right) \xi_{-\bA'}^* , \nn \\
& & Q\bar{\zeta}_{-\bA' L} = \left( \cD_0 -i\cD_1\right) \xi_{-\bA'}^* + G_{-\bA'}^*, \nn \\
& & QG_{-\bA'}^* =  \left( \cD_0 -i\cD_1\right) \bar{\zeta}_{-\bA' R} 
                                + \left(q_{\bA'} \, \tr \, \phi -\twm'_{-\bA'}\right) \bar{\zeta}_{-\bA' L} 
                                  -iq_{\bA'} \, \tr \, (\psi_0 -i\psi_1) \, \xi_{-\bA'}^* , 
\eea 
where the gauginos are renamed as (\ref{rename}).  
The Euclidean action can be rewritten as the $Q$-exact form: 
\bea
S'^{(E)}_{{\rm mat}, +\twm'} & = & Q \int \dd^2x \sum_{\bA =1}^{\ell_+} \frac12 \left[ \bar{\zeta}_{+\bA L} 
                  \left\{ \left(\cD_0+i\cD_1\right)\xi_{+\bA} -G_{+\bA} \right\} \right. \nn \\
   & &  \hspace{2.5cm} +\left\{ \left(\cD_0-i\cD_1\right) \xi_{+\bA}^* -G_{+\bA}^*\right\} \zeta_{+\bA R} \nn \\
  & & \hspace{2.5cm}+\bar{\zeta}_{+\bA R} \left(q_{\bA} \, \tr \, \bar{\phi} -\twm'^*_{+\bA}\right) \xi_{+\bA} 
        -\xi_{+\bA}^* \left( q_{\bA} \, \tr \, \bar{\phi} -\twm'^*_{+\bA} \right) \zeta_{+\bA L} \nn \\
 & &  \hspace{2.5cm}\left. +2i q_{\bA} \, \xi_{+\bA}^* \left(\tr \, \chi \right) \xi_{+\bA}\right], \\
S'^{(E)}_{{\rm mat}, -\twm'} & = &  Q \int \dd^2x \sum_{\bA' =1}^{\ell_-} \frac12 \left[
                      \left\{ \left(\cD_0+i\cD_1\right) \xi_{-\bA'} -G_{-\bA'}\right\} \bar{\zeta}_{-\bA' L} \right. \nn \\
 & & \hspace{2.5cm}+\zeta_{-\bA' R} \left\{ \left(\cD_0 -i\cD_1\right) \xi_{-\bA'}^* -G_{-\bA'}^*\right\} \nn \\
 & & \hspace{2.5cm}-\xi_{-\bA'} \left(q_{\bA'} \, \tr \, \bar{\phi} -\twm'^*_{-\bA'}\right) \bar{\zeta}_{-\bA' R} 
       +\zeta_{-\bA' L} \left(q_{\bA'} \, \tr \, \bar{\phi} -\twm'^*_{-\bA'}\right) \xi_{-\bA'}^* \nn \\
 & & \hspace{2.5cm}\left. -2i q_{\bA'} \, \xi_{-\bA'} \left(\tr \, \chi\right) \xi_{-\bA'}^*\right].
\label{SE_mat_det-twm} 
\eea
When the twisted masses are general, 
the flavor symmetry of the action becomes ${\rm U}(1)^{\ell_+} \times {\rm U}(1)^{\ell_-}$. 
Similarly to the case of the (anti-)fundamental matters, the fields have the $(U(1)_V, U(1)_A)$ charges 
\bea
 & & \zeta_{+\bA L}: (-1, 1), \qquad \zeta_{+\bA R}: (-1, -1), \qquad G_{+\bA}(-2, 0),  \nn \\
 & & \bar{\zeta}_{+\bA L}: (1, -1), \qquad \bar{\zeta}_{+\bA R}: (1,1), \qquad G_{+\bA}^*: (2, 0),    \nn \\
 & & \zeta_{-\bA' L}: (-1, 1), \qquad \zeta_{-\bA' R}: (-1,-1), \qquad G_{-\bA'} : (-2, 0), \nn \\
 & & \bar{\zeta}_{-\bA' L}: (1, -1), \qquad \bar{\zeta}_{-\bA' R}: (1,1), \qquad G_{-\bA'}^*: (2, 0), 
\eea
and the others have $(0,0)$.

\subsection{Lattice Formulation with Overlap Dirac Operator}
Similarly to the case of the (anti-)fundamental matters, we can construct the corresponding lattice theory using the 
overlap Dirac operator. 

First, adding some multiplets to prepare the same number of the $\det^{q_{\bA}}$- and $\det^{-q_{\bA}}$-multiplets 
($\ell_0 =\max(\ell_+, \ell_-)$), 
let us introduce the doublet notations 
\bea
& & \Xi_{\bA} \equiv \left( \begin{array}{c} \xi_{+\bA} \\ \xi_{-\bA}^*\end{array}\right), \qquad 
\Xi_{\bA}^\dagger \equiv  \left(\xi_{+\bA}^*, \xi_{-\bA}\right), \nn \\
& & \cZ_{u\bA} \equiv \left( \begin{array}{c} \zeta_{+\bA L} \\ \bar{\zeta}_{-\bA R} \end{array}\right), \qquad 
\cZ_{d\bA} \equiv \left( \begin{array}{c} \bar{\zeta}_{-\bA L} \\ \zeta_{+\bA R} \end{array} \right), \nn \\
& & \bar{\cZ}_{u\bA} \equiv \left(\zeta_{-\bA R}, \bar{\zeta}_{+\bA L} \right), \qquad 
\bar{\cZ}_{d\bA} \equiv \left(\bar{\zeta}_{+\bA R}, \zeta_{-\bA L}\right), \nn \\
& & G_{\bA} \equiv \left( \begin{array}{c} G_{+\bA} \\ G_{-\bA}^* \end{array} \right), \qquad 
G_{\bA}^\dagger \equiv \left(G_{+\bA}^*, G_{-\bA}\right) \qquad (\bA=1, \cdots, \ell_0), 
\eea 
and the forward (backward) covariant differences $D_\mu$ ($D_\mu^*$): 
\bea
aD_\mu \,\Xi_{\bA}(x) & = & (\det U_\mu (x))^{ q_{\bA}} \, \Xi_{\bA}(x+\hat{\mu}) -\Xi_{\bA}(x), \nn \\
aD_\mu^* \, \Xi_{\bA}(x) & = & \Xi_{\bA}(x) -(\det U_\mu(x-\hat{\mu}))^{- q_{\bA}} \, \Xi_{\bA}(x-\hat{\mu})
\label{D_lat_det1}
\eea
with the same for $\cZ_{u\bA}, \cZ_{d\bA}, G_{\bA}$, and 
\bea
aD_\mu \, \Xi_{\bA}(x)^\dagger & = & (\det U_\mu(x))^{- q_{\bA}} \, \Xi_{\bA}(x+\hat{\mu})^\dagger -\Xi_{\bA}(x)^\dagger, \nn \\ 
aD_\mu^* \, \Xi_{\bA}(x)^\dagger & = & \Xi_{\bA}(x)^\dagger -(\det U_\mu (x-\hat{\mu}))^{q_{\bA}}\, \Xi_{\bA}(x-\hat{\mu})^\dagger   
\label{D_lat_det2}
\eea
with the same for $\bar{\cZ}_{u\bA}, \bar{\cZ}_{d\bA}, G_{\bA}^\dagger$. 
{}From these, the Wilson-Dirac operator $D_W$, the overlap Dirac operator $\hD$ and the chiral projectors $\bar{P}_\pm$ 
are constructed as in the case of the (anti-)fundamental matters.  

$Q(a\hD)$ is expressed as (\ref{QhD}) with
\bea
Q(aD_W) \, {\cal X}(x) & = & \sum_{\mu=0}^1 \left[\frac{\gamma_{\mu}-r}{2} \left(iq_{\bA} \, \tr \, \psi_{\mu}(x)\right) 
                                 \left(\det U_{\mu}(x) \right)^{q_{\bA}} \, {\cal X}(x+\hat{\mu}) \right. \nn \\
 & & \hspace{7mm} \left. + \frac{\gamma_{\mu}+r}{2}\left(iq_{\bA} \, \tr \, \psi_{\mu}(x-\hat{\mu})\right) 
           \left(\det U_{\mu}(x-\hat{\mu}) \right)^{-q_{\bA}} \, {\cal X}(x-\hat{\mu}) \right], \nn \\
Q(aD_W^\dagger) \, {\cal X}(x) & = &  \sum_{\mu=0}^1 \left[-\frac{\gamma_{\mu}+r}{2} \left(iq_{\bA} \, \tr \, \psi_{\mu}(x)\right) 
                                 \left(\det U_{\mu}(x) \right)^{q_{\bA}} \, {\cal X}(x+\hat{\mu}) \right. \nn \\
 & & \hspace{7mm} \left. - \frac{\gamma_{\mu}-r}{2}\left(iq_{\bA} \, \tr \, \psi_{\mu}(x-\hat{\mu})\right) 
           \left(\det U_{\mu}(x-\hat{\mu}) \right)^{-q_{\bA}} \, {\cal X}(x-\hat{\mu}) \right]. \nn \\
 & &           
\eea
${\cal X}(x)$ is any doublet belonging to the $\det^{q_{\bA}}$-representation of $G$, 
for example $\Xi_{\bA}(x)$, $\cZ_{u\bA}(x)$, $\cZ_{d\bA}(x)$, $G_{\bA}(x)$, and their chiral projections.  
The Wilson parameter $r$ is fixed to $r=1$. 

Let us take chiral and anti-chiral variables appearing in $S'^{\rm LAT}_{{\rm mat}, +\twm'}$ as 
\bea
 & & P_+\Xi_{\bA}, \qquad P_+ \cZ_{u\bA}, \qquad \bar{P}_-\cZ_{d\bA}, \qquad \bar{P}_-\gamma_0 G_{\bA},  
\label{chiral_fields_+_det}\\
 & & \Xi_{\bA}^\dagger P_+,  \qquad \bar{\cZ}_{d\bA}P_+, \qquad \bar{\cZ}_{u\bA}\bar{P}_-, 
\qquad G_{\bA}^\dagger \gamma_0\bar{P}_-,
\qquad (\bA =1, \cdots, \ell_+), 
\label{anti-chiral_fields_+_det}
\eea
and chiral and anti-chiral variables appearing in $S'^{\rm LAT}_{{\rm mat}, -\twm'}$ as
\bea
 & & \Xi_{\bA'}^\dagger P_-,  \qquad \bar{\cZ}_{d\bA'}P_-, \qquad \bar{\cZ}_{u\bA'}\bar{P}_+, \qquad 
G_{\bA'}^\dagger \gamma_0\bar{P}_+, 
\label{chiral_fields_-_det}\\
 & & P_-\Xi_{\bA'}, \qquad P_- \cZ_{u\bA'}, \qquad \bar{P}_+\cZ_{d\bA'}, \qquad \bar{P}_+\gamma_0 G_{\bA'}, 
\qquad (\bA' =1, \cdots, \ell_-). 
\label{anti-chiral_fields_-_det}
\eea
Then, the $Q$ supersymmetry transformation to the variables 
\bea
Q(P_+\Xi_{\bA}(x)) & = & -P_+\cZ_{u\bA}(x), \nn \\
Q(P_+\cZ_{u\bA}(x)) & = & -\left(q_{\bA} \, \tr \, \phi(x) -\twm'_{+\bA}\right) P_+\Xi_{\bA}(x), \nn \\
Q(\bar{P}_-\cZ_{d\bA}(x)) & = & a\hD P_+\Xi_{\bA}(x) + \bar{P}_- \gamma_0 G_{\bA}(x) 
                                    +(Q\bar{P}_-) \bar{P}_-\cZ_{d\bA}(x), \nn \\
Q(\bar{P}_-\gamma_0G_{\bA}(x)) & = & \left(\bar{P}_-\left(q_{\bA}\,\tr \,\phi\right) -\twm'_{+\bA}\right)\bar{P}_-\cZ_{d\bA}(x) 
                             +a\hD P_+\cZ_{u\bA}(x) \nn \\
      & & -\bar{P}_-Q(a\hD)P_+\Xi_{\bA}(x) +  (Q\bar{P}_-)\bar{P}_-\gamma_0 G_{\bA}(x) +(Q\bar{P}_-)^2\bar{P}_-\cZ_{d\bA}(x), \nn \\
Q(\Xi_{\bA}(x)^\dagger P_+) & = & -\bar{\cZ}_{d\bA}(x) P_+, \nn \\
Q(\bar{\cZ}_{d\bA}(x) P_+) & = & \left(q_{\bA} \, \tr \, \phi(x) -\twm'_{+\bA}\right) \Xi_{\bA}(x)^\dagger P_+, \nn \\
Q(\bar{\cZ}_{u\bA}\bar{P}_-(x)) & = & \Xi_{\bA}(x)^\dagger P_+a\hD^\dagger +G_{\bA}^\dagger \gamma_0\bar{P}_-(x) 
                                    -\bar{\cZ}_{u\bA}\bar{P}_-(Q\bar{P}_-)(x), \nn \\
Q(G_{\bA}^\dagger\gamma_0\bar{P}_-(x)) & = &  -\bar{\cZ}_{u\bA}\bar{P}_-\left((q_{\bA}\, \tr \, \phi) \bar{P}_- -\twm'_{+\bA}\right)(x) 
               +\bar{\cZ}_{d\bA}(x) P_+a\hD^\dagger \nn \\
       & & -\Xi_{\bA}(x)^\dagger P_+ Q(a\hD^\dagger)\bar{P}_- +G_{\bA}^\dagger\gamma_0\bar{P}_-(Q\bar{P}_-)(x)          
                -\bar{\cZ}_{u\bA}\bar{P}_-(Q\bar{P}_-)^2(x), \nn \\
         & &                                \\
Q(P_-\Xi_{\bA'}(x)) & = & -P_-\cZ_{u\bA'}(x), \nn \\
Q(P_-\cZ_{u\bA'}(x)) & = & -\left( q_{\bA'} \, \tr \, \phi(x) -\twm'_{-\bA'}\right)  P_-\Xi_{\bA'}(x), \nn \\
Q(\bar{P}_+\cZ_{d\bA'}(x)) & = & a\hD P_-\Xi_{\bA'}(x) +\bar{P}_+\gamma_0 G_{\bA'}(x) +(Q\bar{P}_+)\bar{P}_+\cZ_{d\bA'}(x), \nn \\
Q(\bar{P}_+\gamma_0 G_{\bA'}(x)) & = & \left(\bar{P}_+\left(q_{\bA'}\,\tr \,\phi\right) -\twm'_{-\bA'}\right)\bar{P}_+\cZ_{d\bA'}(x) 
                                  + a\hD P_-\cZ_{u\bA'}(x) \nn \\
       & & -\bar{P}_+ Q(a\hD) P_-\Xi_{\bA'}(x) + (Q\bar{P}_+)\bar{P}_+\gamma_0 G_{\bA'}(x) + (Q\bar{P}_+)^2\bar{P}_+\cZ_{d\bA'}(x), \nn \\
Q(\Xi_{\bA'}(x)^\dagger P_-) & = & -\bar{\cZ}_{d\bA'}(x) P_-, \nn \\
Q(\bar{\cZ}_{d\bA'}(x)P_-) & = &  \left(q_{\bA'} \, \tr \, \phi(x) -\twm'_{-\bA'}\right) \Xi_{\bA'}(x)^\dagger P_-, \nn \\
Q(\bar{\cZ}_{u\bA'}\bar{P}_+(x)) & = & \Xi_{\bA'}(x)^\dagger P_-a\hD^\dagger + G_{\bA'}^\dagger \gamma_0\bar{P}_+(x) 
                            -\bar{\cZ}_{u\bA'}\bar{P}_+(Q\bar{P}_+)(x), \nn \\
Q(G_{\bA'}^\dagger\gamma_0\bar{P}_+(x)) & = & -\bar{\cZ}_{u\bA'}\bar{P}_+\left((q_{\bA'}\, \tr \, \phi) \bar{P}_+ -\twm'_{-\bA'}\right)(x)
                                 +\bar{\cZ}_{d\bA'}(x)P_-a\hD^\dagger \nn \\
     & & -\Xi_{\bA'}(x)^\dagger P_- Q(a\hD^\dagger) \bar{P}_+  + G_{\bA'}^\dagger \gamma_0 \bar{P}_+(Q\bar{P}_+)(x) 
             -\bar{\cZ}_{u\bA'}\bar{P}_+ (Q\bar{P}_+)^2(x), \nn \\
       & &        
\eea
is nilpotent in the sense of 
\bea
Q^2 & = & (\mbox{infinitesimal gauge transformation with the parameter $\phi(x)$}) \nn \\
      & & + (\mbox{infinitesimal flavor rotations (\ref{flavor_rot_+_det}) and (\ref{flavor_rot_-_det})})
\eea
with 
\bea
\delta (P_+\Xi_{\bA}) = -\twm'_{+\bA}P_+\Xi_{\bA}, & &    \delta (\Xi_{\bA}^\dagger P_+) = \twm'_{+\bA}\Xi_{\bA}^\dagger P_+, \nn \\
\delta (P_+\cZ_{u\bA}) = -\twm'_{+\bA} P_+\cZ_{u\bA}, & & 
                               \delta (\bar{\cZ}_{u\bA} \bar{P}_-) = \twm'_{+\bA} \bar{\cZ}_{u\bA}\bar{P}_-, \nn \\
\delta (\bar{P}_- \cZ_{d\bA}) = -\twm'_{+\bA} \bar{P}_-\cZ_{d\bA}, & & 
                                   \delta (\bar{\cZ}_{d\bA} P_+) = \twm'_{+\bA}\bar{\cZ}_{d\bA} P_+, \nn \\
\delta (\bar{P}_-\gamma_0 G_{\bA}) = -\twm'_{+\bA} \bar{P}_-\gamma_0 G_{\bA}, 
                                      & & \delta (G_{\bA}^\dagger \gamma_0 \bar{P}_-) = \twm'_{+\bA} G_{\bA}^\dagger \gamma_0 \bar{P}_-, 
\label{flavor_rot_+_det}
\eea
\bea
\delta (\Xi_{\bA'}^\dagger P_-) = \twm'_{-\bA'}\Xi_{\bA'}^\dagger P_-, & & \delta (P_-\Xi_{\bA'}) = -\twm'_{-\bA'} P_-\Xi_{\bA'}, \nn \\
\delta (\bar{\cZ}_{u\bA'} \bar{P}_+) = \twm'_{-\bA'}\bar{\cZ}_{u\bA'} \bar{P}_+, & & 
                                         \delta (P_-\cZ_{u\bA'}) = -\twm'_{-\bA'}P_-\cZ_{u\bA'}, \nn \\
\delta (\bar{\cZ}_{d\bA'} P_-) = \twm'_{-\bA'}\bar{\cZ}_{d\bA'} P_-, & & 
                                          \delta (\bar{P}_+\cZ_{d\bA'}) = -\twm'_{-\bA'} \bar{P}_+\cZ_{d\bA'}, \nn \\
\delta (G_{\bA'}^\dagger \gamma_0 \bar{P}_+) = \twm'_{-\bA'} G_{\bA'}^\dagger \gamma_0 \bar{P}_+, 
                                                & & \delta (\bar{P}_+ \gamma_0 G_{\bA'}) = -\twm'_{-\bA'} \bar{P}_+ \gamma_0 G_{\bA'}. 
\label{flavor_rot_-_det}
\eea   
The $Q$ transformation is closed among each of the ``multiplets" (\ref{chiral_fields_+_det}), (\ref{anti-chiral_fields_+_det}), 
(\ref{chiral_fields_-_det}) and (\ref{anti-chiral_fields_-_det}). 
       
Also, the $Q$-exact action can be expressed as 
\bea
S'^{\rm LAT}_{{\rm mat},\twm'} & = & S'^{\rm LAT}_{{\rm mat}, +\twm'} + S'^{\rm LAT}_{{\rm mat}, -\twm'}, \nn \\
S'^{\rm LAT}_{{\rm mat}, +\twm'} & = & Q\sum_x \sum_{\bA =1}^{\ell_+}\frac12 \left[
          \bar{\cZ}_{u\bA}\bar{P}_-(x) \left(a\hD P_+\Xi_{\bA}(x) -\bar{P}_-\gamma_0 G_{\bA}(x)\right) \right. \nn \\
          & & \hspace{2cm} 
                 +\left(\Xi_{\bA}(x)^\dagger P_+a\hD^\dagger -G_{\bA}^\dagger \gamma_0\bar{P}_-(x)\right) \bar{P}_-\cZ_{d\bA}(x) \nn \\
          & &\hspace{2cm}  -\Xi_{\bA}(x)^\dagger P_+ \left(q_{\bA} \,\tr \, \bar{\phi}(x) -\twm'^*_{+\bA}\right) P_+\cZ_{u\bA}(x) \nn \\
          & & \hspace{2cm} +\bar{\cZ}_{d\bA}(x)P_+ \left(q_{\bA} \,\tr \, \bar{\phi}(x) -\twm'^*_{+\bA}\right) P_+ \Xi_{\bA}(x) \nn \\
          & & \hspace{2cm} \left. +2i q_{\bA} \, \Xi_{\bA}(x)^\dagger P_+\left(\tr \, \chi(x)\right) P_+\Xi_{\bA}(x) \right], \\
S'^{\rm LAT}_{{\rm mat}, -\twm'} & = &  Q\sum_x \sum_{\bA' =1}^{\ell_-}\frac12 \left[    
         \bar{\cZ}_{u\bA'}\bar{P}_+(x) \left(a\hD P_-\Xi_{\bA'}(x) -\bar{P}_+\gamma_0 G_{\bA'}(x) \right) \right. \nn \\
        &  & \hspace{2cm}
               +\left(\Xi_{\bA'}(x)^\dagger P_- a\hD^\dagger -G_{\bA'}^\dagger \gamma_0\bar{P}_+(x)\right) \bar{P}_+\cZ_{d\bA'}(x) \nn \\
       & & \hspace{2cm} -\Xi_{\bA'}(x)^\dagger P_- \left(q_{\bA'} \,\tr \, \bar{\phi}(x) -\twm'^*_{-\bA'}\right) P_- \cZ_{u\bA'}(x) \nn \\
      & & \hspace{2cm} +\bar{\cZ}_{d\bA'}(x) P_-  \left(q_{\bA'} \,\tr \, \bar{\phi}(x) -\twm'^*_{-\bA'}\right) P_- \Xi_{\bA'}(x) \nn \\
      & & \hspace{2cm} \left. -2i q_{\bA'} \, \Xi_{\bA'}(x)^\dagger P_-\left(\tr \, \chi(x)\right) P_-\Xi_{\bA'}(x) \right],      
\label{S_mat_det-twm_hD}      
\eea
and after the $Q$ operation in the r.h.s. it becomes  
\bea
S'^{\rm LAT}_{{\rm mat}, +\twm'} & = & \sum_x \sum_{\bA =1}^{\ell_+}\left[
          a^2 \Xi_{\bA}(x)^\dagger P_+\hD^\dagger\hD P_+\Xi_{\bA}(x) 
           -\left(G_{\bA}^\dagger \gamma_0\bar{P}_-(x)\right)\left(\bar{P}_-\gamma_0 G_{\bA}(x)\right) \right. \nn \\
 & & \hspace{1.5cm}
          +\bar{\cZ}_{u\bA}\bar{P}_-(x)a\hD P_+\cZ_{u\bA}(x) -\bar{\cZ}_{d\bA}(x)P_+a\hD^\dagger \bar{P}_-\cZ_{d\bA}(x) \nn \\
 & & \hspace{1.5cm}  +\Xi_{\bA}(x)^\dagger P_+ \left(q_{\bA}\,\tr\,\phi(x) -\twm'_{+\bA}\right)
                  \left(q_{\bA}\,\tr\,\bar{\phi}(x) -\twm'^*_{+\bA}\right) P_+\Xi_{\bA}(x) \nn \\
 & & \hspace{1.5cm}  -q_{\bA} \, \Xi_{\bA}(x)^\dagger P_+ \, \tr \left(D(x) +\frac12 \whPhi(x)\right)P_+\Xi_{\bA}(x) \nn \\
 & & \hspace{1.5cm}  +\bar{\cZ}_{u\bA}\bar{P}_-(x)\left(q_{\bA}\,\tr\,\phi(x) -\twm'_{+\bA}\right)\bar{P}_-\cZ_{d\bA}(x) \nn \\
 & & \hspace{1.5cm}  +\bar{\cZ}_{d\bA}(x) P_+ \left(q_{\bA}\,\tr\,\bar{\phi}(x) -\twm'^*_{+\bA}\right) P_+\cZ_{u\bA}(x) \nn \\
 & & \hspace{1.5cm}  -\bar{\cZ}_{u\bA}\bar{P}_-(x) Q(a\hD) P_+ \Xi_{\bA}(x) 
                        + \Xi_{\bA}(x)^\dagger P_+ Q(a\hD^\dagger) \bar{P}_-\cZ_{d\bA}(x) \nn \\
 & & \hspace{1.5cm}  -q_{\bA} \, \bar{\cZ}_{d\bA}(x) P_+ \,\tr\left(\frac12\eta(x) +i\chi(x)\right) P_+\Xi_{\bA}(x)  \nn \\
 & & \hspace{1.5cm}   -q_{\bA} \, \Xi_{\bA}(x)^\dagger P_+ \,\tr\left(\frac12\eta(x) -i\chi(x)\right) P_+\cZ_{u\bA}(x) \nn \\
 & & \hspace{1.5cm}  \left. +\bar{\cZ}_{u\bA}\bar{P}_-(x) (Q\bar{P}_-)^2 \bar{P}_-\cZ_{d\bA}(x)\right], \\
S'^{\rm LAT}_{{\rm mat}, -\twm'} & = &  \sum_x \sum_{\bA' =1}^{\ell_-}\left[                                  
       a^2 \Xi_{\bA'}(x)^\dagger P_-\hD^\dagger\hD P_-\Xi_{\bA'}(x) 
           -\left(G_{\bA'}^\dagger \gamma_0\bar{P}_+(x)\right)\left(\bar{P}_+\gamma_0 G_{\bA'}(x)\right) \right. \nn \\                  
 & & \hspace{1.5cm}  
              +\bar{\cZ}_{u\bA'}\bar{P}_+(x)a\hD P_-\cZ_{u\bA'}(x) -\bar{\cZ}_{d\bA'}(x)P_-a\hD^\dagger \bar{P}_+\cZ_{d\bA'}(x) \nn \\
 & & \hspace{1.5cm}  +\Xi_{\bA'}(x)^\dagger P_- \left(q_{\bA'}\,\tr\,\phi(x) -\twm'_{-\bA'}\right)
                  \left(q_{\bA'}\,\tr\,\bar{\phi}(x) -\twm'^*_{-\bA'}\right) P_-\Xi_{\bA'}(x) \nn \\
 & & \hspace{1.5cm}  +q_{\bA'} \, \Xi_{\bA'}(x)^\dagger P_- \, \tr \left(D(x) +\frac12 \whPhi(x)\right)P_-\Xi_{\bA'}(x) \nn \\
 & & \hspace{1.5cm}  +\bar{\cZ}_{u\bA'}\bar{P}_+(x)\left(q_{\bA'}\,\tr\,\phi(x) -\twm'_{-\bA'}\right)\bar{P}_+\cZ_{d\bA'}(x) \nn \\
 & &  \hspace{1.5cm}  +\bar{\cZ}_{d\bA'}(x) P_- \left(q_{\bA'}\,\tr\,\bar{\phi}(x) -\twm'^*_{-\bA'}\right) P_-\cZ_{u\bA'}(x) \nn \\
 & & \hspace{1.5cm}  -\bar{\cZ}_{u\bA'}\bar{P}_+(x) Q(a\hD) P_- \Xi_{\bA'}(x) 
                        + \Xi_{\bA'}(x)^\dagger P_- Q(a\hD^\dagger) \bar{P}_+\cZ_{d\bA'}(x) \nn \\
 & & \hspace{1.5cm}  -q_{\bA'} \, \bar{\cZ}_{d\bA'}(x) P_- \,\tr\left(\frac12\eta(x) -i\chi(x)\right) P_-\Xi_{\bA'}(x)  \nn \\
 & & \hspace{1.5cm}  -q_{\bA'} \, \Xi_{\bA'}(x)^\dagger P_- \,\tr\left(\frac12\eta(x) +i\chi(x)\right) P_-\cZ_{u\bA'}(x) \nn \\
 & & \hspace{1.5cm}  \left. +\bar{\cZ}_{u\bA'}\bar{P}_+(x) (Q\bar{P}_+)^2 \bar{P}_+\cZ_{d\bA'}(x)\right]. 
\eea
These are valid for general $\ell_\pm$ and general twisted masses. 
As shown in appendix~\ref{app:adm_det}, for $\hD$ to be well-defined, 
it is sufficient to choose $\epsilon$ in the admissibility condition (\ref{admissibility}) as 
\be
0 < \epsilon < \frac{1}{8Nq} \qquad \mbox{with} \qquad q \equiv \max_{\bA =1,\dots, \ell_0}(|q_{\bA}|). 
\label{adm_det}
\ee
We assumed the charges $q_{\bA}\in {\bf Z}_{\neq 0}$ for $\bA =1, \cdots, \ell_0$. 
Since this is stronger than the constraint from the SYM sector (\ref{epsilon_vac}) or (\ref{adm_theta}), 
eq.~(\ref{adm_det}) gives the condition to the total system.  

On the lattice, the ${\rm U}(1)_A$ and ${\rm U}(1)_V$ transformations are defined for the matter variables as \\
${\rm U}(1)_A$: 
\bea
P_+ \cZ_{u\bA}(x) \to \e^{i\alpha} P_+\cZ_{u\bA}(x), & & 
       \bar{\cZ}_{u\bA}\bar{P}_- (x) \to \bar{\cZ}_{u\bA}\bar{P}_- (x) \e^{-i\alpha}, \nn \\
\bar{P}_-\cZ_{d\bA}(x) \to \e^{-i\alpha} \bar{P}_- \cZ_{d\bA}(x), & & 
       \bar{\cZ}_{d\bA}(x)P_+ \to \bar{\cZ}_{d\bA}(x)P_+ \e^{i\alpha}, \nn \\
P_- \cZ_{u\bA'}(x) \to \e^{i\alpha} P_-\cZ_{u\bA'}(x), & & 
       \bar{\cZ}_{u\bA'}\bar{P}_+ (x) \to \bar{\cZ}_{u\bA'}\bar{P}_+ (x) \e^{-i\alpha}, \nn \\
\bar{P}_+\cZ_{d\bA'}(x) \to \e^{-i\alpha} \bar{P}_+ \cZ_{d\bA'}(x), & & 
       \bar{\cZ}_{d\bA'}(x)P_- \to \bar{\cZ}_{d\bA'}(x)P_- \e^{i\alpha},        
\eea
${\rm U}(1)_V$:
\bea
\bar{P}_-\gamma_0 G_{\bA}(x) \to \e^{- i2\alpha} \bar{P}_-\gamma_0 G_{\bA}(x), & & 
      G_{\bA}^\dagger \gamma_0\bar{P}_- (x) \to G_{\bA}^\dagger \gamma_0\bar{P}_- (x) \e^{ i2\alpha}, \nn \\
P_+ \cZ_{u\bA}(x) \to \e^{- i\alpha} P_+ \cZ_{u\bA}(x), & & 
       \bar{\cZ}_{u\bA}\bar{P}_- (x) \to  \bar{\cZ}_{u\bA}\bar{P}_- (x) \e^{ i\alpha}, \nn \\
\bar{P}_- \cZ_{d\bA}(x) \to \e^{- i\alpha}  \bar{P}_- \cZ_{d\bA}(x), & &
       \bar{\cZ}_{d\bA}(x) P_+ \to \bar{\cZ}_{d\bA}(x) P_+ \e^{ i\alpha}  , \nn \\
\bar{P}_+\gamma_0 G_{\bA'}(x) \to \e^{i2\alpha} \bar{P}_+\gamma_0 G_{\bA'}(x), & & 
      G_{\bA'}^\dagger \gamma_0\bar{P}_+ (x) \to G_{\bA'}^\dagger \gamma_0\bar{P}_+ (x) \e^{ -i2\alpha}, \nn \\
P_- \cZ_{u\bA'}(x) \to \e^{ i\alpha} P_- \cZ_{u\bA'}(x), & & 
       \bar{\cZ}_{u\bA'}\bar{P}_+ (x) \to  \bar{\cZ}_{u\bA'}\bar{P}_+ (x) \e^{ -i\alpha}, \nn \\
\bar{P}_+ \cZ_{d\bA'}(x) \to \e^{ i\alpha}  \bar{P}_+ \cZ_{d\bA'}(x), & &
       \bar{\cZ}_{d\bA'}(x) P_- \to \bar{\cZ}_{d\bA'}(x) P_- \e^{ -i\alpha}        
\eea
with the others unchanged.

\paragraph{Superpotentials}
Contribution from the superpotentials 
\be
W=W(P_+\Xi_{\bA}, \Xi_{\bA'}^\dagger P_-), \qquad \bar{W} = \bar{W}(\Xi_{\bA}^\dagger P_+, P_-\Xi_{\bA'})
\ee
can be written as the $Q$-exact form: 
\bea
S'^{\rm LAT}_{\rm pot} & = & Q\sum_x \sum_{\bA=1}^{\ell_+}\left[
          -\frac{\der W}{\der (P_+\Xi_{\bA}(x))}\left(\gamma_0\bar{P}_-\cZ_{d\bA}(x)\right) 
          -\left(\bar{\cZ}_{u\bA}\bar{P}_-(x)\gamma_0\right) \frac{\der \bar{W}}{\der (\Xi_{\bA}(x)^\dagger P_+)}\right] \nn \\
       & &\hspace{-3mm} +Q\sum_x\sum_{\bA' =1}^{\ell_-}\left[
          -\frac{\der \bar{W}}{\der (P_-\Xi_{\bA'}(x))}\left(\gamma_0\bar{P}_+\cZ_{d\bA'}(x)\right) 
          -\left(\bar{\cZ}_{u\bA'}\bar{P}_+(x)\gamma_0\right) \frac{\der W}{\der (\Xi_{\bA'}(x)^\dagger P_-)}\right].  \nn \\
        & &    
\eea 
Chirally projected variables with the parentheses, say 
$(P_+\Xi_{\bA}(x))$, $\left(\gamma_0\bar{P}_-\cZ_{d\bA}(x)\right)$, $(P_-\Xi_{\bA'}(x))$, 
$\left(\gamma_0\bar{P}_+\cZ_{d\bA'}(x)\right)$ etc, are understood as independent degrees of freedom of the doublets 
projected by $P_\pm$ or $\bar{P}_\pm$.  
Concerning the dependence on the matter sector variables, 
since terms from $W$ depends only on the chiral variables (\ref{chiral_fields_+_det}) and (\ref{chiral_fields_-_det}), 
and terms from $\bar{W}$ depends only on the anti-chiral variables (\ref{anti-chiral_fields_+_det}) and 
(\ref{anti-chiral_fields_-_det}), 
holomorphy is exactly realized on the lattice.

\paragraph{Path-integral Measure}
The path-integral measure is defined by 
\bea
\left( \dd \mu'_{\rm mat}\right)  & = & \left(\prod_{\bA =1}^{\ell_+} \dd\mu'_{{\rm mat}, +\bA}\right) 
                                        \left(\prod_{\bA' =1}^{\ell_-} \dd\mu'_{{\rm mat}, -\bA'}\right) , \nn \\
\dd\mu'_{{\rm mat}, +\bA} & \equiv & \prod_x \dd (P_+\Xi_{\bA}(x)) \, \dd (\Xi_{\bA}(x)^\dagger P_+) \, 
                 \dd (\bar{P}_-\gamma_0 G_{\bA}(x)) \, \dd (G_{\bA}^\dagger \gamma_0\bar{P}_-(x)) \nn \\
 & & \hspace{4mm} \times \dd (P_+\cZ_{u\bA}(x)) \, \dd (\bar{\cZ}_{u\bA}\bar{P}_-(x)) \, \dd (\bar{P}_-\cZ_{d\bA}(x)) \, 
                      \dd (\bar{\cZ}_{d\bA}(x) P_+), \\
\dd\mu'_{{\rm mat}, -\bA'} & \equiv & \prod_x \dd (P_-\Xi_{\bA'}(x)) \, \dd (\Xi_{\bA'}(x)^\dagger P_-) \, 
                 \dd (\bar{P}_+\gamma_0 G_{\bA'}(x)) \, \dd (G_{\bA'}^\dagger \gamma_0\bar{P}_+(x)) \nn \\
 & & \hspace{4mm} \times \dd (P_-\cZ_{u\bA'}(x)) \, \dd (\bar{\cZ}_{u\bA'}\bar{P}_+(x)) \, \dd (\bar{P}_+\cZ_{d\bA'}(x))  
                      \dd (\bar{\cZ}_{d\bA'}(x) P_-),  
\label{measure_mat-det}                      
\eea
together with the SYM part (\ref{measure_SYM}). 
Similarly to the case of the (anti-)fundamental matters, it can be seen that each of $\dd\mu'_{{\rm mat}, +\bA}$ and 
$\dd\mu'_{{\rm mat}, -\bA'}$ is invariant under gauge, $Q$ and ${\rm U}(1)_V$ transformations. 
The ${\rm U}(1)_A$ with the infinitesimal parameter $\alpha$ transforms as 
\bea
\dd\mu'_{{\rm mat}, +\bA} & \to &  \left[1+i\alpha \, \Tr\left(\gamma_3 a\hD\right)\right] \dd\mu'_{{\rm mat}, +\bA} \nn \\
                 & \simeq & \left[1+i\alpha \, \frac{q_{\bA}}{\pi} \int \dd^2 x \, \tr\, F_{01}\right] \dd\mu'_{{\rm mat}, +\bA}      
                                        \qquad (a\to 0),  \\
\dd\mu'_{{\rm mat}, -\bA'} & \to & \left[1-i\alpha \, \Tr\left(\gamma_3 a\hD\right)\right] \dd\mu'_{{\rm mat}, -\bA'} \nn \\
                 & \simeq & \left[1-i\alpha \, \frac{q_{\bA'}}{\pi} \int \dd^2 x \, \tr\, F_{01}\right] \dd\mu'_{{\rm mat}, -\bA'}      
                                          \qquad (a\to 0) ,  
\label{U(1)_A_measure-det}                                          
\eea
where we assumed that the gauge fields are smooth in evaluating the $a\to 0$ limit. 

\setcounter{equation}{0}
\section{Application to Gauged Linear Sigma Models}
\label{sec:GLS}
Taking $G={\rm U}(N)$ and $q_{\bA'}\in {\bf Z}_{>0}$, we briefly explain gauged sigma models~\cite{witten2,hori-tong}, 
where $n_+$ fundamental matters and $\ell_-$ matters in the $\det^{-q_{\bA'}}$-representations are coupled. 
The action of the continuum theory is 
\be
S^{(E)} _{\rm GLS} =  S^{(E)}_{{\rm mat}, +\twm} + S'^{(E)}_{{\rm mat}, -\twm'} + S^{(E)}_{\rm 2DSYM} +S^{(E)}_{{\rm FI}, \vartheta} 
     + S^{(E)}_{\cW},  
\ee  
where the first four terms were given by (\ref{SE_mat+twm}), (\ref{SE_mat_det-twm}), (\ref{SE_2DSYM}), 
(\ref{FI_theta_cont}), respectively. 
The superpotential term $S^{(E)}_{\cW}$ will be given shortly. 

When $n_+\geq N$, baryonic chiral superfields 
\be
B_{I_1\cdots I_N} \equiv  \epsilon_{i_1\cdots i_N} \Phi_{+I_1i_1}\cdots \Phi_{+I_Ni_N} , \qquad 
B_{I_1\cdots I_N}^*  \equiv \epsilon_{i_1\cdots i_N} \Phi_{+I_1i_1}^*\cdots \Phi_{+I_Ni_N}^*  
\ee
do not all vanish, and they transform under the gauge transformation $g(x)\in G$ as 
\be
B_{I_1\cdots I_N}(x) \to (\det g(x)) B_{I_1\cdots I_N}(x), \qquad B_{I_1\cdots I_N}(x)^* \to (\det g(x))^{-1} B_{I_1\cdots I_N}(x)^*. 
\ee 
Let $\cG_{\bA'}(B)$ ($\bar{\cG}_{\bA'}(B^*)$) be a homogeneous polynomial of degree $q_{\bA'}$ 
with respect to $B_{I_1\cdots I_N}$ ($B_{I_1\cdots I_N}^*$). Then, the gauge-invariant superpotentials 
\be
\cW = \sum_{\bA'=1}^{\ell_-} \, \Xi_{-\bA'} \, \cG_{\bA'}(B), \qquad 
\overline{\cW} = \sum_{\bA' =1}^{\ell_-} \, \Xi_{-\bA'}^* \, \bar{\cG}_{\bA'}(B^*)
\ee
contribute to the action as 
\be
S^{(E)}_{\cW} = -\int \dd^2 x \left(\thth{\cW} + \tbtb{\overline{\cW}}\right). 
\ee

After eliminating the auxiliary fields, potentials for the bosonic fields become 
\bea
U & = & \sum_{\bA' =1}^{\ell_-} \left|\cG_{\bA'}(b)\right|^2 
      + \sum_{I=1}^{n_+} \sum_{i=1}^N\left|\sum_{\bA' =1}^{\ell_-} \xi_{-\bA'}\sum_{I_1<\cdots < I_N} 
                   \frac{\der \cG_{\bA'}(b)}{\der b_{I_1\cdots I_N}} \frac{\der b_{I_1\cdots I_N}}{\der \phi_{+Ii}}\right|^2 \nn \\
   & & + \frac{g^2}{4} \, \tr\left\{ \left[\sum_{I=1}^{n_+} \phi_{+I}\phi_{+I}^\dagger 
                -\left(\sum_{\bA' =1}^{\ell_-}q_{\bA'}\xi_{-\bA'}^*\xi_{-\bA'}+\kappa\right) \id_N\right]^2\right\} \nn \\
   & & +\frac{1}{4g^2} \, \tr \left([\phi, \bar{\phi}]^2\right) 
          + \sum_{I=1}^{n_+} \frac12 \, \phi_{+I}^\dagger \left\{ \phi -\twm_{+I}, \bar{\phi}-\twm_{+I}^*\right\} \phi_{+I} \nn \\
   & & + \sum_{\bA' =1}^{\ell_-} \left|q_{\bA'}\, \tr \, \phi -\twm'_{-\bA'}\right|^2 \xi_{-\bA'}\xi_{-\bA'}^*, 
\eea
where $b_{I_1\cdots I_N}$ represent the lowest components of the chiral superfields $B_{I_1\cdots I_N}$.  
The first and second lines come from the F-term and D-term conditions, respectively. 
When 
\be
n_+ - \sum_{\bA' =1}^{\ell_-} q_{\bA'} =0
\label{U(1)_A_anomaly_candellation}
\ee 
is satisfied, the ${\rm U}(1)_A$ anomaly cancels, and the model is considered to flow in the infra-red limit 
to a nonlinear sigma model whose target space, including Calabi-Yau manifolds, 
is determined by the F-term and D-terms conditions~\cite{witten2,hori-tong,hanany-hori}.

\subsection{Lattice Formulation}
It is straightforward to latticize the continuum gauged linear sigma models using the construction in the previous sections. 
The lattice action is given by 
\be
S^{\rm LAT} _{\rm GLS} =  S^{\rm LAT}_{{\rm mat}, +\twm} + S'^{\rm LAT}_{{\rm mat}, -\twm'} + S^{\rm LAT}_{\rm 2DSYM} 
                  +S^{\rm LAT}_{{\rm FI}, \vartheta}  + S^{\rm LAT}_{\cW}, 
\ee
where first four terms have been constructed as (\ref{S_mat+twm_hD_a}), (\ref{S_mat_det-twm_hD}), 
(\ref{S_lat_SYM1}) and (\ref{S_lat_SYM2}), (\ref{FI_theta_lat}), respectively.   

In order to present $S^{\rm LAT}_{\cW}$, we define the baryonic variables 
\bea
b_{I_1\cdots I_N}(x) & \equiv & 
             \sum_{i_1, \cdots, i_N=1}^N \epsilon_{i_1\cdots i_N} (P_+\Phi_{I_1}(x))_{i_1}\cdots (P_+\Phi_{I_N}(x))_{i_N}, \nn \\
b_{I_1\cdots I_N}(x)^* & \equiv & 
             \sum_{i_1, \cdots, i_N=1}^N \epsilon_{i_1\cdots i_N} (\Phi_{I_1}(x)^\dagger P_+)_{i_1}\cdots (\Phi_{I_N}(x)^\dagger P_+)_{i_N}, 
\eea
and take the superpotentials 
\bea
\cW & = & \cW (\Xi_{\bA'}^\dagger P_-, P_+\Phi_I) 
                   = \sum_{\bA' =1}^{\ell_-} \, (\Xi_{\bA'}(x)^\dagger P_-) \, \cG_{\bA'}(b), \nn \\
\overline{\cW} & = & \overline{\cW} (P_-\Xi_{\bA'}, \Phi_I^\dagger P_+) 
                           = \sum_{\bA' =1}^{\ell_-} \, (P_- \Xi_{\bA'}(x)) \, \bar{\cG}_{\bA'}(b^*). 
\eea
Then, $S^{\rm LAT}_{\cW}$ is written as the $Q$-exact form: 
\bea
S^{\rm LAT}_{\cW} & = & Q \sum_x \sum_{i=1}^N \sum_{I=1}^{n_+}\left[
                 -\frac{\der \cW}{\der (P_+\Phi_I(x))_i} \left(\gamma_0\bar{P}_-\Psi_{dI}(x)\right)_i 
                 -\left(\bar{\Psi}_{uI}\bar{P}_-(x)\gamma_0\right)_i\frac{\der \overline{\cW}}{\der (\Phi_I(x)^\dagger P_+)_i}\right] \nn \\
        & & \hspace{-3mm} +Q\sum_x\sum_{\bA' =1}^{\ell_-} \left[
                 -\frac{\der \overline{\cW}}{\der (P_- \Xi_{\bA'}(x))} \left(\gamma_0\bar{P}_+\cZ_{d\bA'}(x)\right) 
                 -\left(\bar{\cZ}_{u\bA'}\bar{P}_+(x) \gamma_0\right) \frac{\der \cW}{\der (\Xi_{\bA'}(x)^\dagger P_-)}\right] \\
         & = & Q\sum_x \sum_{i=1}^N \sum_{I=1}^{n_+} \sum_{\bA' =1}^{\ell_-} \left[ 
                       -(\Xi_{\bA'}(x)^\dagger P_-) \frac{\der \cG_{\bA'}(b)}{\der (P_+\Phi_I(x))_i} \left(\gamma_0\bar{P}_-\Psi_{dI}(x)\right)_i 
                           \right. \nn \\        
         & & \hspace{3.5cm} \left. -\left(\bar{\Psi}_{uI}\bar{P}_-(x) \gamma_0\right)_i (P_-\Xi_{\bA'}(x)) 
                      \frac{\der \bar{\cG}_{\bA'}(b^*)}{\der (\Phi_I(x)^\dagger P_+)_i} \right] \nn \\
         & & \hspace{-3mm} +Q\sum_x\sum_{\bA' =1}^{\ell_-} \left[-\bar{\cG}_{\bA'}(b^*) \left(\gamma_0\bar{P}_+\cZ_{d\bA'}(x)\right)
                              -\left(\bar{\cZ}_{u\bA'}\bar{P}_+(x)\gamma_0\right) \cG_{\bA'}(b)\right].                  
\eea

The path-integral measure is given by 
\bea
\dd\mu^{\rm (GLS)} & = & \left( \dd\mu_{\rm 2DSYM} \right) \left( \dd\mu^{\rm (GLS)}_{\rm mat} \right), \\
\left( \dd\mu^{\rm (GLS)}_{\rm mat}\right)  &\equiv &  \left(\prod_{I=1}^{n_+} \dd\mu_{{\rm mat}, +I} \right) 
                                                     \left(\prod_{\bA' =1}^{\ell_-} \dd \mu'_{{\rm mat}, -\bA'}\right). 
\label{measure_GLS_mat}                                                     
\eea
For the first and second factors in the r.h.s. of (\ref{measure_GLS_mat}), see (\ref{measure_mat+I}) and (\ref{measure_mat-det}). 
{}From (\ref{U(1)_A_measure_+I}), (\ref{gamma_hD_cont}) and (\ref{U(1)_A_measure-det}), $\left( \dd\mu^{\rm (GLS)}_{\rm mat}\right)$ 
changes under the infinitesimal ${\rm U}(1)_A$ transformation as 
\be
\left( \dd\mu^{\rm (GLS)}_{\rm mat}\right)  \to \left[1+i\alpha \, \frac{1}{\pi} \left(n_+-\sum_{\bA' =1}^{\ell_-} q_{\bA'}\right) 
                                \int \dd^2x \, \tr \, F_{01} \right] \left(\dd\mu^{\rm (GLS)}_{\rm mat} \right) 
\ee
in the $a\to 0$ limit, reproducing the ${\rm U}(1)_A$ anomaly cancellation in the continuum (\ref{U(1)_A_anomaly_candellation}). 

Comparing (\ref{adm_U(N)_theta}) with (\ref{adm_det}), we find the admissibility condition (\ref{admissibility}) with 
$\epsilon$ chosen in the range 
\be
0 < \epsilon < \frac{1}{8Nq} \qquad \mbox{with} \qquad q \equiv \max_{\bA'=1, \cdots, \ell_-}(q_{\bA'}) . 
\ee

\setcounter{equation}{0}
\section{Summary and Discussion}
\label{sec:summary}
In this paper, we have presented lattice formulations for two-dimensional SQCD 
with $n_+$ fundamental and $n_-$ anti-fundamental matter multiplets. 
The formulations preserve one of the supersymmetry $Q$ on the lattice, 
and furthermore the chiral flavor symmetry, which enhances maximally to ${\rm U}(n_+)\times {\rm U}(n_-)$, is exactly realized 
by virtue of the Ginsparg-Wilson formulation. 

A lattice model for the same theory was constructed in~\cite{sugino_sqcd}, which can be regarded in the doublet notation as 
a formulation with the Wilson-Dirac operator $D_W$ used. 
The lattice action there was defined for the cases of $n_+=n_-$ and $\twm_{+I}=\twm_{-I}$ due to the Wilson terms breaking the 
chiral flavor symmetry. Here, we have successfully constructed the lattice action for general $n_\pm$ and general twisted masses 
thanks to the Ginsparg-Wilson formulation. 

As another merit of this formulation, superpotential terms can be introduced with 
holomorphic or anti-holomorphic structure preserved. 
The terms are expected to receive no radiative correction from loops of the matter variables in perturbative computation. 
For the case of two-dimensional Wess-Zumino models, analogous argument is presented in~\cite{kikukawa-nakayama}.  
As a difference from that case, 
the superpotentials depend on the SYM variables through the chiral projectors $\bar{P}_\pm$ or their $Q$ transformations. 
Thus, there might arise some radiative corrections from loops of the SYM variables. 
It will be interesting to perform the perturbative computation and check whether such radiative corrections are relevant or not.  

Due to the exact chiral structure, we can separately couple matter multiplets belonging to different representations of $G$ 
to different chiral sectors of the model. 
As a example, we have discussed matters of the $\det^q$-representation, and have applied it to 
formulate gauged linear sigma models on the lattice. 

Two-dimensional $\cN =(2,2)$ SQCD models with various superpotentials have been analytically investigated based on 
the effective twisted superpotentials~\cite{witten2,hori-tong,hanany-hori}. 
The number of the vacua or the Witten index of the models has been computed for various $N, n_{\pm}$, 
and an analog of the Seiberg duality in four dimensions has been discussed based on the symmetry of Grassmannians. 
Some insights have been obtained with respect to the property of sigma models on Calabi-Yau manifolds via correspondence 
between gauged linear sigma models and nonlinear sigma models. 
It will be worth confirming those properties and exploring new aspects, which are not yet investigated there,  
from the first principle computation using this lattice formulation. 
 
Actually, numerical investigation of the two-dimensional lattice SYM models has been developed 
in~\cite{catterall3,suzuki,kanamori-ss,kanamori-suzuki}. 
In particular, ref.~\cite{kanamori-ss} discusses the numerical method to observe dynamical supersymmetry breaking for 
general supersymmetric lattice models, which exactly preserve at least one supercharge. 
Also, ref.~\cite{kanamori-suzuki} numerically shows restoration of full supersymmetry in the continuum limit 
of the two-dimensional lattice SYM model~\cite{sugino2}, which preserves one supercharge at the lattice level.      
It is intriguing to extend these methods to the present SQCD systems and explore the above-mentioned physical aspects. 

Use of the overlap Dirac operator yields another possibility of the FI and $\vartheta$-terms (\ref{FI_theta_lat}).  
Thanks to the Ginsparg-Wilson relation, $\Tr\left(\gamma_3 a\hD\right)$ is a topological quantity 
on the lattice~\cite{luscher2,Narayanan:1993ss,Hasenfratz:1998ri,Luscher:1998kn}
and it follows that 
\be
\delta \,\Tr\,\left(\gamma_3 a\hD\right) =\sum_x \, \delta \,\tr\,\left[\tr_{\rm spin}\left(\gamma_3 a\hD\right)(x,x) \right] =0
\ee
for any variations of the admissible gauge field\footnote{It can be quickly seen from 
$
\delta \,\Tr \, \left(\bar{P}_\pm \right) =\delta \,\Tr \, \left(\bar{P}_\pm^2 \right) 
        = 2\,\Tr \, \left(\bar{P}_\pm \delta\bar{P}_\pm \right) = 0
$
by $\bar{P}_\pm=\bar{P}_\pm^2$ and by the identity analogous to (\ref{identity_bar}).}.  
Then, 
instead of (\ref{FI_theta_lat}), we may use 
\be
S^{\rm LAT}_{{\rm FI}, \vartheta \, (\hD)} \equiv 
Q \kappa \sum_x \tr \left(-i\chi(x)\right) -\frac{\vartheta-2\pi i\kappa}{2\pi} ia^2\sum_x \tr \, \hF_{01}(x)
\label{FI_theta_lat_hD}
\ee
with 
\be
\hF_{01}(x) \equiv  \frac{\pi}{a} \,\tr_{\rm spin} \left(\gamma_3 \hD\right) (x,x) ,    
\ee
which becomes $F_{01}$ in the continuum limit. 
Furthermore, 
if one could also replace $\whPhi(x)$ with $2a^2 \hF_{01}(x)$ in our formulation, 
the $\kappa$-dependent topological term would exactly cancel, serving an improvement. 
Interestingly, in such case, 
the gauge field action and a part of the gaugino action 
(the part $2i\chi [\cD_\mu \psi_\nu-\cD_\nu \psi_\mu]$)
would descend from the overlap 
Dirac operator.\footnote{
In \cite{Horvath:2006md,Horvath:2006az}, Horv\'{a}th proposed 
a gauge field action defined with the scalar part of  the overlap Dirac operator $\Tr \{\hD\}$ and 
gave a ``coherent'' formulation of lattice QCD. Classical continuum limit of the scalar part was examined 
in detail~\cite{Alexandru:2008fu} and numerical algorithm for the formulation was discussed~\cite{Liu:2006wa}.  
In~\cite{Liu:2007hq, Liu:2006wa}, the possiblity to define $F_{\mu\nu}(x)$ from  the tensor 
part of the overlap Dirac operator $\tr_{\rm spin} \{\sigma_{\mu\nu} \hD\}(x,x)$ was also considered and classical 
continuum limit of the  tensor part was examined in detail. 
Our attempt here is in the two-dimensional version of the second approach.} 
As explained in appendix~\ref{app:tensor}, however, the latter attempt fails because of doubler modes appearing 
in the gauge or gaugino kinetic terms.  

The ${\rm U}(1)_V$ R-symmetry, 
which originates from the ${\rm U}(1)_R$ symmetry of the four-dimensional SYM, is not preserved
in the present formulation, as discussed in section~\ref{sec:lat_sqcd_hD}.
Since $Q = -(Q_L+\bar Q_R)/\sqrt{2}$ is a sum of the two supercharges of the 
${\rm U}(1)_V$ charges $\pm 1$,  
it transforms non-trivially under the ${\rm U}(1)_V$ and yields another supercharge, 
$Q'\equiv (Q_L-\bar Q_R)/\sqrt{2}$, which is not preserved here. 
Therefore, it seems highly nontrivial
to preserve both the $Q$-supersymmetry and the ${\rm U}(1)_V$ R-symmetry. 
In fact, by ${\rm U}(1)_V$ R-symmetry, two parts of the gaugino action 
$2i\chi [\cD_\mu \psi_\nu-\cD_\nu \psi_\mu]$ and 
$i\eta [ \cD_\mu \psi_\mu ]$ should interrelate, while,  in the present formulation, 
the plaquette structure of 
$i\chi(x) Q\whPhi(x)$ seems totally different from the lattice-difference structure of 
$i\eta(x) \nabla_\mu^\ast \psi^{}_\mu(x)$.\footnote{$\nabla_\mu^*$ ($\nabla_\mu$) 
represent the backward (forward) covariant differences for adjoint field variables.}

%

Once the exact  $Q$-supersymmetry is given up, it is possible to preserve ${\rm U}(1)_V$, as 
discussed in~\cite{suzuki-taniguchi}. In our case with the non-compact bosonic field variables 
$\phi(x)$ and $\bar \phi(x)$, the action may be given by 
\bea
S^{\rm LAT}_{\rm 2DSYM} & = & \frac{1}{g_0^2} \sum_x\tr\left[
\frac{1}{4} \whPhi(x)^2 - D(x)^2  + \frac14 [\phi(x), \bar{\phi}(x)]^2 \right.\nn \\
&& \hspace{1.5cm}
 + 2 \left(\chi(x), \, \frac12 \eta(x)\right)  \sigma_2 a \widetilde D
 \left( \begin{array}{c} \psi_0(x) \\ \psi_1(x) \end{array} \right)
\nn\\
 & & \hspace{1.5cm} \left. 
- \chi(x) [ \phi(x), \chi(x) ] - \frac14 \eta(x) [ \phi(x), \eta(x) ]  
- 2 \sum_{\mu=0}^1\tilde\psi_\mu(x) \tilde\psi_\mu(x) \bar{\phi}(x)
 \right], \nn\\
\eea
where $\widetilde D$ is the overlap Dirac operator for adjoint field variables  
in Majorana representation ($\tilde \gamma_0=\sigma_3, \tilde \gamma_1=\sigma_1, \tilde \gamma_3=\sigma_2$). 
$\tilde \psi_\mu(x)$ are defined by 
\begin{equation}
 \left( \begin{array}{c} \tilde\psi_0(x) \\ \tilde\psi_1(x) \end{array} \right)
 \equiv \left(1- \frac12 a\widetilde D\right)  \left( \begin{array}{c} \psi_0(x) \\ \psi_1(x) \end{array} \right). 
\end{equation}
This action is invariant under the transformation:
\begin{eqnarray}
\left(\begin{array}{c} \psi_0(x) \\ \psi_1(x)\end{array}\right) \to 
          \e^{i\alpha\widehat{\sigma}_2}\left(\begin{array}{c} \psi_0(x) \\ \psi_1(x)\end{array}\right), 
 \quad
 \left(\chi (x), \, \frac12\eta (x)\right)
 \rightarrow  \left( \chi (x), \, \frac12\eta (x)\right) \e^{i\alpha\sigma_2} 
\label{U(1)_V_lat_SYM} 
\end{eqnarray}
with $\widehat{\sigma}_2 \equiv \sigma_2 (1-a\wt{D})$. 
Note $\tilde{\psi}_\mu(x)$ then change as 
\be
\left(\begin{array}{c} \tilde{\psi}_0(x) \\ \tilde{\psi}_1(x)\end{array}\right) \to 
          \e^{i\alpha\sigma_2}\left(\begin{array}{c} \tilde{\psi}_0(x) \\ \tilde{\psi}_1(x)\end{array}\right). 
\ee          
The matter-part action, which is invariant under (\ref{U(1)_V_lat_SYM}), (\ref{U(1)_V_lat_1}) and (\ref{U(1)_V_lat_2}), 
 may be given by eqs.~(\ref{S_mat+twm_hD_a2}) 
and (\ref{S_mat-twm_hD_a2}) with 
$Q(a \hD)$ replaced by  $i \gamma_\mu \tilde \psi_\mu(x)$ and 
the lattice artifact terms omitted.

\vspace{1cm}
\noindent
{\bf Acknowledgments}\\
The authors thank the Yukawa Institute for Theoretical Physics at Kyoto University, 
where this work was initiated during the YITP-W-08-04 
on ``Development of Quantum Field Theory and String Theory''.

\vspace{1cm}

\appendix
\section{Cancellation of the ``Pauli terms''}
\label{app:pauli}
\setcounter{equation}{0}
In this appendix, we show that the boson kinetic kernel $\hD^\dagger\hD$ in (\ref{S_mat+twm_hD2}) and (\ref{S_mat-twm_hD2}) 
cancels with the ``Pauli terms'' containing $\whPhi(x)$ and leaves the covariant Laplacian in the continuum limit. 
This is analogous to the continuum case, because 
\be
\Slash{\cD}^\dagger\Slash{\cD} = -\cD_\mu\cD_\mu + \gamma_3 F_{01} 
\ee
and $\whPhi(x) \simeq 2a^2F_{01}(x)$ ($a\to 0$). 

We assume for any variable $\varphi(x)$ belonging to the fundamental representation 
\be
\sum_y \, {\rm tr}_{\rm spin}\left\{\gamma_3 a^2 \hD^\dagger \hD  \right\}(x,y)\varphi(y) 
= Y(x) \,  \varphi(x) + R(x), 
\label{Y_R}
\ee
where $R(x)$ gives higher derivative terms of $\varphi(x)$ in the continuum limit.  
The suffix ``spin'' in the trace means taking the trace over the Dirac indices. From the continuum result, 
we expect $Y(x) \simeq 2a^2 F_{01}(x)$ $(a\to 0)$. 

Let us compute $Y(x)$. 
Since $\hD^\dagger = \gamma_3 \hD \gamma_3$, from the Ginsparg-Wilson relation (\ref{GW_relation}) 
\be
\tr_{\rm spin} \left(\gamma_3 a^2\hD^\dagger \hD\right) = 2 \, \tr_{\rm spin} \left(\gamma_3 a\hD\right). 
\ee
Using the integral representation of $\hD$ 
\bea
a\hD & = &  1-X\int^\infty_{-\infty} \frac{\dd t}{\pi} \frac{1}{t^2 +X^\dagger X} \nn \\
        & = & 1+ \int^\infty_{-\infty} \frac{\dd t}{\pi} \, \gamma_3 \frac{1}{i\gamma_3 t -1+aD_W} \gamma_3
\label{hD_int}        
\eea
and $D_W$ in terms of the plane wave basis
\bea
aD_W(x,y) &=&  \int_{-\pi}^\pi \frac{\dd^2 k }{(2\pi)^2} \,  \e^{i k \cdot (x-y)} \left(a\wt{D}_W(k) +\Delta_k(x)\right), \\
a\wt{D}_W(k) & \equiv & \sum_{\mu=0}^1 \left[\frac{\gamma_\mu -1}{2} \left(\e^{ik_\mu}-1\right) 
                                                       + \frac{\gamma_\mu +1}{2} \left(1- \e^{-ik_\mu}\right)  \right], \nn \\
\Delta_k (x) & \equiv &
 \sum_{\mu=0}^1 \left[ \frac{\gamma_\mu -1}{2} \, \e^{i k_\mu }\left( U_\mu(x) \, \e^{\partial_\mu} -1 \right) 
                           +\frac{\gamma_\mu +1}{2} \, \e^{-i k_\mu } \left(1-\e^{-\partial_\mu} U_\mu(x)^{-1} \right) \right] \nn
\eea
($\e^{\pm \der_\mu}$ are forward, backward displacement operators by $\hat{\mu}$),  
\bea
(\mbox{l.h.s. of (\ref{Y_R})}) & = & 2\int^\infty_{-\infty} \frac{\dd t}{\pi} \int_{-\pi}^\pi \frac{\dd^2 k }{(2\pi)^2} \, \e^{i k \cdot x}
\, \wt{\varphi}(k) \nn \\
 & & \hspace{2cm} \times \tr_{\rm spin} \left(\frac{1}{i\gamma_3 t -1 +a\wt{D}_W(k) +\Delta_k(x)}\, \gamma_3\right), 
\eea
where $\e^{\pm \der_\mu}$ in $\Delta_k (x)$ act only on the link variables $U_\mu(x)$, $U_\mu(x)^{-1}$. 
Because $k_\mu$'s in $\tr_{\rm spin}(\cdots)$ become $x$-derivatives of $\varphi(x)$ in the continuum limit, 
$Y(x)$ is obtained from $\tr_{\rm spin}(\cdots)$ set to $k=0$, i.e. 
\be
Y(x) = 2\int^\infty_{-\infty} \frac{\dd t}{\pi}\, \tr_{\rm spin} \left(\frac{1}{i\gamma_3 t -1 +\Delta_0(x)}\, \gamma_3\right). 
\ee
In turn, $Y(x)$ can also be written as 
\be
Y(x) = 2 \sum_y \tr_{\rm spin} \left(\gamma_3 a\hD\right) (x,y).
\label{Y_noncov}
\ee
Note that this is not a covariant expression\footnote{A possible way to make $Y(x)$ gauge-covariant 
is to fix the gauge field $U_\mu$ in the complete axial gauge with respect to the point $x$.}, 
different from the covariant quantity leading to the chiral anomaly (\ref{tr_spin_anomaly}) : 
\be
\tr_{\rm spin}\left(\gamma_3 a\hD\right) (x,x) \simeq \frac{1}{\pi} a^2 F_{01}(x) \qquad (a\to 0). 
\ee
As we see shortly, however, (\ref{Y_noncov}) converges to a covariant quantity in the continuum limit. 

Since $\Delta_0(x) = a\Slash{\cD} + \cO(a^2)$ ($a\to 0$), 
\bea
Y(x) & \simeq & 2\int^\infty_{-\infty} \frac{\dd t}{\pi}\, \tr_{\rm spin}
\left[\left(\frac{1}{i\gamma_3 t-1} a\Slash{\cD} \right)^2 \frac{1}{i\gamma_3 t-1} \gamma_3\right] \nn \\
 & = & 2a^2 F_{01}(x) \qquad (a\to 0),  
\eea
which coincides with our expectation.

\section{Admissibility Condition for $\det^q$-matters}
\label{app:adm_det}
\setcounter{equation}{0}
In this appendix, we consider the admissibility condition for the overlap Dirac operator $\hD$ 
to be well-defined for matters in the $\det^q$-representation of $G={\rm U}(N)$. 
We focus on the case $q_{\bA}\in {\bf Z}_{\neq 0}$ for $\bA =1, \cdots, \ell_0$. 

First, when 
\be
||1-U_{01}(x)|| < \epsilon, 
\label{adm_app}
\ee
let us evaluate $|1-(\det U_{01}(x))^{q_{\bA}}|$. 
We diagonalize $U_{01}(x)$ as 
\be
U_{01}(x) = \Omega(x) \left(\begin{array}{ccc} \e^{i\theta_1(x)} &  &   \\   & \ddots &   \\    &   & \e^{i\theta_N(x)} \end{array} \right) 
                 \Omega(x)^\dagger \qquad (\Omega(x) \in {\rm SU}(N))
\ee
with $-\pi \leq \theta_j(x) < \pi$, to rewrite (\ref{adm_app}) as 
\be
\frac{\epsilon}{2} > \left[ \frac{1}{N}\sum_{j=1}^N \sin^2\left(\frac{\theta_j(x)}{2}\right)\right]^{1/2} . 
\label{adm_app2}
\ee
{}From $ |\sin \frac{\theta}{2}| \geq |\frac{2}{\pi}\frac{\theta}{2}|$ for $-\pi \leq \theta < \pi$, 
\be
(\mbox{r.h.s. of (\ref{adm_app2})}) \geq \left[\frac{1}{N}\sum_{j=1}^N \left(\frac{2}{\pi}\frac{\theta_j(x)}{2}\right)^2\right]^{1/2}. 
\ee
Further, using the Schwarz inequality 
\be
\left[N\sum_{j=1}^N \left(\frac{\theta_j(x)}{2}\right)^2\right]^{1/2}  \geq \left|\sum_{j=1}^N \frac{\theta_j(x)}{2}\right| 
\ee
and $ |x| \geq |\sin x| $ for $\forall x\in {\bf R}$, we have 
\be
\frac{\pi}{4} \epsilon N  > \left|\sin\left(\sum_{j=1}^N \frac{\theta_j(x)}{2}\right)\right|. 
\ee
Then, 
\bea
\left|1-(\det U_{01}(x))^{q_{\bA}}\right| & \leq & |q_{\bA}| |1-\det U_{01}(x)|  
                          = 2|q_{\bA}| \left|\sin\left(\sum_{j=1}^N \frac{\theta_j(x)}{2}\right)\right| \nn \\
                          & < & \frac{\pi}{2}\epsilon N  |q_{\bA}|  \, (\equiv \epsilon'_{\bA}). 
\eea

{}From a similar argument to the case of the (anti-)fundamental matters, we obtain the bound 
\be
||X^\dagger X|| \geq 1-5\epsilon'_{\bA}
\ee
for matters of the charge $q_{\bA}$. For $\hD$ to be well-defined, we require $0<\epsilon'_{\bA}<\frac15$, that is 
\be
0 < \epsilon < \frac{2}{5\pi} \frac{1}{N |q_{\bA}|}. 
\ee
Therefore, it is sufficient to choose 
\be
0 < \epsilon < \frac{1}{8Nq} \qquad \mbox{with} \qquad q \equiv \max_{\bA =1, \cdots, \ell_0}(|q_{\bA}|). 
\label{adm_det_app}
\ee

\section{An Attempt to Introduce $\hD$ to SYM Sector}
\label{app:tensor}
\setcounter{equation}{0}
If the replacement of $\whPhi(x)$ with $2a^2 \hF_{01}(x)$ was allowed in our lattice formulation, 
we could improve the FI and $\vartheta$-terms (\ref{FI_theta_lat_hD}) 
so that the $\kappa$-dependent topological terms exactly cancel.  
We here examine the possibility of the replacement. The result is negative. 

In the integral representation used in (\ref{hD_int}) 
\be
\hF_{01}(x) = \frac{\pi}{a^2}\int^\infty_{-\infty}\frac{\dd t}{\pi} \int^\pi_{-\pi}\frac{\dd^2 k}{(2\pi)^2} \, \tr_{\rm spin} 
\left(\frac{1}{i\gamma_3 t -1 +a\wt{D}_W(k) +\Delta_k(x)}\, \gamma_3\right),
\ee
we expand the link variables around 1 as $U_\mu(x) = 1+ iaA_\mu(x) + \cdots$. 
Then, we have 
\bea
\Delta_k(x) & = & \left.\Delta_k(x)\right|_{U=1} +\delta\Delta_k(x) +\cO(A^2), \nn \\
\delta\Delta_k(x) &\equiv &  ia \sum_{\mu=0}^1\left[\frac{\gamma_\mu-1}{2}\, \e^{ik_\mu} A_\mu(x) \, \e^{\der_\mu} 
                 +\frac{\gamma_\mu+1}{2} \, \e^{-ik_\mu}A_\mu(x-\hat{\mu}) \, \e^{-\der_\mu}\right], 
\eea
and 
\bea
\hF_{01}(x) & = & -\frac{\pi}{a^2}\int^\infty_{-\infty}\frac{\dd t}{\pi} \int^\pi_{-\pi}\frac{\dd^2 k}{(2\pi)^2} \, \tr_{\rm spin} 
\left[\frac{1}{i\gamma_3 t -1 +a\wt{D}_W(k) +\left.\Delta_k(x)\right|_{U=1}} \right. \nn \\
  & & \left. \hspace{3.7cm} \times \delta\Delta_k(x) \, \frac{1}{i\gamma_3 t -1 +a\wt{D}_W(k)}\, \gamma_3\right] +\cO(A^2) \nn \\
  & = & \int^\pi_{-\pi}\frac{\dd^2\ell}{(2\pi)^2} \, \e^{i\ell\cdot x} \sum_{\mu=0}^1 f_\mu(\ell) \wt{A}_\mu(\ell) +\cO(A^2), 
\eea
where the ``form factors'' $f_\mu(\ell)$ are expressed as 
\bea
f_\mu(\ell) & \equiv & -i\frac{2\pi}{a} \, \e^{-i\frac12 \ell_\mu} 
                  \int^\infty_{-\infty}\frac{\dd t}{\pi} \int^\pi_{-\pi}\frac{\dd^2 k}{(2\pi)^2} \,
                \frac{1}{t^2 +\left|-1+a\wt{D}_W(k)\right|^2} \, \frac{1}{t^2 +\left|-1+a\wt{D}_W(k+\ell)\right|^2} \nn \\
                & & \hspace{4cm} \times\left\{\left(1-\sum_\rho \cos(k_\rho+\ell_\rho) \right)
                          \cos\left( k_\mu + \frac{\ell_\mu}{2} \right) \sum_\nu \epsilon_{\mu\nu} \sin k_\nu  \right. \nn \\
                & & \hspace{4.5cm} -\left(1-\sum_\rho \cos k_\rho \right) \cos\left( k_\mu+\frac{\ell_\mu}{2} \right) 
                              \sum_\nu \epsilon_{\mu\nu} \sin (k_\nu +\ell_\nu) \nn \\
                & & \hspace{4.5cm} \left. -\sin \left( k_\mu+\frac{\ell_\mu}{2} \right) \sum_{\rho, \nu} 
                           \epsilon_{\rho\nu} \sin (k_\rho +\ell_\rho) \sin k_\nu \right\}
\eea        
with 
\be
 \left|-1+a\wt{D}_W(k)\right|^2 = \sum_{\mu=0}^1\left(\sin k_\mu \right)^2 + \left(1-\sum_{\mu=0}^1 \cos k_\mu \right)^2. 
\ee 
Note that $\left|-1+a\wt{D}_W(k)\right|^2$ and $\left|-1+a\wt{D}_W(k+\ell)\right|^2$ are strictly positive. 
It can be seen that $f_\mu(\ell)$ vanish at $(\ell_0, \ell_1) = (0,0), (0,\pi), (\pi,0), (\pi,\pi)$, 
meaning that $\hF_{01}(x)$ or $Q\hF_{01}(x)$ appearing in the gauge or gaugino kinetic terms contains doubler modes. 
Thus, it is not allowed to replace $\whPhi(x)$ with $\hF_{01}(x)$. 

A similar argument shows that the replacement of $\whPhi(x)$ with $Y(x)$ of (\ref{Y_noncov}) is not allowed either.


\end{document}